\documentclass[preprint,12pt,authoryear]{elsarticle}

\usepackage{amsthm}
\usepackage{amssymb}
\usepackage{amsmath}
\usepackage{amsmath}
\usepackage{subfigure}
\usepackage{url}
\usepackage{multirow} 
\usepackage{booktabs} 
\usepackage{amsthm}
\usepackage{xcolor}
\newtheorem{definition}{Definition}
\usepackage{siunitx}
\usepackage{booktabs}
\usepackage{siunitx}
\usepackage{caption} 
\usepackage{bibunits}

\journal{Elsevier}

\begin{document}

\begin{frontmatter}



\title{HOI-Brain: a novel multi-channel transformers framework for brain disorder diagnosis by accurately extracting signed higher-order interactions from fMRI data} 


\author[label1]{Dengyi Zhao}
\ead{zhaodengyi@mail.sdu.edu.cn}

\author[label1,label2]{Zhiheng Zhou}
\ead{zhouzhiheng@amss.ac.cn}

\author[label2]{Guiying Yan}
\ead{yangy@amt.ac.cn}

\author[label3]{Dongxiao Yu}
\ead{dxyu@sdu.edu.cn}

\author[label1]{Xingqin Qi\corref{cor1}}
\ead{qixingqin@sdu.edu.cn}

\cortext[cor1]{Corresponding author.}


\affiliation[label1]{organization={School of Mathematics and Statistics, Shandong~University}, 
            city={Weihai},
            postcode={264209}, 
            state={Shandong},
            country={China}}
\affiliation[label2]{organization={Academy of Mathematics and Systems Science, Chinese Academy of Sciences}, 
            city={Beijing},
            postcode={100190}, 
            state={Beijing},
            country={China}}      
\affiliation[label3]{organization={School of Computer Science and Technology, Shandong~University}, 
            city={Qingdao},
            postcode={266000}, 
            state={Shandong},
            country={China}}  

\begin{abstract}
Accurately characterizing the higher-order interactions of brain regions and effectively extracting the interpretable higher-order organizational patterns from functional Magnetic Resonance Imaging (fMRI) data are crucial for the diagnosis of brain diseases. However, current graph models mainly focus on pairwise patterns, as well as triadic patterns within the brain while overlooking more higher-order patterns with signs, limiting an integrated understanding of brain-wide communication. To address these challenges, we propose HOI-Brain (Higher-Order Interaction in Brain Network), a novel computational framework that enables the utilization of signed higher-order interactions and signed organizational patterns in fMRI data for the diagnosis of brain diseases. Specifically, we present a new calculation of co-fluctuations based on Multiplication of Temporal Derivatives to detect higher-order interactions with adequate temporal resolution. Next, we further distinguish positively and negatively synergistic higher-order interactions and encode them in the monotonic weighted simplicial complexes, which can offer detailed insights into the communication within the brain. Moreover, we employ Persistent Homology in the monotonic weighted simplicial complexes of the brain to extract signed higher-dimensional neural organizations from a spatiotemporal perspective. Finally, a multi-channel transformers architecture is proposed to holistically integrate information from heterogeneous topological features. Comprehensive experiments across Alzheimer’s disease (AD), Parkinson’s disease (PD), and Autism Spectrum Disorder (ASD) datasets demonstrate the superiority, effectiveness, and interpretability of our framework. These key regions and higher-order patterns align with existing evidence: concordant HOIs decrease in AD but increase in PD and ASD, providing mechanistic clues to disorder-specific network changes.

\end{abstract}
\begin{graphicalabstract}
\begin{figure}[h]
    \centering
    \includegraphics[scale=0.40]{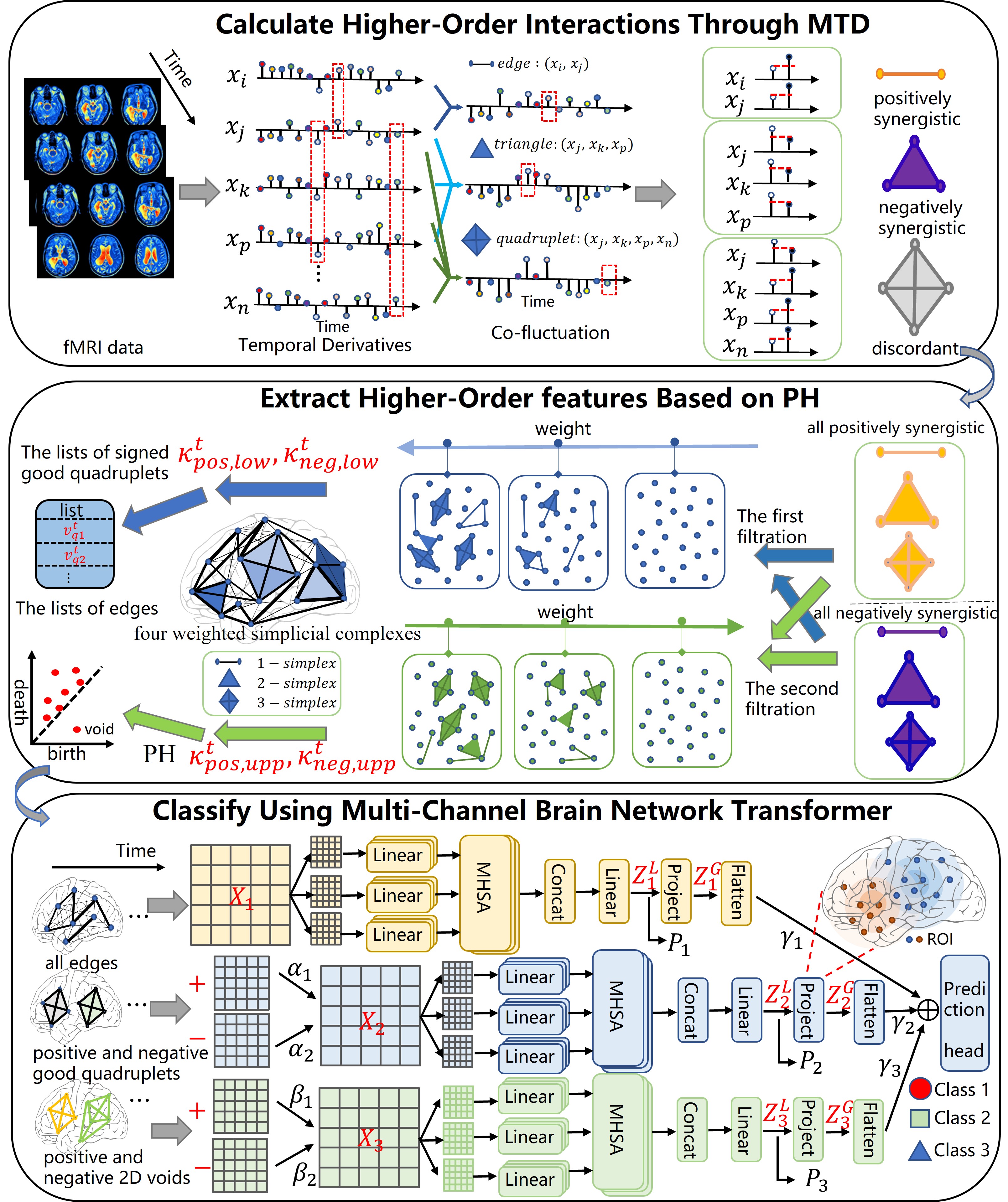}
\end{figure}
\end{graphicalabstract}

\begin{highlights}
\item Introduce a novel metric (MTD) capturing higher-order interactions.

\item Distinguish positive and negative synergistic higher-order interactions in the brain. 

\item Define monotonic weighted simplicial complexes to extract higher-order features.

\item Propose a multi-channel transformers architecture integrating heterogeneous topological features.

\item 
Identify HOI patterns: concordant HOIs decline in AD but increase in PD/ASD.
\end{highlights}


\end{frontmatter}



\section{Introduction}
\label{sec1}
Functional magnetic resonance imaging (fMRI) represents a pivotal instrument in the field of neuroscience, serving to identify potential neuroimaging biomarkers. These biomarkers are instrumental in the automated diagnosis of various brain disorders, including Alzheimer's disease (AD), autism spectrum disorder (ASD), and Parkinson's disease (PD) \citep{1,2}. Specifically, resting state fMRI measures functional connectivity between brain regions via blood-oxygen-level-dependent (BOLD) signals. This process naturally models the brain as a network, defining regions of interest (ROIs) as nodes and functional connections between ROIs as edges \citep{3,4}. Abnormalities in this network at the whole-brain level are hypothesised to yield disease-related biomarker signatures \citep{5}.

A significant number of graph data mining methodologies are then applied to such brain networks for brain disorder diagnosis, with the goal of understanding and analysing the elements and interactions of neurological systems from a network perspective. Examples of such models include graph neural network (GNN)-based models \citep{6}, brain Transformer-based models \citep{20}, hypergraph neural network (HGNN)-based models \citep{8}, and Persistent Homology (PH)-based models \citep{9}. Despite the evidence from these studies demonstrating the considerable potential of network-based approaches to elucidating the complexities of the brain and diagnosing brain disorders, graph analysis of brain networks remains an emerging field in its infancy \citep{10}.

In recent developments, graph neural network (GNN) techniques have been employed to analyse brain networks characterised by pairwise interactions. The utilisation of these methodologies enables the extraction of potential topological features, which are instrumental in the diagnosis of brain disorders \citep{101,16}. A number of GNN-based models have been proposed for brain networks, including GroupINN \citep{11}, BrainGNN \citep{12}, FBNetGen \citep{13}, BPI-GNN \citep{14}, and ASD-HNet \citep{15}. These models have demonstrated the potential to achieve favourable diagnostic performance. It is widely accepted that GNN-based models utilise a message passing mechanism, whereby the embedding of a brain ROI is updated by aggregating information from its neighbouring ROIs. This process facilitates the learning of a discriminative graph-level representation of the brain connectivity network. However, these methods are limited by the underlying assumption that interactions between nodes are strictly two-way. Recent studies have indicated a growing body of evidence that suggests the complexity of brain-region interactions extends far beyond the established pairwise connections. There is now a growing consensus that widespread co-fluctuations occur in groups of nodes that evolve over time \citep{17,18}. Consequently, the limited expressiveness of traditional GNNs prevents them from capturing higher-order interactions (HOIs) in brain networks, and the topological features they extract are typically regarded as lower-order.

Transformer-based brain network models have achieved considerable success in the diagnosis of brain disorders, due to their capacity to capture global patterns. Representative examples include Graph Transformer \citep{19}, Brain Network Transformer \citep{20}, Transformer and Snowball Encoding Networks (TSEN) \citep{21}, and Long-range Brain Transformer \citep{22}. It is evident that these models naturally construct fully connected graphs and, through a powerful global attention mechanism, adaptively learn pairwise interaction relationships for brain disorder diagnosis. In practice, each ROI is treated as a token, and self-attention learns data-driven weights over all ROI pairs, enabling the model to capture long-range dependencies without relying on local neighbourhood aggregation. By stacking attention layers and pooling the resulting ROI embeddings, Transformer-based models produce a discriminative brain-level representation for classification. However, the prevailing approach involves the utilisation of either the raw time series features of brain regions or the functional connectivity matrix as node-level input, thereby overlooking significant higher-order topological information among ROIs.

A number of studies have concentrated on the representation of HOIs in brain networks using more sophisticated models, with a particular focus on hypergraphs. This approach has been shown to enhance diagnostic performance for brain diseases, as evidenced by citations in the literature \citep{23,24,25,26,8}. Concretely, hypergraph methods define hyperedges that connect one node to a set of nodes (i.e., a group), and then perform incidence-based message passing or hypergraph convolution to aggregate information among nodes within the same hyperedge, thereby extending pairwise edges to group relations. Despite the encouraging outcomes demonstrated by HGNN-based models in characterising HOIs through hyperedges generated via \textit{k}-nearest neighbour or \textit{k}-hop neighbourhoods, this node-centric construction scheme appears to be incompatible with the concept of group dependence. This is due to its inability to capture authentic simultaneous interactions among groups of ROIs. Furthermore, it has been demonstrated that higher-order features extracted by HGNN-based models can obscure their relationship to specific topological or neurobiological phenomena \citep{27,28}.

In order to extract higher-order topological features that are more readily interpretable for the purpose of brain-disorder diagnosis, recent studies have adopted Persistent Homology, a topological approach capable of reconstructing HOI structures and delivering state-of-the-art performance in characterising brain topological profiles \citep{29,30}. Briefly, persistent homology builds a filtration of simplicial complexes from weighted brain networks and tracks the birth and death of topological features across thresholds, yielding persistence summaries (e.g., persistence diagrams) that are robust to noise and can be coupled with downstream classifiers. However, the majority of investigations into higher-order structures in brain networks have focused on 0-dimensional (connected components) and 1-dimensional (cycles) topological profiles formed by nodes and edges. These dimensions do not link HOIs in the brain to higher-order organizations that would provide a truly higher-dimensional perspective \citep{31,32,33}. In order to address the aforementioned limitation, researchers have recently employed synchronisation phenomena to construct more accurate simplicial complexes for modelling HOIs. Furthermore, they have utilised Persistent Homology to capture 1-dimensional cycles that reflect triplet interactions \citep{34}. Comprehensive analyses demonstrate that methods incorporating inferred HOIs among three ROIs outperform traditional pairwise approaches, thereby offering new insights into the higher-order organisation of fMRI time-series data.
Nonetheless, the extant evidence indicates that triplet interactions can be decomposed into linear combinations of pairwise interactions, provided that said interactions are linearly decomposable. This finding suggests that some triplets may not in fact represent genuine higher-order phenomena \citep{35,36}. Furthermore, extant measures of instantaneous co-fluctuation are contingent on extended Pearson correlation, a method that is deficient in temporal resolution when it comes to detecting spatiotemporal interactions among groups of regions \citep{37}. Finally, the role of signed HOIs, which have the potential to provide valuable diagnostic information in cases of neurological conditions, has been largely overlooked in these studies.

\begin{figure}
    \centering
    \includegraphics[scale=0.4]{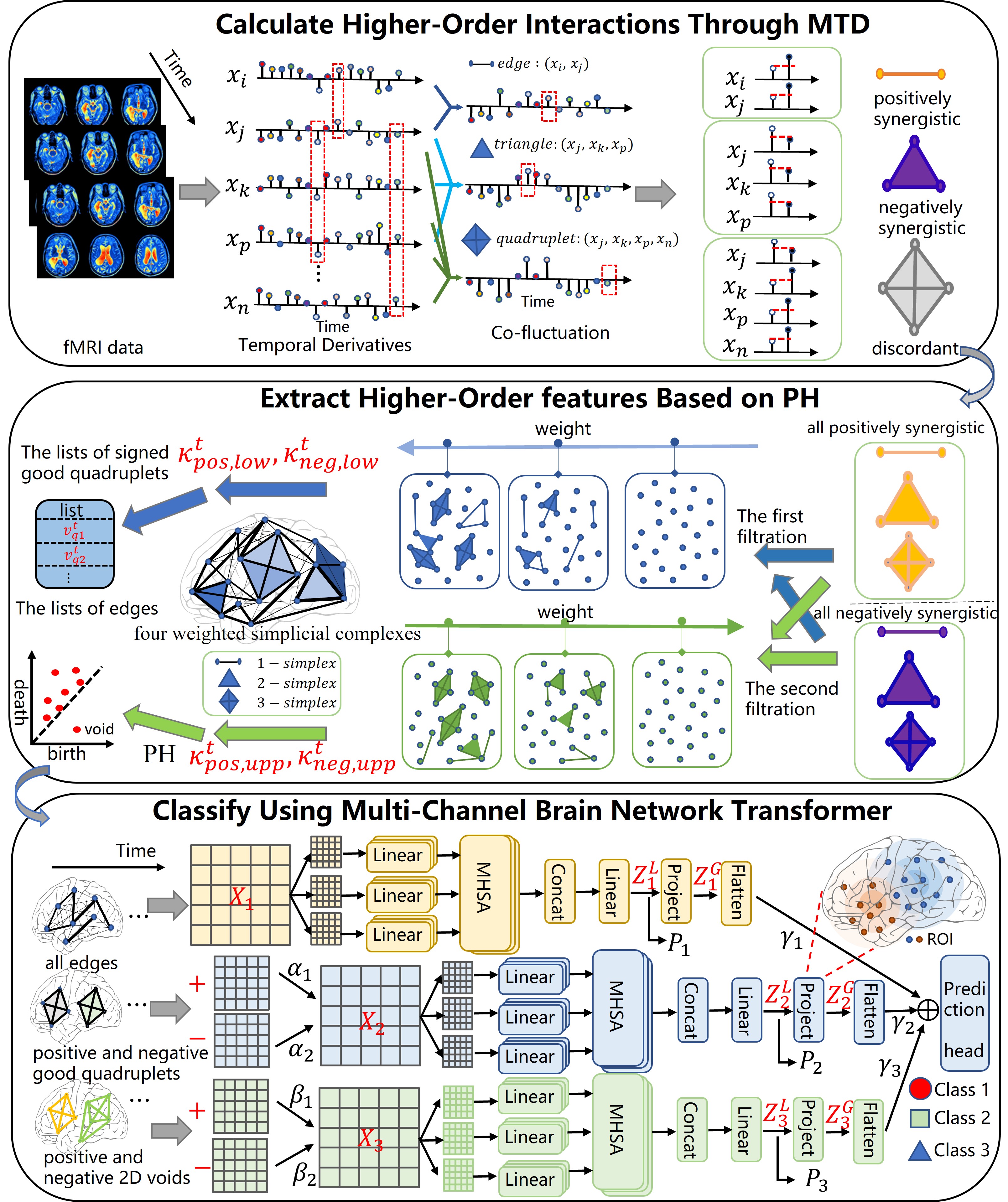}
    \caption{Overall framework of HOI-Brain. Individual fMRI data are transformed into \( N \) original fMRI signals through a preprocessing pipeline. A novel metric - Multiplication of Temporal Derivatives (MTD) - is used to quantify dynamic functional co-fluctuations of group ROIs. Then, at each timepoint \textit{t}, some instantaneous \textit{k}-order co-fluctuations  are encoded into four signed monotonic weighted simplicial complexes based on filtrations, respectively. Persistent Homology is applied to analyze these signed monotonic weighted simplicial complexes at each \textit{t}. Five feature matrices are generated by extracting all edges, signed good quadruplets, and signed 2D voids. These matrices are temporally averaged across all timepoints to stabilize feature representations. By incorporating the lower-order and higher-order features,  a multi-channel brain network Transformer is used to produce the final classification of the disease status (e.g., Classes 1, 2 and 3 correspond to CN, MCI and AD in ADNI).
}
    \label{fig1}
\end{figure}

Therefore, these limitations motivate the need for (i) a time-resolved measure that captures genuine groupwise co-fluctuations beyond pairwise correlations, (ii) an explicit and interpretable representation that links signed higher-order interactions to higher-dimensional organisation, and (iii) a principled way to fuse heterogeneous lower- and higher-order cues within a single predictive model. 
To this end, we propose HOI-Brain (Fig.~\ref{fig1}), a unified framework that quantifies instantaneous \(k\)-node co-fluctuations using the Multiplication of Temporal Derivatives (MTD), encodes them as signed monotonic weighted simplicial complexes from which quadruplet signatures and 2D void descriptors are extracted via persistent homology, and finally integrates these higher-order topological descriptors together with lower-order edge features through a multi-channel brain network Transformer for end-to-end classification.
Overall, this study makes five main contributions as follows.

\begin{enumerate}
\item   We introduce a novel metric - Multiplication of Temporal Derivatives (MTD) - for quantifying dynamic functional co-fluctuations of group ROIs. This approach employs element-wise products of temporal derivatives of blood oxygen level-dependent (BOLD) signals to represent instantaneous co-fluctuation magnitudes of \textit{k-}node interactions, which can achieve reliable identification of genuine higher-order neural interactions compared to Pearson correlation.

\item For the first time, we define the monotonic weighted simplicial complex for brain networks, extracting quadruplet-level interaction signatures and two-dimensional void descriptors via persistent homology. This provides a higher-dimensional perspective for studying higher-order brain structures compared to one-dimensional cycles.

\item To the best of our knowledge, this is the first attempt to distinguish between positively and negatively synergistic HOIs, which may offer valuable insights into the complex coordination and communication within the brain, thereby improving the effectiveness of brain disease diagnosis.

\item We propose a novel multi-channel brain network Transformer to synergistically integrate lower-order edge features with the higher-order topological invariants.

\item We identify disease-stage-dependent higher-order organizational patterns, showing gradually weakened concordant  HOIs from healthy controls to AD patients, while revealing an opposite trend in PD and ASD. 

\end{enumerate}

The contributions made in this work illustrate the potential of quadruplet or greater-order region interactions in improving the diagnostic efficacy of brain diseases and advancing the understanding of brain networks and neurodegenerative diseases.

\section{Related work}
\label{sec2}

\subsection{Graph Models Based on Deep Learning}

A significant number of graph models are then applied to brain networks for the purpose of brain disorder diagnosis, with the objective being to understand and analyse the elements and interactions of neurological systems from a network perspective. Graph neural networks (GNNs) have been demonstrated to be particularly effective in modelling brain connectomes, due to their ability to effectively capture subtle, latent representations and nonlinear relationships in graph-structured data. A number of GNN-based models, specifically designed for brain networks, have been proposed, and these have achieved a promising diagnostic performance. For instance, GroupINN \citep{11} incorporates the concept of node grouping into the neural network and designs a random-walk-based variant of graph convolutional layer. BrainGNN \citep{12} has developed a novel ROI-aware graph convolutional (Ra-GConv) layer that utilises the topological and functional information of fMRI. FBNETGEN \citep{13} employs a task-aware GNN-based framework for fMRI analysis via functional brain network generation, which generates the brain connectivity matrices and predicts clinical outcomes simultaneously from fMRI BOLD signal series. BPI-GNN \citep{14} utilises the prototype learning method to analyse fMRI. ASD-HNNet \citep{15} is a hybrid neural network model for the identification of autism spectrum disorder (ASD). This model extracts features from functional brain networks at three distinct levels: local ROI, community, and global representation. However, these methodologies are fundamentally predicated on a two-way network to model pairwise interactions and mine low-order topological features for the diagnosis of brain disorders, overlooking the influence of HOIs in the brain.

A number of studies have been conducted on the construction of HOIs in the brain with a view to enhancing the performance of brain disorder diagnosis based on hypergraphs. For instance, a novel framework is proposed to estimate the hyper-connectivity network of brain functions for the diagnosis of brain disease \citep{23}. MHL-Hypergraph \citep{24} proposes a multi-hypergraph learning-based method to compute a unified hypergraph similarity matrix from multi-paradigm fMRI data to represent an FCN for each subject. An evolving hypergraph convolutional network \citep{25} for the dynamic hyper-brain network is proposed, which adds the attention mechanism to further enhance the ability of representation learning. In order to take the temporal characteristics of longitudinal data into consideration, a weighted hypergraph convolution network (WHGCN) has been designed to utilise the internal correlations among different time points and to leverage higher-order relationships between subjects for the purpose of AD detection \citep{26}. CcSi-MHAHGEL \citep{8} pioneered a novel hypergraph convolution network framework for extracting multiatlas-based FCN embeddings, with the objective of facilitating multisite ASD identification. In this framework, hyperedge-aware HGCN was developed to capture complex higher-order information in brain networks. The majority of these studies employ \textit{k}-nearest neighbour or \textit{k}-hop neighbour constructions in order to generate hypergraphs. However, it has been demonstrated that these constructions are inconsistent with the definition of group dependence in brain regions. Moreover, higher-order features extracted using hypergraph neural networks (HGNNs) are frequently regarded as being less interpretable.

\subsection{Transformer-based brain network models}
With the development of Transformer architectures, Transformer-based methodologies are being increasingly incorporated into brain graph structures. The Graph Transformer \citep{19}, a sophisticated machine learning algorithm, can be applied to brain networks to learn the strength of the connections between ROIs across individuals. Subsequently, the Brain Network Transformer (BNT) \citep{20} employs the distinctive characteristics inherent in brain network data to maximise the efficacy of Transformer-based models for brain network analysis, thereby circumventing the necessity for time-consuming computations of eigenvalues or eigenvectors. Transformer and Snowball Encoding Networks (TSEN) \citep{21} pioneered the incorporation of snowball graph convolution as position embedding within the Transformer structure, a methodology that has been proven to be both straightforward and efficacious in the context of capturing local patterns of brain activity in a natural manner. Long-range Brain Transformer \citep{22} injects the long-range embeddings into a Transformer framework, integrating both short-range and long-range dependencies between ROIs using the self-attention mechanism. It is evident that these works disregard substantial higher-order topological information among ROIs.

\subsection{Persistent Homology on Brain Connectome}
Persistent homology (PH) is a widely utilised, effective algebraic topological tool for the analysis of the brain connectome, enabling the capture of more interpretable, higher-order topological features. In lieu of endeavouring to ascertain a solitary optimum threshold, researchers \citep{29} propose the examination of the topological changes in the brain network as the threshold is increased continuously, based on Persistent Homology. Researchers \citep{30} also highlight the importance of cliques and cavities in the human connectome, and locate topological cavities of different dimensions, around which information may flow in either diverging or converging patterns. In order to address the challenging issue of comparing functional connectivity networks (FCNs) across different spatiotemporal resolutions, researchers \citep{38} have developed a novel network comparison framework based on Persistent Homology. The purpose of this framework is to observe the change of 0-dimensional homology groups. Furthermore, researchers \citep{32} propose an adversarially trained Persistent Homology-based graph convolutional network (ATPGCN) to capture disease-specific brain connectome patterns for brain disorder diagnosis. Researchers \citep{33} utilise the persistent topological features of connected components $H_0$ and loops $H_1$ for the identification of brain disorders, which are quantified by two methods:  Vietoris-Rips filtration and graph filtration. Nevertheless, the aforementioned studies have overlooked the significance of two-dimensional topological profiles (i.e., voids) and have failed to establish a correlation between HOIs in the brain and these higher-order organizations. Recently, researchers \citep{34} have adopted certain methodologies grounded in the synchronisation phenomenon to construct more accurate simplicial complexes for modeling HOIs and use PH to capture 1-dimensional cycles related to triplet interactions. However, this can easily generate a large number of false positive HOIs and overlook higher-order quadruplet interactions with adequate temporal resolution. Moreover, the influence of signed HOIs was overlooked in these studies, which could offer a broader diagnostic view of neurological conditions.

\section{Preliminaries and Mathematical Foundations}
In this section, we first present the problem formulation of this paper. Then, we propose a general method for mining HOIs in time series data using Multiplication of Temporal Derivatives. Finally, we introduce a new definition of monotonic weighted simplicial complex. These will provide the theoretical basis for the methods of brain disease diagnosis in the following text.
\subsection{Problem Formulation}
The task of diagnosing brain disorders primarily involves inferring specific properties (represented as class labels) from fMRI data in the form of graphs. Given a labeled dataset $\mathcal{D} = \{(\mathcal{G}, \mathcal{Y})\} = \{(G_i, y_i)\}_{i=1}^N$, where each graph $G_i \in \mathcal{G}$ corresponds to a label $y_i \in \mathcal{Y}$, the goal is to extract features to train a mapping function $f_\theta: \mathcal{G} \rightarrow \mathcal{Y}$ that generalizes to unobserved graphs. For example, in studies of Alzheimer's disease using brain networks, the label space comprises three diagnostic categories: CN (Cognitively Normal), MCI (Mild Cognitive Impairment) and AD (Alzheimer's Disease). The objective is to extract both lower-order and higher-order topological features from annotated brain network data in order to learn a robust classifier $f_\theta$, ensuring its effectiveness on novel, unseen brain connectomes.

\subsection{Multiplication of Temporal Derivatives for Processing Multivariate Time Series}
\label{3.2}
To better capture HOIs in multivariate time series data with adequate temporal resolution, we propose a novel method for estimating instantaneous co-fluctuation magnitudes between several variables.

Let \( \mathbf{x}_i = [x_i(1), x_i(2), \ldots, x_i(T)] \) represent the generic time series recorded from variable \( i \).
 We first calculate the temporal derivative \( \dot{x}_i \) of each time series \( \mathbf{x}_i \) by performing a first-order differencing:
\begin{equation}
\dot{x}_i(t-1) = x_i(t) - x_i(t-1)
\end{equation}
for \( t = 2, 3, \ldots, T \). Then we normalize each data point by dividing the temporal derivative \( \dot{x}_i(t) \) by its standard deviation \( \sigma_{\dot{x}_i} \), computed over the entire time course. This normalization yields the new time series \( q_i(t) \) as:
\begin{equation}
q_i(t) = \frac{\dot{x}_i(t)}{\sigma_{\dot{x}_i}}
\end{equation}
Then the new representation vector for the \textit{i}th variable is denoted as  $\mathbf{q_i}$.

Subsequently, the generic element at time \( t \) of the \( z \)-scored \( k \)-order co-fluctuation among \( k+1 \) new time series is calculated as:
\begin{equation}
\xi_{0 \ldots k}(t) = \frac{\prod_{m=0}^{k}{q_{m}}(t) - \mu\left[\bigodot_{m=0}^{k} \mathbf{q_m}\right]}{\sigma\left[\bigodot_{m=0}^{k} \mathbf{q_m}\right]},
\end{equation}
where \( \mu[\cdot] \) and \( \sigma[\cdot] \) represent the time-averaged mean and standard deviation functions, respectively; \(\prod\) denotes the product of the elements; \(\bigodot\) represents the element-wise product of vectors.

To differentiate concordant group interactions from discordant ones in a \( k \)-order product, concordant signs are always positively mapped, while discordant signs are negatively mapped. Formally,
\begin{equation}
\text{sign} \left[ \xi_{0 \ldots k}(t) \right] =
\begin{cases}
+1 & \text{if } q_{_0}(t), \ldots, q_{_k}(t) \text{ are all non-negative or non-positive}, \\
-1 & \text{otherwise}.
\end{cases}
\end{equation}
In other words, the weight \( w_{0 \ldots k}(t) \) at time \( t \) of the \( k \)-order co-fluctuations is defined as
\begin{equation}
w_{0 \ldots k}(t) = \text{sign} \left[ \xi_{0 \ldots k}(t) \right] \left| \xi_{0 \ldots k}(t) \right|.
\end{equation}

It is noteworthy that our proposed method, by computing the product of first-order difference sequences of multivariate time series, enables the investigation of higher-order dynamic interactions, thereby providing a viable approach for exploring complex nonlinear coupling mechanisms in multivariate time series data. In particular, when applied to fMRI time series, individual fMRI data are transformed into \( N \) original fMRI signals through a preprocessing pipeline, where each variable corresponds to one ROI. 
If we compute all possible products up to order \( k \), this will result in \( \binom{N}{k+1} \) different co-fluctuation time series for each order \( k \). 
In this paper, we focus on co-fluctuations up to dimension \( k = 3 \) to capture quadruplet interactions. 

\subsection{The Monotonic Weighted Simplicial Complex}
\label{3.3}

\begin{figure}
    \centering
    \includegraphics[scale=0.5]{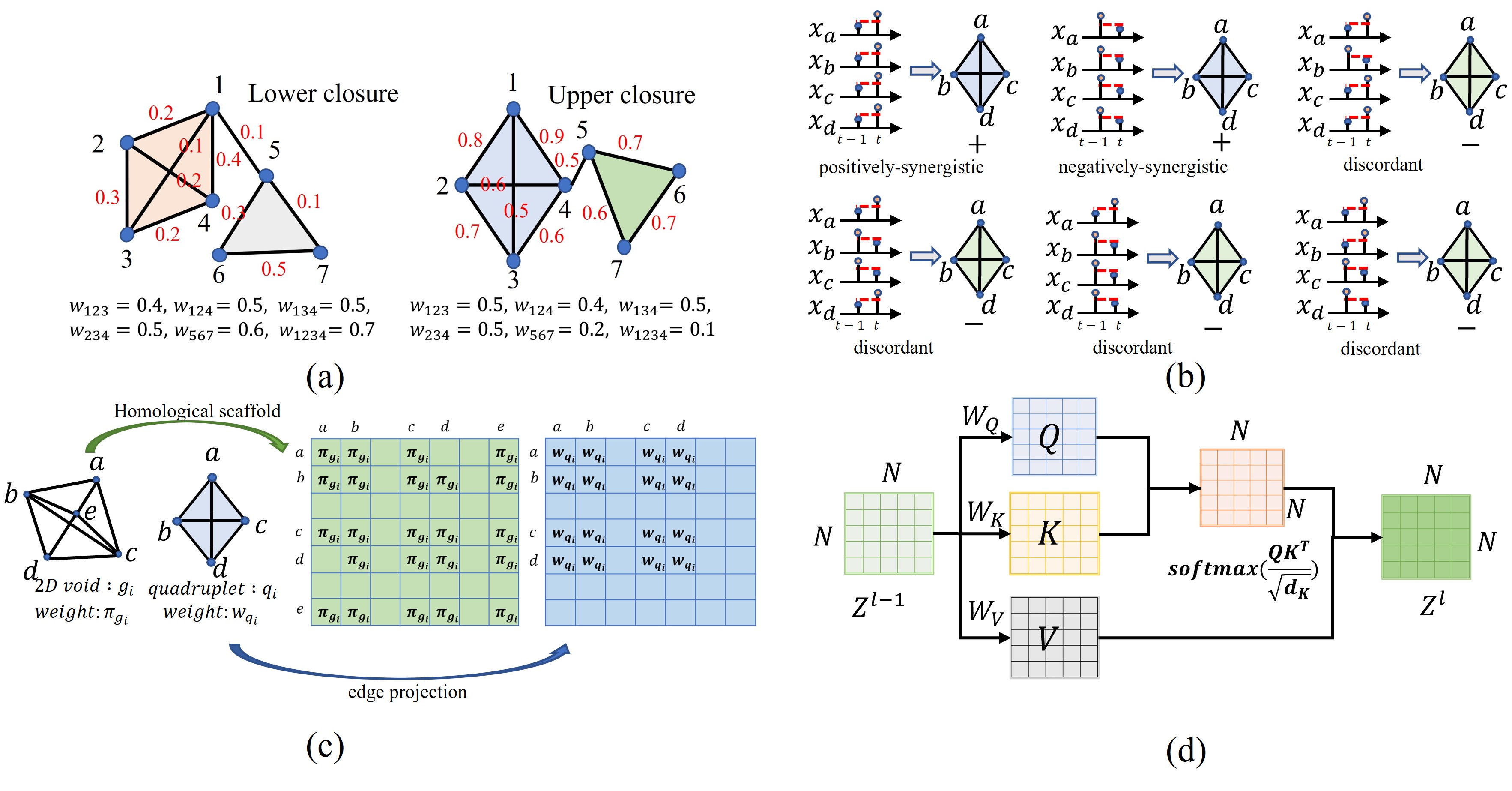}
   \caption{(a) illustrates two examples of monotonic weighted simplicial complexes (left) that satisfy the lower-closure condition and one example (right) that satisfies the upper-closure condition. The red numbers denote the weights of 1-simplices, and the text below indicates the weights of 2-simplices and 3-simplices. (b) illustrates six types of signed quadruplet higher-order interactions. For example, positively synergistic quadruplet interactions indicate that the activation values of the four brain regions increase simultaneously at time \textit{t} relative to time \textit{t}-1. We focus exclusively on the first two interaction types with concordant positive signs (all non-negative or all non-positive). (c) illustrates the construction of the feature matrices: 2D-void features are encoded from the homological scaffold, whereas quadruplet signatures are encoded by projecting signed quadruplets onto edges to obtain edge-based matrices that retain higher-order interaction information. (d) illustrates a single-channel, single-head self-attention layer, where the input node feature matrix is linearly projected into queries, keys, and values, and the output is obtained by aggregating value vectors weighted by the scaled dot-product attention.}
    \label{fig_2}
\end{figure}

\noindent
\begin{definition} (\textbf{Simplicial Complex})
An abstract simplicial complex $\mathcal{K}$ on a finite vertex set $V$ is a collection of some non-empty subsets of $V$ satisfying:
\begin{enumerate}
    \item  For every $v \in V$, the singleton $\{v\}$ belongs to $\mathcal{K}$.
    \item If $\sigma \in \mathcal{K}$ and $\tau \subseteq \sigma$ is a non-empty subset, then $\tau \in \mathcal{K}$.
\end{enumerate}
\end{definition}
Each element $\sigma \in \mathcal{K}$ is called a simplex. A simplex with $|\sigma| = k + 1$ vertices is a $k$-simplex, representing a $k$-order interaction. A subset $\tau \subseteq \sigma$ satisfying $|\tau| = k$ is called a \emph{face} of $\sigma$.

Previous studies assigned arbitrary weights to simplices without imposing closure conditions on the weights \citep{76}. To explore some special higher-order structures in the simplicial complex, here we propose a new definition of weighted simplicial complexes. 

\noindent

\begin{definition}(\textbf{Monotonic Weighted Simplicial Complex}) Let $\mathcal{K}$ be an abstract simplicial complex on vertex set $V$, where each simplex $\sigma \in \mathcal{K}$ is a non-empty finite subset of $V$. A monotonic weighted simplicial complex is a pair $(\mathcal{K}, w)$ with a weight function $w: \mathcal{K} \to \mathbb{R}$ satisfying the following lower closure condition:
\begin{itemize}
\item (\textbf{Lower closure}) For any $\sigma \in \mathcal{K}$, any face $\tau \subseteq \sigma$, \( w(\tau) \leq w(\sigma) \).
\end{itemize}

In addition, a monotonic weighted simplicial complex is also a pair $(\mathcal{K}, w)$ with a weight function $w: \mathcal{K} \to \mathbb{R}$ satisfying the following upper closure condition:
\begin{itemize}
\item (\textbf{Upper closure}) For any $\sigma \in \mathcal{K}$, any face $\tau \subseteq \sigma$, \( w(\tau) \geq w(\sigma) \).
\end{itemize}
\end{definition}

As shown in  Figure \ref{fig_2}(a), it illustrates two types of monotonic weighted simplicial complexes. Specifically, in these simplicial complexes, a 0-simplex is a node, a 1-simplex is an edge associated with pairwise interaction, a 2-simplex is a triangle associated with triplet interaction, a 3-simplex corresponds to a quadruplet interaction. The new definition allows us to explore some special higher-order structures satisfying the monotonicity closure conditions in the weighted simplicial complex, providing new insights.

\section{Method}
In this section, we introduce the method of HOI-Brain for diagnosing brain diseases, which captures signed high-order interactions and signed higher-order organizational patterns in fMRI data. HOI-Brain consists of four parts: calculating signed high-order interactions through Multiplication of Temporal Derivatives, constructing monotonic weighted simplicial complexes, extracting  higher-order and low-order organizations in fMRI data, and classifying with a multi-channel brain network Transformer.
\subsection{Calculate Signed Higher-Order Interactions Through Multiplication of Temporal Derivatives.}
\label{4.1}
Recently, some computational methods have focused on capturing dynamic higher-order interactions in the brain \citep{39,40,34}, but these methods based on extended Pearson correlation have generally been limited by the lack of adequate temporal resolution, especially when capturing more higher-order interactions, such as quadruplets \citep{37}.
To better capture higher-order interactions in the brain, we utilize the designed Multiplication of Temporal Derivatives on fMRI data, which is described in Section \ref{3.2}. 



In this paper, we compute signed \(k\)-order co-fluctuation weights \(w_{i_0 i_1 \cdots i_k}(t)\) following Section \ref{3.2}, where the sign mapping explicitly distinguishes concordant group interactions (\(w_{i_0 i_1 \cdots i_k}(t) > 0\)) from discordant ones (\(w_{i_0 i_1 \cdots i_k}(t) < 0\)). We focus exclusively on concordant signs, eschewing the examination of discordant signs. This is due to the potential for discordant signs to represent a confusing redundancy of information across multiple brain regions \citep{40}. 
Therefore, \(w_{i_0 i_1 \cdots i_k}(t) > 0\) indicates a concordant interaction at time \(t\), which we further separate into positively and negatively synergistic interactions according to the signs of the underlying temporal derivatives \(q_{i_j}(t)\), as shown in  Figure \ref{fig_2}(b):


\noindent
\textbf{1) Positively synergistic interactions} are indicative of multiple brain regions that exhibit simultaneous activation at a given moment relative to the preceding one. These interactions are mathematically defined by the conditions:
\begin{equation}
w_{i_0 i_1 \cdots i_k} > 0 \quad \text{and} \quad q_{i_0} > 0,\ q_{i_1} > 0,\ \ldots,\ q_{i_k} > 0
\end{equation}
where $w_{i_0 i_1 \cdots i_k}$ denotes the weight assigned to the $k$-simplex $\sigma = \{i_0, i_1, \dots, i_k\}$, $q_{i_j}$ denotes the first-order temporal difference of the activity value at node $i_j$.
Let $S^{+}$ denote the set of all \textit{k}-order positively synergistic interactions ($k<4$), with their weights $w^{+}$ .

\noindent
\textbf{2) Negatively synergistic interactions} indicate that these regions collectively exhibit inhibition at the current moment compared to the prior moment. These interactions are mathematically defined by the conditions:
\begin{equation}
w_{i_0 i_1 \cdots i_k} > 0 \quad \text{and} \quad q_{i_0} < 0,\ q_{i_1} < 0,\ \ldots,\ q_{i_k} < 0
\end{equation}
where $w_{i_0 i_1 \cdots i_k}$ denotes the weight assigned to the $k$-simplex $\sigma = \{i_0, i_1, \dots, i_k\}$, $q_{i_j}$ denotes the first-order temporal difference of the activity value at node $i_j$. Let $S^{-}$ denote the set of all \textit{k}-order negatively synergistic interactions ($k<4$), with their weights $w^{-}$.

Collectively, distinguishing these interactions can offer valuable insights into the complex coordination and communication within the brain, thus improving the effectiveness of brain disease diagnosis.

\subsection{Construct Monotonic Weighted Simplicial Complexes }
Recent advances in neuroimaging analysis have demonstrated the utility of representing brain as simplicial complexes \citep{30, 40}, which  preserves the inherent higher-order dependencies in neural systems that are critical for understanding cognitive processes and pathological mechanisms compared to two-way networks \citep{42}. Based on this, we further propose the concept of monotonic weighted simplicial complex in Section \ref{3.3} to explore some special higher-order structures in the brain and model neurobiological principles where functional coherence depends on hierarchical organization.


To construct these monotonic weighted simplicial complexes at each temporal instance $t$ using the concordant interactions $S^{+}$ and $S^{-}$ obtained in Section \ref{4.1}, we leverage the concept of filtration from Persistent Homology theory \citep{31}, a computational topology technique adept at analyzing high-dimensional datasets. Specifically, we design two types of filtrations to generate two monotonic weighted simplicial complexes that satisfy the lower closure condition, and two monotonic weighted simplicial complexes that satisfy the upper closure condition. The procedure is summarized as follows:

\textbf{1) The ascending filtration:} 
Sort the weights $w^{+}$ in the set of all positively synergistic simplices $S^{+}$ in ascending order and use the parameter \( \epsilon_l \in \mathbb{R} \) to scan this sequence $S^{+}$. At each step \( l \), we include all 1-simplices, 2-simplices, and 3-simplices that satisfy the lower closure condition of the monotonic weighted simplicial complex to generate a monotonic weighted simplicial complex $\mathcal{K}^t_{positive,lower}$. The set of all negatively synergistic simplices $S^{-}$ is processed in the same way to generate a monotonic weighted simplicial complex $\mathcal{K}^t_{negative,lower}$. 
These are defined as:
\begin{equation}
\begin{split}
\mathcal{K}^t_{\mathrm{positive,lower}} = \left\{ \sigma \in S^{+} \mid 
for \ any\ face \ \tau  \ of\ \sigma ,\, w(\tau) \leq w(\sigma) \right\},\\
\mathcal{K}^t_{\mathrm{negative,lower}} = \left\{ \sigma \in S^{-} \mid 
for \ any \ face \ \tau \  of \ \sigma ,\, w(\tau) \leq w(\sigma) \right\}.
\end{split}
\end{equation}

{\bf 2) The descending filtration:} Sort the weights $w^{+}$ in the set of all positively synergistic simplices $S^{+}$ in descending order and use the parameter \( \eta_l \in \mathbb{R} \) to scan this sequence $S^{+}$. At each step \( l \), we include all 1-simplices, 2-simplices, and 3-simplices that satisfy the upper closure condition of the monotonic weighted simplicial complex to generate a monotonic weighted simplicial complexes $\mathcal{K}^t_{positive,upper}$. The set of all negatively synergistic simplices $S^{-}$ is processed in the same way to generate a weighted simplicial complexes $\mathcal{K}^t_{negative,upper}$. These are defined as:
 \begin{equation}
\begin{split}
\mathcal{K}^t_{\mathrm{positive,upper}} = \left\{ \sigma \in S^{+} \mid 
for\  any\  face\  \tau \  of\  \sigma ,\, w(\tau) \geq w(\sigma) \right\},\\
\mathcal{K}^t_{\mathrm{negative,upper}} = \left\{ \sigma \in S^{-} \mid 
for \ any \ face \ \tau \  of\  \sigma,\, w(\tau) \geq w(\sigma) \right\}.
\end{split}
\end{equation}

These four weighted simplicial complexes at each temporal instance $t$ encode the brain from different perspectives, providing the potential to further explore higher-order organizational patterns in the brain.

\subsection{Topological Lower-Order and Higher-Order Organizations of fMRI Signals}
\label{4.2}
Many studies have explored higher-order structures in brain networks by focusing on loops (1D cycles) formed by nodes and edges \citep{32,33}, with limited attention paid to more complex higher-order organizational patterns. Moreover, the influence of signed HOIs was overlooked in these studies, which could offer a broader diagnostic view of neurological conditions. The new definition of weighted simplicial complexes allows researchers to explore more complex signed higher-order structures, such as quadruplet structures and voids (2D holes) formed by triangular faces. This advance is significant because it further refines the HOIs by defining sign and links HOIs in the brain to higher-order organizations, and provides a higher-dimensional perspective for studying higher-order structures in the brain.

\textbf{1) Capture lower-order edges:} To explore brain activity patterns for diagnosis from a global perspective, we extract all the 1-simplices in the monotonic weighted simplicial complex $\mathcal{K}^t_{positive,lower}$ and the monotonic weighted simplicial complex $\mathcal{K}^t_{negative,lower}$ representing the lower-order organizations in the brain and add them into a list $\Delta^{edge}={\{(i,j),w_{ij}}\}$.

In this paper, we focus on the impact of signed HOIs, so we do not further explore the signs of lower-order edges. We construct a weighted matrix \(  \widetilde{A}^{edge} \in \mathbb{R}^{N \times N} \) from the list $\Delta^{edge}$ and assign a weight $w_{ij}$ equal to the weight of the edge for each element $(i, j)$ in matrix.

\textbf{2) Capture  higher-order good quadruplets:} To mine the topological higher-order of fMRI signals, we first extract all 3-simplices from the monotonic weighted simplicial complexes $\mathcal{K}^t_{positive,lower}$and $\mathcal{K}^t_{negative,lower}$ and add them into two lists, respectively. These structures represent co-fluctuations that are not observable through lower-order interactions alone and represent extremely good higher-order organizations. 

To quantify these good quadruplets at the edge levels from the lists, we use edge projections \citep{42}. For each edge \( (i, j) \), we assign a weight \( w_{ij} \) equal to the average sum of the weights of the quadruplets defined by that edge. The feature matrices of quadruplet signatures \( A^{lower} \in \mathbb{R}^{N \times N} \) using edge projections are illustrated in Figure \ref{fig_2}(c).

\textbf{3) Capture higher-order 2D voids:} Similarly, we can extract all 3-simplices from the monotonic weighted simplicial complexes $\mathcal{K}^t_{positive,upper}$ and $\mathcal{K}^t_{negative,upper}$ to extract structures that represent extremely bad higher-order organizations. To some extent, these higher-order organizations can represent the 2-dimensional voids in the brain that are extracted by Persistent Homology, which reflect the intrinsic shape of the data from a higher-dimensional perspective. Therefore, we employ Persistent Homology \citep{31} to extract 2-dimensional voids, which not only capture extremely bad higher-order organization but also help characterize the intrinsic geometric organization of neural activity from the perspective of a higher-dimensional manifold. For more details about  Persistent Homology theory and 2D voids, please refer to \ref{Persistent Homology}. The core concept involves constructing a filtration—a sequence of simplicial complexes \(\{S^l\}\) that progressively approximate the original weighted simplicial complex with increasing precision:

\begin{equation}
\emptyset = S^0 \subset S^1 \subset \ldots \subset S^l \subset \ldots \subset S^n .
\end{equation}

Finally, the $S^n$ is equal to the monotonic weighted simplicial complexes $\mathcal{K}^t_{positive,upper}$ or $\mathcal{K}^t_{negative,upper}$. Persistent homology examines how higher-order topological features evolve through the filtration \(\{ \mathcal{S}^l \}\), offering a measure of their robustness across scales. It is worth noting that calculating Persistent Homology requires defining the distance among multiple nodes in \textit{k}-simplex, which can be easily achieved by computing $1-w_{i_{0}\dots i_{k}}$(for $k<4$), where $w_{i_{0}\dots i_{k}}$ represents the weight of the \textit{k}-simplex. 

By focusing on 2D voids within the homology group \(H_2\), we track these higher-order organizations. Specifically, we apply Persistent Homology to the monotonic weighted simplicial complexes $\mathcal{K}^t_{positive,upper}$ and $\mathcal{K}^t_{negative,upper}$ to generate two two-dimensional persistence diagrams. In each diagram, each point \((b_g, d_g)\) signifies a void \( g \) that emerges during the filtration process. The persistence \(\pi_g = d_g - b_g\) quantifies the void's lifespan, indicating its importance.

To better quantify the higher-order topological features captured by persistence homology, we employ a homological scaffold \citep{41}, a weighted network representing the topological features in the persistence diagram. This scaffold comprises all voids corresponding to generators \( g_i \), weighted by their persistence \( \pi_{g_i} \). Specifically, if an edge \( e \) belongs to multiple two-dimensional voids \( g_0, g_1, \ldots, g_s \), its weight \( \bar{w}_e^\pi \) is defined as the sum of the persistences of these generators:
\begin{equation}
\bar{w}_e^\pi = \sum_{g_i \mid e \in g_i} \pi_{g_i}.
\end{equation}
The homological scaffold reveals the roles of different links in shaping the system's homological properties. A larger total persistence \( \bar{w}_e^\pi \) for a link \( e \) indicates its function as a locally strong bridge within the space of coherent and decoherent co-fluctuations. The feature matrices of 2D voids \( A^{upper} \in \mathbb{R}^{N \times N} \) using homological scaffold are illustrated in Figure \ref{fig_2}(c).

At each time point \( t \), five distinct weighted networks are constructed: one  \( \widetilde{A}^{edge}\) derived from edge structures representing lower-order interactions, two \( A^{lower}_1, A^{lower}_2\)  from signed good quadruplet structures signifying HOIs, and another two \( A^{upper}_1, A^{upper}_2\)     from signed 2D holes illustrating the data's intrinsic geometric shape from a higher-dimensional perspective. By averaging across the entire time period, the resulting networks exhibit enhanced stability and mitigate the impact of fMRI noise.

\subsection{Multi-Channel Brain Network Transformer}
\label{4.3}
To holistically integrate complementary information from heterogeneous topological features while preserving channel-specific properties, we propose a multi-channel brain network Transformer architecture with four coherently designed components. 

\subsubsection{Signed Higher-Order Features Decoupling Mechanism}
Higher-order features that are positive are related to patterns of simultaneous activation in multiple brain regions at a given moment, while higher-order features that are negative are related to patterns of collective inhibition in these regions at the current moment. These features can offer valuable information about complex coordination and communication within the brain. We further explore the influence of signed higher-order features using higher-order features decoupling mechanism. Specifically, for each type of higher-order feature, we consider two signed weighted matrices \( A^{lower}_{j}   \ or \ A^{upper}_{j} \in \mathbb{R}^{N \times N} \) (\( j = 1, 2\)). For each of these matrices, we extract its upper triangular part and flatten it to construct a feature vector \(  h^{lower}_{j}\ or\ h^{upper}_{j} \in \mathbb{R}^{N(N-1)/2} \). We generate two new higher-order topological features by adaptively learning weights to aggregate positive and negative information:
\begin{align}
    \alpha_1,\alpha_2 &= \text{Softmax}({f}^{lower} (h^{lower}_{1},h^{lower}_{2})) \\
    {\widetilde{A}}^{lower} &= \alpha_1 \odot A^{lower}_{1}+\alpha_2 \odot A^{lower}_{2}\\
      \beta_1,\beta_2 &= \text{Softmax}({f}^{upper} (h^{upper}_{1},h^{upper}_{2})) \\
    {\widetilde{A}}^{upper} &= \beta_1 \odot A^{upper}_{1}+\beta_2 \odot A^{upper}_{2}
\end{align}
where \( f^{lower} \ and \ f^{upper}\) denote two separate two-layer MLPs, and \( \widetilde{A}^{lower} \ and\ \widetilde{A}^{upper}\) denote, respectively, features of quadruplet structures and features of 2D voids that aggregate positive and negative information, respectively. The attention scores \(\mathbf{\alpha}_i,{\beta}_i\)  emphasize discriminative signed features while suppressing redundant signed information.

\subsubsection{Multi-Head Self-Attention Module of Multiple Channels}
The feature matrices extracted above actually represent the low-order and higher-order topological information associated with each brain region separately. Simply feeding the concatenated low-order and higher-order information into the model would overlook the heterogeneity of the related information. Therefore, inspired by Brain network Transformer \citep{20}, we leverage an $L$-layer non-linear mapping module, namely Multi-Head Self-Attention (MHSA), to generate more expressive node features using the low-order and higher-order topological features of multiple channels ${Z}^L_{i} = \text{MHSA}({\widetilde{A}_i}) \in \mathbb{R}^{N \times N},(i=1,2,3)$. Specifically, as shown in  Figure \ref{fig_2}(d), for each channel $i$, the output of each layer ${Z_{{i}}}^l$ is obtained by
\begin{equation}
{Z_{{i}}}^l = (\left\|_{m=1}^M {h_{\textit{i}}}^{l,m} \right\|) {W}_{O,i}^l
\end{equation}
\begin{equation}
{h_i}^{l,m} = \text{Softmax} \left( \frac{ \left( {W}_{Q,i}^{l,m} {Z_{i}}^{l-1} \right) \left( {W}_{K,i}^{l,m} {Z_i}^{l-1} \right)^\top }{ \sqrt{d_{K,i}^{l,m}} } \right) {W}_{V,i}^{l,m} {Z_i}^{l-1}
\end{equation}
where ${Z_i}^{l=0} = {\widetilde{A_i}}, \left\| \cdot \right\|$ is the concatenation operator, $M$ is the number of heads, $m$ is the head index, $l$ is the layer index, ${{W}}_O^l$, ${W}_{Q,i}^{l,m}$, ${W}_{K,i}^{l,m}$, ${W}_{V,i}^{l,m}$ are learnable model parameters of each channel \textit{i}, and $d_{K,i}^{l,m}$ is the first dimension of ${W}_{K,i}^{l,m}$.

\subsubsection{Orthonormal Clustering Readout of Multiple Channels}
Research has shown that low-level and high-level feature patterns in the brain often exhibit distinct functional modular structures, thereby supporting the richest functional interactions \citep{43}. Therefore, to leverage the inherent properties of brain networks, where nodes within the same functional module tend to exhibit similar behaviors and form clustered representations, we design an orthonormal clustering readout of multiple channels inspired by the reference \citep{20}. This module is intended to capture modular-level similarities between ROIs across different brain network patterns, where nodes are softly assigned to well-defined clusters through an unsupervised process. 

Formally, given \( K \) cluster centers where each center has \( V \) dimensions in a given channel \textit{i}, \({E_{i}} \in \mathbb{R}^{K \times V}\), a Softmax projection operator calculates the probability \( {P}^{jk}_i \) of assigning node \( j \) to cluster \( k \):
\begin{equation}
{P}^{j,k}_{i} = \frac{
  e^  {\langle {Z}^{L,j}_i, {E}^{k}_i \rangle }
}{
  \sum\limits_{k=1}^{K} e^  {\langle {Z}^{L,j}_i, {E}^{k}_i} \rangle }
\end{equation}
where \( \langle \cdot, \cdot \rangle \) denotes the inner product, $k$ denotes the cluster index, \( {Z}^{L,j}_i \) is the learned embeddings of node \textit{j} from the last \textit{L}-th layer of the MHSA module in given channel \textit{i}, and \( {E}^{k}_i \) denotes the embeddings of cluster \textit{k} in given channel \textit{i}. Using this soft assignment \( {P_i} \in \mathbb{R}^{N \times K} \) (where \( N \) is the number of nodes), the graph-level embedding \( {Z}^{G}_i \) is obtained by:
\begin{equation}
{Z}^{G}_i = {P_i}^{\top} {Z}^{L}_i.
\end{equation}
To obtain representative soft assignment \( {P_i} \), the initialization of cluster centers \( {E_i} \) is  generated by orthonormal initialization \citep{20}.

\subsubsection{Attention-Guided Feature Fusion Mechanism for Classification} 
For embedding at the graph level \( {F}_i \) of each channel \textit{i} obtained by flattening \( {Z}^{G}_i \) (\( i = 1, 2, 3 \)), we propose an attention-guided feature fusion mechanism to adaptively determine the weights of various channels depending on the disorder, dataset and the individual variability, thereby emphasizing discriminative features while suppressing redundant information. The formula can be described as:
\begin{align}
    w_i &= \sigma({q}^T_{i} \tanh({W}_i {Z}^{G}_i)) \\
   \gamma_{i} &= \frac{
  \ e^ {w_{i}}
}{\ e^ {w_{1}}+e^ {w_{2}}+e^ {w_{3}}}\\
    \widetilde{F} &= \gamma_{1} \odot {F}_1 \,\|\, \gamma_{2} \odot {F}_2 \,\|\, \gamma_{3} \odot {F}_3
\end{align}
where \(\sigma(\cdot)\) denotes sigmoid function, \(\odot\) is element-wise multiplication, and \( {q}^{T}_i \), \( {W}_i \) are learnable model parameters. Subsequently, a three-layer multi-layer perceptron (MLP) is employed using \( \widetilde{F}\) as input for the prediction head. Finally, the cross-entropy loss function is used as the loss function.

\section{Experiments}
\subsection{Experimental Settings}
\subsubsection{Datasets and Preprocessing}

\begin{table}[t]
\centering
\caption{Class distribution of brain network datasets.}
\vspace{2pt}
\label{tab:class_distribution}
\scalebox{0.74}{
\begin{tabular}{@{}lllc@{}}
\toprule
Dataset & Class & \# Subjects & Disease Type \\
\midrule

\multirow{3}{*}{\parbox{3cm}{ ADNI}} & AD & 90 & \multirow{3}{*}{\parbox{5cm}{\centering Alzheimer's Disease}} \\
     & MCI & 76 &  \\
     & CN & 96 &  \\
\midrule
\multirow{2}{*}{\parbox{3cm}{ TaoWu}} & PD & 20 & \multirow{2}{*}{\parbox{5cm}{\centering Parkinson's Disease}}  \\
      & NC & 20 &  \\
       \midrule
\multirow{2}{*}{\parbox{3cm}{ PPMI}} & PD & 53 & \multirow{2}{*}{\parbox{5cm}{\centering Parkinson's Disease}} \\
     &  prodromal & 53 &  \\ 
     \midrule
\multirow{2}{*}{\parbox{3cm}{ ABIDE }} &ASD & 488 & \multirow{2}{*}{\parbox{5cm}{\centering Autism Spectrum Disorder }}  \\
      & NC & 537 &  \\
 \bottomrule

\end{tabular}
}
\end{table}
To comprehensively evaluate the proposed method, we employ four fMRI datasets with varying sample sizes. The class distribution of brain network datasets is presented in Table \ref{tab:class_distribution}. 
1) \textit{\textbf{ADNI dataset}} \citep{44}: The Alzheimer's Disease Neuroimaging Initiative (ADNI) primarily aims to investigate whether the combination of serial magnetic resonance imaging (MRI), positron emission tomography (PET), other biological markers, along with clinical and neuropsychological assessments, can effectively measure the progression of mild cognitive impairment (MCI) and early Alzheimer's disease (AD). The dataset comprises 90 Alzheimer's disease (AD) patients, 96 cognitively normal (CN) subjects, and 76 individuals with mild cognitive impairment (MCI).
2) \textit{\textbf{TaoWu dataset}} \citep{45}: Released by the International Conference on Imaging (ICI) \citep{45}, the TaoWu dataset represents one of the earliest neuroimaging collections for Parkinson's disease research, containing 20 Parkinson's disease (PD) patients and 20 normal controls (NC).
3) \textit{\textbf{PPMI dataset}} \citep{45}: The Parkinson's Progression Markers Initiative (PPMI) is a landmark longitudinal study designed to identify biomarkers of Parkinson's disease risk, onset, and progression. Our study utilizes data from 53 Parkinson's disease (PD) patients and 53 prodromal cases obtained from the PPMI database.
4) \textit{\textbf{ABIDE dataset}} \citep{48}: The Autism Brain Imaging Data Exchange (ABIDE) initiative facilitates Autism Spectrum Disorder (ASD) research by aggregating functional brain imaging data from multiple international sites. Our analysis includes 488 ASD patients and 537 normal controls (NC) from the ABIDE database.

All resting-state fMRI data were preprocessed following the pipeline described in \citep{46}, which includes motion correction, realignment, field unwarping, normalization, bias field correction, and brain extraction. For more details about data preprocessing using fMRIPrep, please refer to \ref{Data preprocessing}. The preprocessed images were parcellated into 90 regions of interest (ROIs) using the Automated Anatomical Labeling (AAL) atlas \citep{47}. The blood-oxygen-level-dependent (BOLD) time series for each ROI were obtained by averaging the time series of all constituent voxels. We additionally provide classification results on the ABIDE dataset under Schaefer-100
parcellation setting \citep{81} in \ref{Model Comparison}, to evaluate the robustness of our method to the choice of parcellation.

\subsubsection{Methods for Comparison}
The selected baselines correspond to five categories. The first category consists of traditional machine learning models by the scikit-learn library \citep{49} using edge features, including Multilayer Perceptron (MLP), Support Vector Machine Classifier (SVM), Logistic Regression, and Random Forest.
The second category consists of GNN-based models that learn lower-order topological features from a two-way network. These models include GCN \citep{50}, GraphSAGE \citep{51}, GAT \citep{52}, GroupINN \citep{11},  BrainGNN \citep{12}, FBNetGen \citep{13}, BPI-GNN \citep{14}.
The third category consists of Transformer-based models that utilize the Transformer architecture to perceive the entire brain network, including Graph Transformer \citep{19}, Brain Network Transformer \citep{20}, TSEN \citep{21}, Long-range Brain Transformer \citep{22}.
The fourth category consists of higher-order methods based on hypergraph, including HGCN \citep{26}, HGAT \citep{25}. 
The fifth category consists of  Persistent Homology (PH)-based models, including Brain-HORS \citep{34}, PH-MCI \citep{33}, ATPGCN \citep{32}.

\subsubsection{Implementation Details }
For all experiments, we evaluated the performance in terms of diagnosis accuracy, recall, precision, and F1-score. For dataset partitioning, we adopt an 8:1:1 ratio for training, validation and testing for the ADNI dataset and ABIDE dataset, while using a 3:1:1 ratio for the TaoWu and PPMI datasets. When evaluating model performance, we implement 10-fold cross-validation for the ADNI dataset and ABIDE dataset while 5-fold cross-validation for the others. All models are trained using the Adam optimiser \citep{54} with an initial learning rate of \(10^{-4}\) and a L2 weight decay of \(10^{-4}\), in conjunction with the ReduceLROnPlateau scheduler \citep{55}. During training, we applied a dropout rate of 0.5 in the Transformer layers. The batch size is set to 16 for the ADNI dataset and ABIDE dataset while 4 for others, with a maximum number of epochs of 100. After each epoch, the model performance is evaluated using the validation set. We keep the model with the highest F1 score for testing. All experiments are run within the PyTorch framework \citep{55}. Further implementation details are summarized below:

\begin{itemize}
    \item In HOI-Brain, we employ a Julia-based implementation of the Ripserer and PersistenceDiagrams libraries \citep{53} to compute Persistent Homology and homological scaffold, extracting higher-order topological features. 
 
     \item For traditional ML models, feature vectors were derived from the flattened upper triangle of Pearson correlation matrices computed from BOLD time series. An MLP with three fully connected layers is applied for classification. An SVM with a radial basis function kernel is applied for classification. A LogisticRegression with \texttt{solver=lbfgs} is applied for classification. A RandomForest with \texttt{n\_estimators=300} is applied for classification.
 
     \item For GNN-based models and Transformer-based models, the correlation matrix is calculated using Pearson correlation for all time series. Node features are derived from the columns of the Pearson correlation matrix. The adjacency matrix in the graph is obtained by thresholding (0.5) the correlation matrix. Generally, two graph convolutional layers and an average pooling layer are used for classification.  
 
     \item For hypergraph-based higher-order methods, specifically, for HGCN, a graph is first constructed through the correlation matrix, where the correlation of each edge of the graph is greater than 0.7. Each node constructs 2-hyperedges, 3-hyperedges, and 4-hyperedges  by selecting its \textit{k}-hop neighbors in the graph. For HGAT, the feature of the central node corresponding to each hyperedge is used as the hyperedge feature. During information aggregation, attention coefficients between hyperedges and nodes are computed for classification. Note that more HGNN-based methods applied to brain data have been proposed recently. However, the code of these models is not publicly available, and they require significant engineering which is difficult to reproduce faithfully based solely on the papers. 
     
     \item For PH-based models (e.g., PH-MCI and ATPGCN), we use the persistent topological features of connected components $H_0$ and loops $H_1$ extracted by Ripserer tool. For Brain-HORS, following the method, we extract the matrix of violating triangles, the matrix of one-dimensional loops and the matrix of lower-order edge features, which are then flattened and classified using an SVM. 
\end{itemize}

The original codes shared by the authors of these baselines are used for the comparative analysis. We adapt their open-source codes, strictly follow the parameters provided in the papers, and modify them to fit our datasets.

\begin{table}[t]
  \centering
  \caption{Performance comparison with five categories of baselines on the ADNI dataset (\%). The best results are marked in bold and the
second-best results are underlined.}
  \scalebox{0.74}{
  \begin{tabular}{@{}cccccc@{}}
    \toprule
   Type & Method & Accuracy & Precision & Recall & F1-score \\
    \midrule
    \multirow{4}{*}{\parbox{3cm}{\centering Traditional Machine Learning Models}}
    & MLP    & 66.3$\pm$5.8 & 67.4$\pm$6.4 & 68.3$\pm$6.4 & 69.1$\pm$5.8   \\
    & SVM     & 71.1$\pm$5.1 & 71.6$\pm$4.8 & 71.7$\pm$4.8 & 70.7$\pm$5.0  \\
    & Logistic Regression  & 70.7$\pm$5.9 & 73.3$\pm$7.7 & 71.7$\pm$7.1 & 70.5$\pm$7.2   \\
    & RandomForest  & 64.8$\pm$8.0 & 66.8$\pm$7.2 & 65.4$\pm$8.3 & 64.9$\pm$7.8   \\
    \midrule
    \multirow{7}{*}{\parbox{3cm}{\centering GNN-based models}}
    & GCN & 60.9$\pm$11.8 & 61.6$\pm$12.5 & 61.1$\pm$11.9 & 60.3$\pm$12.1   \\
    & GraphSAGE & 64.5$\pm$8.9& 65.6$\pm$9.1 & 64.5$\pm$8.9 & 64.0$\pm$9.0   \\
    & GAT & 60.0$\pm$10.1 & 61.8$\pm$10.6 & 60.0$\pm$10.1 & 59.2$\pm$10.3   \\
    & GroupINN & 57.3$\pm$8.8 & 61.0$\pm$9.6 & 57.4$\pm$9.8 & 56.9$\pm$9.1  \\
     & FBNETGEN & 66.4$\pm$7.0 & 67.7$\pm$7.5& 67.4$\pm$7.1 & 60.8$\pm$5.9   \\
    & BPI-GNN & 51.4$\pm$8.5 & 49.2$\pm$14.6 & 52.1$\pm$8.8 & 48.0$\pm$11.1   \\
    & BrainGNN & 58.9$\pm$7.7 & 60.5$\pm$8.8 & 58.9$\pm$7.7 & 58.2$\pm$8.0   \\
    \midrule
    \multirow{4}{*}{\parbox{3cm}{\centering Transformer-based models}}
    & Graph Transformer & 64.0$\pm$6.5 & 65.6$\pm$7.4 & 64.8$\pm$6.5 & 64.1$\pm$6.7  \\
    & BrainnetTransformer & 69.6$\pm$6.2 & 72.2$\pm$6.2 & 69.5$\pm$6.8 & 69.2$\pm$6.3   \\
    & LR-BrainTransformer & \underline{72.0$\pm$8.5} & \underline{76.5$\pm$8.4} &  \underline{72.2$\pm$8.0} &  \underline{72.0$\pm$9.1}   \\
    & TSEN & 62.5$\pm$7.3 & 67.5$\pm$8.9 & 63.2$\pm$8.0 & 62.4$\pm$7.4   \\
   \midrule
    \multirow{2}{*}{\parbox{3cm}{\centering HGNN-based models}} 
    & HGCN & 58.9$\pm$7.6 & 61.2$\pm$7.0 & 60.1$\pm$7.7& 59.5$\pm$7.3  \\
    & HGAT & 54.5$\pm$7.5 & 55.4$\pm$7.7 & 55.3$\pm$6.7 & 54.7$\pm$6.9   \\
      \midrule      
    \multirow{3}{*}{\parbox{3cm}{\centering PH-based models}} 
    & PH-MCI  & 64.8$\pm$6.9 & 67.8$\pm$7.3 & 64.3$\pm$5.7 & 63.5$\pm$6.2 \\
    & ATPGCN & 70.8$\pm$6.7 & 72.6$\pm$7.3 & 70.6$\pm$6.3 & 70.8$\pm$6.7   \\
     & Brain-HORS & 64.0$\pm$7.2 & 66.2$\pm$7.0 & 64.0$\pm$7.2 & 64.1$\pm$6.8  \\
    \midrule
    \multirow{1}{*}{Our Framework} 
    & HOI-Brain & \textbf{75.9$\pm$8.6} & \textbf{80.0$\pm$7.9} & \textbf{75.6$\pm$9.1} & \textbf{75.5$\pm$9.2}  \\
    \bottomrule
  \end{tabular}
  \label{tab:performance_comparison on ADNI} 
  }
\end{table}

\begin{table}[t]
  \centering
  \caption{Performance comparison with five categories of baselines on the TaoWu dataset (\%). The best results are marked in bold and the
second-best results are underlined.}
  \scalebox{0.74}{
  \begin{tabular}{@{}cccccc@{}}
    \toprule
   Type & Method & Accuracy & Precision & Recall & F1-score \\
    \midrule
    \multirow{4}{*}{\parbox{3cm}{\centering Traditional Machine Learning Models}}
    & MLP    & 63.6$\pm$15.3 & 74.4$\pm$17.9 & 65.3$\pm$26.7 & 64.6$\pm$15.4   \\
    & SVM     & 57.5$\pm$12.8 & 65.0$\pm$22.6 & 40.0$\pm$20.0 & 46.8$\pm$17.1  \\
    & Logistic Regression  & 62.5$\pm$23.7 & 63.7$\pm$23.9 & 60.0$\pm$25.5 & 61.2$\pm$24.0   \\
    & RandomForest  & 67.5$\pm$16.0 & 71.3$\pm$17.3 & 70.0$\pm$18.7 & 67.9$\pm$11.1   \\
    \midrule
    \multirow{7}{*}{\parbox{3cm}{\centering GNN-based models}}
    & GCN &  \underline{72.5$\pm$18.4} & 72.7$\pm$21.7 & 75.0$\pm$25.0 & 70.6$\pm$21.8   \\
    & GraphSAGE & 67.5$\pm$20.3 & 68.3$\pm$21.3 & 60.0$\pm$25.5 & 63.6$\pm$23.5   \\
    & GAT & 62.5$\pm$11.2 & 63.3$\pm$11.3 & 55.0$\pm$18.7 & 58.1$\pm$15.8   \\
    & GroupINN & 62.5$\pm$7.9 & 59.7$\pm$19.0 & 62.5$\pm$7.9 & 58.5$\pm$13.3  \\
     & FBNETGEN & 57.5$\pm$12.8 & 68.7$\pm$26.0 & 50.0$\pm$22.4 & 52.4$\pm$15.4   \\
    & BPI-GNN & 55.0$\pm$12.8 & 56.7$\pm$15.8 & 55.0$\pm$12.8 & 53.7$\pm$12.7   \\
    & BrainGNN & 60.0$\pm$16.6 & 57.7$\pm$19.5 & 57.6$\pm$22.3 & 60.0$\pm$16.6   \\
    \midrule
    \multirow{4}{*}{\parbox{3cm}{\centering Transformer-based models}}
    & Graph Transformer & 67.5$\pm$6.1 &  \underline{75.8$\pm$6.8}& 67.5$\pm$6.1 & 64.6$\pm$8.1  \\
    & BrainnetTransformer & 70.0$\pm$12.8 & 72.7$\pm$16.7 & \textbf{80.0$\pm$24.5} & 71.9$\pm$11.2   \\
    & LR-BrainTransformer & 65.0$\pm$12.2 & 73.3$\pm$9.4 & 71.2$\pm$14.0 &  \underline{74.0$\pm$12.1}   \\
    & TSEN & 57.5$\pm$10.0 & 54.3$\pm$19.1 & 57.5$\pm$10.0 & 53.7$\pm$13.4   \\
   \midrule
    \multirow{2}{*}{\parbox{3cm}{\centering HGNN-based models}} 
    & HGCN & 55.0$\pm$20.3 & 58.7$\pm$23.1 & 55.0$\pm$20.3 & 52.4$\pm$20.2  \\
    & HGAT & 57.5$\pm$15.0 & 59.0$\pm$17.4 & 57.5$\pm$15.0 & 56.3$\pm$15.3   \\
      \midrule      
    \multirow{3}{*}{\parbox{3cm}{\centering PH-based models}} 
    & PH-MCI  & 55.8$\pm$6.1 & 46.8$\pm$24.5 & 60.0$\pm$37.4 & 49.5$\pm$25.1 \\
    & ATPGCN & 57.5$\pm$12.8 & 58.3$\pm$12.3 & 60.0$\pm$12.3 & 58.7$\pm$11.5   \\
     & Brain-HORS & 57.5$\pm$20.3 & 58.3$\pm$21.1 & 57.5$\pm$20.3 & 57.2$\pm$20.2  \\
    \midrule
    \multirow{1}{*}{Our Framework} 
    & HOI-Brain & \textbf{77.5$\pm$12.3} & \textbf{82.4$\pm$9.3} &  \underline{77.5$\pm$12.3} & \textbf{75.9$\pm$13.9}  \\
    \bottomrule
  \end{tabular}
  \label{tab:performance_comparison on TaoWu} 
  }
\end{table}

\begin{table}[t]
  \centering
  \caption{Performance comparison with five categories of baselines on the PPMI dataset (\%). The best results are marked in bold and the
second-best results are underlined.}
  \scalebox{0.74}{
  \begin{tabular}{@{}cccccc@{}}
    \toprule
   Type & Method & Accuracy & Precision & Recall & F1-score \\
    \midrule
    \multirow{4}{*}{\parbox{3cm}{\centering Traditional Machine Learning Models}}
    & MLP    & 61.3$\pm$5.3 &  \underline{65.3$\pm$11.0 }& 54.5$\pm$11.4 & 58.1$\pm$6.4   \\
    & SVM     &  \underline{64.2$\pm$8.3} & 58.4$\pm$10.4 & 58.4$\pm$10.4 & 61.8$\pm$8.7   \\
    & Logistic Regression  & 60.5$\pm$7.0 & 62.4$\pm$7.2 & 54.7$\pm$16.8 & 56.8$\pm$11.0   \\
    & RandomForest  & 59.5$\pm$5.3 & 61.7$\pm$6.7 & 56.7$\pm$16.3 & 57.3$\pm$7.5   \\
    \midrule
    \multirow{7}{*}{\parbox{3cm}{\centering GNN-based models}}
    & GCN & 57.5$\pm$6.8 & 58.7$\pm$10.1 & 62.4$\pm$5.1 & 59.6$\pm$3.7   \\
    & GraphSAGE & 60.4$\pm$5.6 & 61.0$\pm$8.1 & 60.6$\pm$10.7 & 60.2$\pm$6.5   \\
    & GAT & 61.3$\pm$6.8 & 61.3$\pm$7.3 &  64.4$\pm$8.1 & 60.4$\pm$6.3   \\
    & GroupINN & 55.6$\pm$6.1 & 55.7$\pm$6.2 & 55.6$\pm$6.0 & 55.4$\pm$6.2   \\
    & FBNETGEN & 59.4$\pm$8.4 & 60.6$\pm$10.9 & 62.4$\pm$5.1 & 60.8$\pm$5.9   \\
     & BPI-GNN & 52.3$\pm$14.7 & 50.5$\pm$18.9 & 49.4$\pm$15.1 & 47.5$\pm$14.9   \\
    & BrainGNN & 60.3$\pm$11.3 & 59.8$\pm$11.8 & 60.5$\pm$11.6 & 60.3$\pm$11.3   \\
    \midrule
    \multirow{4}{*}{\parbox{3cm}{\centering Transformer-based models}}
    & Graph Transformer & 59.5$\pm$8.5 & 60.5$\pm$8.5 & 60.0$\pm$8.3 & 58.6$\pm$9.2  \\
    & BrainnetTransformer & 60.4$\pm$7.4 & 65.1$\pm$15.9 & 58.6$\pm$15.4 & 59.1$\pm$7.8   \\
    & LR-BrainTransformer & 58.6$\pm$7.0 & 59.8$\pm$10.7 & 64.4$\pm$8.5 & 60.9$\pm$4.7   \\
    & TSEN & 58.5$\pm$7.6 & 59.3$\pm$7.9 & 58.6$\pm$7.5 & 57.8$\pm$7.6   \\
   \midrule
    \multirow{2}{*}{\parbox{3cm}{\centering HGNN-based models}} 
    & HGCN &  64.2$\pm$10.0 &  65.2$\pm$10.3  &  64.3$\pm$10.3 &  \underline{63.4$\pm$10.5}  \\
    & HGAT & 61.3$\pm$5.3 & 61.9$\pm$5.7 & 61.4$\pm$5.4 & 61.0$\pm$5.3   \\
      \midrule      
    \multirow{3}{*}{\parbox{3cm}{\centering PH-based models}} 
    & PH-MCI  & 53.6$\pm$4.1 & 58.8$\pm$6.5 & 60.0$\pm$8.4 & 55.5$\pm$6.6 \\
    & ATPGCN & 61.2$\pm$3.5 & 62.1$\pm$3.4 &  \underline{64.5$\pm$3.3 } & 62.9$\pm$3.3   \\
     & Brain-HORS & 60.4$\pm$7.9 & 60.8$\pm$7.6 & 60.4$\pm$7.9 & 60.0$\pm$8.4   \\
    \midrule
    \multirow{1}{*}{Our Framework} 
    & HOI-Brain & \textbf{66.1$\pm$4.0} & \textbf{69.0$\pm$4.0} & \textbf{66.3$\pm$4.2} & \textbf{64.7$\pm$4.7}  \\
    \bottomrule
  \end{tabular}
  \label{tab:performance_comparison on PPMI} 
  }
\end{table}

\begin{table}[t]
  \centering
  \caption{Performance comparison with five categories of baselines on the ABIDE dataset (\%). The best results are marked in bold and the second-best results are underlined.}
  \scalebox{0.74}{
  \begin{tabular}{@{}cccccc@{}}
    \toprule
   Type & Method & Accuracy & Precision & Recall & F1-score \\
    \midrule
    \multirow{4}{*}{\parbox{3cm}{\centering Traditional Machine Learning Models}}
    & MLP    & 58.3$\pm$6.3 & 60.3$\pm$5.0 & 59.6$\pm$5.4 & 58.7$\pm$6.4   \\
    & SVM     & 61.4$\pm$5.5 & 63.8$\pm$5.6 & 61.4$\pm$5.1 & 62.5$\pm$5.0   \\
    & Logistic Regression  & 63.1$\pm$5.9 & 65.5$\pm$5.9 & 63.1$\pm$5.3 & 64.2$\pm$5.3  \\
    & RandomForest  & 60.7$\pm$5.2 & 61.0$\pm$4.1 & \textbf{68.3$\pm$8.4} & 64.4$\pm$5.5   \\
    \midrule
    \multirow{7}{*}{\parbox{3cm}{\centering GNN-based models}}
    & GCN & 60.4$\pm$4.5 & 61.8$\pm$4.3 & 64.4$\pm$7.5 & 62.9$\pm$4.7   \\
    & GraphSAGE & 61.2$\pm$3.6 & 64.0$\pm$4.1 & 59.3$\pm$5.6 & 61.5$\pm$5.5   \\
    & GAT & 59.3$\pm$3.8 & 57.9$\pm$5.3 & 61.1$\pm$3.5 & 59.5$\pm$4.8   \\
    & GroupINN & 57.1$\pm$4.4 & 58.4$\pm$5.0 & 57.1$\pm$4.2 & 55.7$\pm$4.3   \\
    & FBNETGEN & 58.0$\pm$4.2 & 59.8$\pm$4.1 & 61.1$\pm$8.6 & 60.1$\pm$5.2   \\
     & BPI-GNN & 52.5$\pm$4.7 & 51.7$\pm$6.9 & 54.6$\pm$5.5 & 53.5$\pm$4.5   \\
    & BrainGNN & 59.8$\pm$5.5 & 59.5$\pm$5.5 & 59.9$\pm$5.7 & 59.8$\pm$5.5  \\
    \midrule
    \multirow{4}{*}{\parbox{3cm}{\centering Transformer-based models}}
    & Graph Transformer & 57.5$\pm$4.2 & 57.5$\pm$4.3 & 57.3$\pm$4.2 & 57.1$\pm$4.2  \\
    & BrainnetTransformer & 63.1$\pm$5.8 & 64.6$\pm$5.8 &  \underline{66.1$\pm$6.0} &  \underline{65.2$\pm$5.2}   \\
    & LR-BrainTransformer &  \underline{63.1$\pm$4.3} &  \underline{65.5$\pm$4.9} & 63.3$\pm$6.0 & 64.2$\pm$4.3   \\
    & TSEN & 60.5$\pm$4.6 & 59.3$\pm$3.9 & 59.1$\pm$5.1 & 59.8$\pm$3.6   \\
   \midrule
    \multirow{2}{*}{\parbox{3cm}{\centering HGNN-based models}} 
    & HGCN & 61.0$\pm$3.0 & 55.2$\pm$5.3 & 62.3$\pm$4.6 & 60.4$\pm$4.3  \\
    & HGAT & 58.3$\pm$4.3 & 51.9$\pm$5.7 & 61.4$\pm$3.4 & 59.6$\pm$6.3   \\
      \midrule      
    \multirow{3}{*}{\parbox{3cm}{\centering PH-based models}} 
    & PH-MCI  & 57.9$\pm$5.2 & 58.4$\pm$4.6 & 61.5$\pm$4.5 & 59.3$\pm$6.2 \\
    & ATPGCN & 62.2$\pm$4.5 & 62.4$\pm$5.3 & 60.6$\pm$5.3 & 61.8$\pm$4.6   \\
     & Brain-HORS & 63.0$\pm$3.4 & 63.1$\pm$3.5& 63.4$\pm$3.4 & 62.9$\pm$3.5   \\
    \midrule
    \multirow{1}{*}{Our Framework} 
    & HOI-Brain & \textbf{65.6$\pm$3.5} & \textbf{66.2$\pm$3.8} & 65.6$\pm$3.4 & \textbf{65.3$\pm$3.5}  \\
    \bottomrule
  \end{tabular}
  \label{tab:performance_comparison on ABIDE} 
  }
\end{table}
\subsection{Model Comparison}

Tables \ref{tab:performance_comparison on ADNI}--\ref{tab:performance_comparison on ABIDE} present the comparative results across the ADNI, TaoWu, PPMI, and ABIDE datasets. Performance metrics (mean ± standard deviation) from 10-fold cross-validation are reported, with the best and second-best results highlighted in bold and underlined, respectively.

Several key observations emerge from our comparative analysis. First, general-purpose GNN-based models demonstrate comparable performance to traditional machine learning methods in most cases, without showing significant advantages. This can be attributed to neural networks' increased parameter count when utilizing edge features, rendering them more prone to overfitting while maintaining better interpretability. Second, Transformer-based models consistently outperform GNN-based approaches across all four datasets, suggesting their superior capacity for capturing global connectome structures. Third, HGNN-based models show limited performance advantages, with notable results only on the PPMI dataset. We attribute this to: (1) their excessive parameterization increasing overfitting risk, and (2) their \textit{k}-hop neighbor modeling failing to adequately represent co-fluctuation patterns among brain regions, thus poorly capturing HOIs. Notably, PH-based models (e.g., ATPGCN and Brain-HORS) demonstrate significant superiority over both GNN-based and traditional machine learning approaches by incorporating persistent homology-derived higher-order loop and connected component features. This underscores the critical importance of higher-order topological features in brain disease diagnosis.

Against this backdrop, HOI-Brain achieves competitive or superior performance compared with all 20 baselines on all four datasets. In particular, when we focus on the strongest baseline for each dataset (e.g., LR-BrainTransformer on ADNI), HOI-Brain attains comparable or slightly higher accuracy and F1-score, with consistent though sometimes modest absolute gains.
When compared with the strongest baseline (LR-BrainTransformer) on the ADNI dataset, HOI-Brain achieves average improvements of approximately 3.6(\%) in accuracy, precision, recall, and F1-score. Relative to the strongest baseline (BrainnetTransformer) on the TaoWu dataset, the corresponding average gains are about 4.7(\%) across these metrics. On the PPMI dataset, HOI-Brain outperforms the strongest baseline (HGCN) by roughly 2.9(\%) on average, while on the ABIDE dataset it yields average improvements of around 1.3(\%) over the strongest baseline (BrainnetTransformer).
Finally, when compared to PH-based models that already leverage persistent homology, HOI-Brain still yields noticeable improvements in accuracy, precision, recall and F1-score (e.g., compared with ATPGCN on the ADNI, HOI-Brain achieves average improvements of approximately 5.1(\%). We attribute these gains partly to linking signed higher-order interactions to higher-order organizations in a principled way, and partly to exploring more complex structures such as signed quadruplet interactions and two-dimensional voids formed by triangular faces and quadruplets. 

Overall, these experimental results demonstrate that HOI-Brain offers a consistent, interpretable and competitively accurate framework for brain disorder diagnosis across multiple cohorts, rather than dominating all baselines by a large margin on a single dataset.

\subsection{Model analysis}
\subsubsection{Ablation Study}

\begin{table}[t]
  \centering
  \caption{Performance comparison with varying combinations of features on the ADNI dataset (\%). The best results are marked in bold and the second-best results are underlined.}
  \scalebox{0.74}{
  \begin{tabular}{@{}cccccc@{}}
    \toprule
  \multirow{2}{*}{Method} & \multicolumn{4}{c}{ADNI} \\
    \cmidrule(lr){2-5} 
     & Accuracy & Precision & Recall & F1-score &   \\
    \midrule
     edge   & 68.4$\pm$5.5 & 69.9$\pm$5.4 & 69.3$\pm$6.0 & 68.1$\pm$5.7 \\
     edge+violating triangles+1D loops  & 59.7$\pm$7.5 & 61.8$\pm$9.5 & 60.6$\pm$7.6 & 59.0$\pm$7.9  \\
    edge+violating triangles+good quadruplets  & 73.9$\pm$7.8 & 74.6$\pm$7.2 & 73.8$\pm$8.0 & 73.7$\pm$7.8 \\
     edge+1D loops+2D voids  & 62.9$\pm$6.6 & 68.2$\pm$10.2 & 63.9$\pm$6.4 & 61.7$\pm$6.9 \\
     edge+good quadruplets+2D voids & \underline{74.7$\pm$7.4} & \underline{76.2$\pm$7.8}& \underline{75.5$\pm$7.4}& \underline{74.6$\pm$7.6}\\
     edge+signed higher-order features+ extended Pearson & {72.3$\pm$6.5} & {75.5$\pm$6.4} &  {73.2$\pm$7.0} &  {72.5$\pm$7.1}   \\
    
     edge+signed good quadruplets+signed 2D voids & \textbf{75.9$\pm$8.6} & \textbf{80.0$\pm$7.9} & \textbf{75.6$\pm$9.1} & \textbf{75.5$\pm$9.2} \\
    \bottomrule
  \end{tabular}
  \label{tab:performance_comparison_ADNI} 
   }
\end{table}

\begin{table}[t]
  \centering
  \caption{Performance comparison with varying combinations of features on the ABIDE datasets (\%). The best results are marked in bold.}
  \scalebox{0.74}{
  \begin{tabular}{@{}cccccc@{}}
    \toprule
  \multirow{2}{*}{Method} & \multicolumn{4}{c}{ABIDE} \\
    \cmidrule(lr){2-5} 
     & Accuracy & Precision & Recall & F1-score &   \\
    \midrule
     edge   & 58.3$\pm$5.3 & 58.3$\pm$5.4 & 59.1$\pm$4.1 & 58.1$\pm$5.5 \\
     edge+violating triangles+1D loops  & 61.6$\pm$4.4 & 60.0$\pm$4.2 & 63.5$\pm$5.9 & 62.4$\pm$4.1   \\
     edge+violating triangles+good quadruplets  & 60.5$\pm$4.2 & 62.4$\pm$4.7 & 59.4$\pm$4.3 & 61.7$\pm$4.3  \\
     edge+1D loops+2D voids  & \underline{64.3$\pm$3.2} & 63.3$\pm$6.2 & 60.9$\pm$4.7 & \underline{63.6$\pm$5.3 }  \\
     edge+good quadruplets+2D voids & 63.7$\pm$4.5 & \underline{63.4$\pm$2.5}& {61.4$\pm$4.4}& 63.5$\pm$4.1   \\
     edge+signed higher-order features+ extended Pearson & {61.9$\pm$3.6} & {62.0$\pm$4.9} & \underline{63.6$\pm$4.1} & {63.2$\pm$5.2} \\
     edge+signed good quadruplets+signed 2D voids & \textbf{65.6$\pm$3.5} & \textbf{66.2$\pm$3.8} & \textbf{65.6$\pm$3.4} & \textbf{65.3$\pm$3.5}  \\
    \bottomrule
  \end{tabular}
  \label{tab:performance_comparison_ABIDE} 
   }
\end{table}
To evaluate the soundness and effectiveness of the proposed model’s architecture, we conducted a series of ablation experiments to assess the impact of different components on overall performance.
Our first aim was to evaluate the effectiveness of more complex signed quadruplet structures and 2D voids extracted by MTD methods.
Specifically, following the method in \citep{34}, we extracted lower-order edge features, violating triangles, and one-dimensional loops. In addition, we extracted unsigned good quadruplets and unsigned two-dimensional voids based on our MTD methods.  We implement it within different combinations of these features on all datasets. We also extracted signed good quadruplets and signed two-dimensional voids based on extended Pearson correlation methods. The experimental results presented in Tables \ref{tab:performance_comparison_ADNI}, 
\ref{tab:performance_comparison_ABIDE}, 
\ref{tab:performance_comparison_Taowu}, and
\ref{tab:performance_comparison_PPMI} indicate that integrating both higher-order and lower-order interactions proves to be more effective for brain disorder diagnosis compared to using lower-order interactions alone. Furthermore, quadruplet interactions captured by HOI-Brain have been demonstrated to be more effective in improving diagnosis than triplet interactions. This may be because a significant proportion of triplet interactions can be decomposed into a linear combination of pairwise interactions, which is insufficient to capture the true HOIs. Additionally, distinguishing between positively and negatively synergistic HOIs to generate signed higher-order structures, as opposed to unsigned ones, can further improve the effectiveness of brain disease diagnosis, which may be due to the fact that it can provide detailed insights into the communication within the brain. 
Finally, Across all datasets, the MTD-derived higher-order features consistently yield stronger classification performance than extended Pearson correlation method. This indicate that by operating on temporal derivatives, captures the dynamic relationships among ROIs more accurately that is more informative for downstream diagnosis, thereby offering improved temporal sensitivity in practice.
\begin{figure}
    \centering
    \includegraphics[scale=0.46]{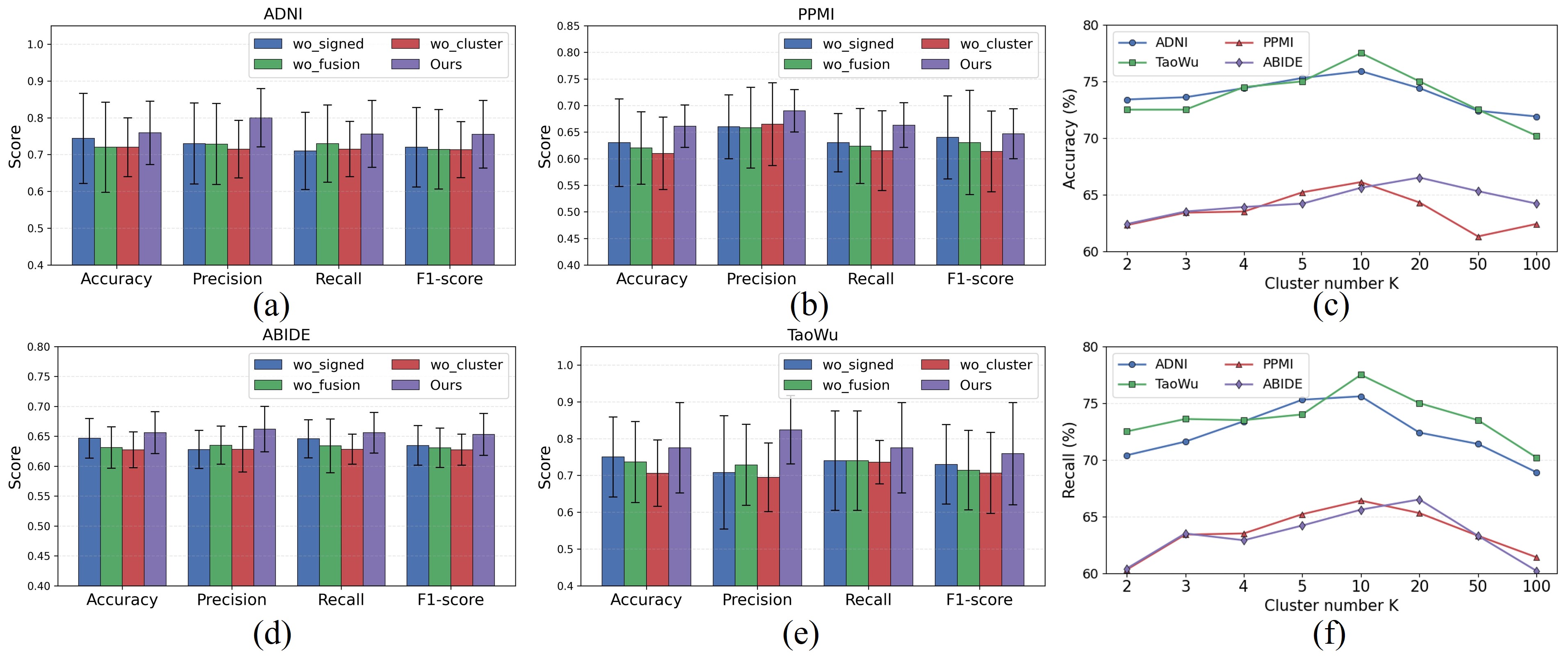}
   \caption{(a), (b), (d), and (e) present the ablation study of the multi-channel brain network Transformer in HOI-Brain across all datasets, comparing the full model with three degenerated variants: wo-signed (removing the signed higher-order feature decoupling mechanism), wo-fusion (removing the attention-guided feature fusion mechanism), and wo-cluster (removing the orthonormal clustering readout across multiple channels). (c) and (f) examine the impact of a key hyperparameter, the number of clusters K, on model performance, reported in terms of accuracy and recall, respectively. The results show that the model’s performance improves as the number of clusters K increases from 2 to 10 or 20 but declines when K rises from 10 or 20 to 100.
    \label{fig_3}}
\end{figure}

Our second aim was to evaluate the effectiveness of different components in multi-channel brain network Transformer. The different variants are: not adding signed higher-order features decoupling mechanism (wo-signed), not adding attention-guided feature fusion mechanism (wo-fusion), and not adding orthonormal clustering readout of multiple channels (wo-cluster). From Figure \ref{fig_3}(a), (b), (d), (e), we can see that adding signed higher-order features decoupling mechanism and attention-guided feature fusion mechanism is always helpful as HOI-Brain outperforms all variants, including (wo-signed) and (wo-fusion). This result is consistent with our assumption that by introducing the attention mechanism, our model can adaptively fuse both positive/negative information and multi-channel information, thereby emphasizing discriminative features while suppressing redundant information. Incorporating orthonormal clustering readout across multiple channels (wo-cluster) is highly beneficial, as this approach effectively captures both lower-order and higher-order feature patterns, which often reside within distinct functional modular structures. Overall, these results demonstrate the effectiveness of the multi-channel brain network Transformer designed in HOI-Brain.

\begin{figure}
\centering
\includegraphics[scale=0.6]{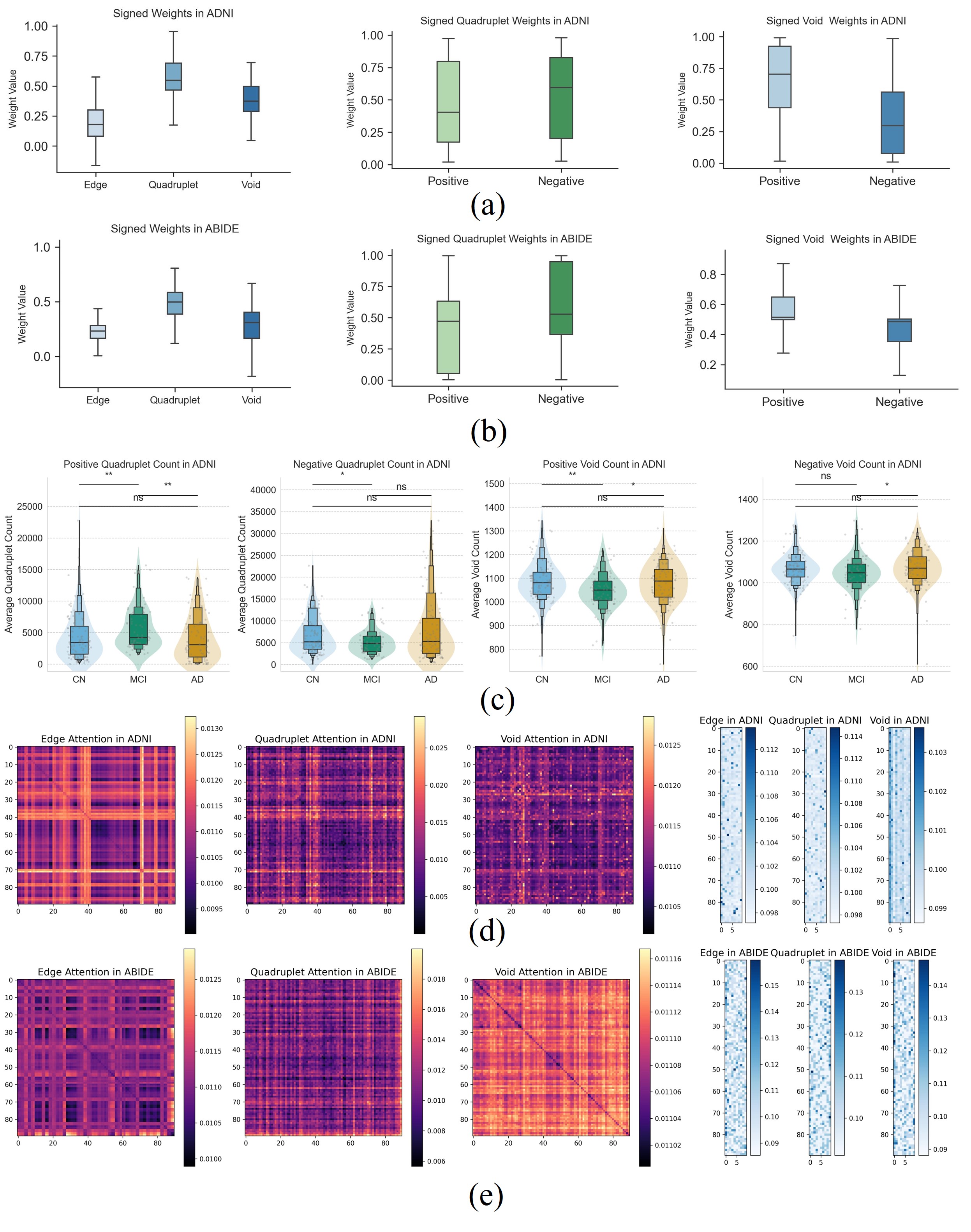}
\caption{(a) and (b) show, for the ADNI and ABIDE datasets respectively, the distributions of the learned weights for the three channels (left), the distributions of the learned signed weights for the quadruplet channel (middle), and the distributions of the learned signed weights for the void channel (right).
(c) presents the group-difference analysis on ADNI using Mann–Whitney U tests. Violin plots (with embedded box summaries) depict the average counts of positive/negative signed quadruplet interaction signatures and positive/negative signed two-dimensional void descriptors across the CN, MCI, and AD groups. Pairwise significance after FDR correction is denoted as: $^{***}p<0.001$, $^{**}p<0.01$, $^{*}p<0.05$, and ns $p\ge 0.05$.
(d) and (e) show, for the ADNI and ABIDE datasets, the attention patterns from the first layer of the multi-head self-attention module for the three channels (edge, quadruplet, and void), together with the corresponding cluster-assignment matrices produced by the multi-channel clustering readout. }
\label{fig_4}
\end{figure}

\subsubsection{Hyperparameter Analysis}
To further demonstrate how the design of orthonormal clustering readout across multiple channels affects our model's performance, we investigate a key hyperparameter: the number of clusters K. Specifically, we evaluate the method with K set to 2, 3, 4, 5, 10, 20, 50, and 100. The results of accuracy and recall metrics  when tuning this hyperparameter across all datasets are presented in Figure \ref{fig_3}(c), (f). The results of other metrics are presented in Figure \ref{fig:enter-label1}. We observe that the model's performance improves as K increases from 2 to 10 or 20 but declines when K rises from 10 or 20 to 100. This suggests that the optimal number of clusters is relatively small, reducing computational cost while aligning with the fact that the typical number of functional modules is fewer than 25, a finding consistent with previous studies \citep{20}.
\subsection{Model Interpretability }
\subsubsection{In-depth Analysis of Attention Mechanism}

To better understand the mechanism of our model, which focuses on the most discriminative information in the brain for the diagnosis of brain diseases, we first performed a visualization analysis of the attention scores $\gamma_{1}$, $\gamma_{2}$ of the attention-guided feature fusion mechanism for all patients of each type of disease, using ADNI and ABIDE as examples. Figure \ref{fig_4}(a), (b) show that our model is capable of adaptively learning weights according to the characteristics of different diseases, thereby integrating higher-order and lower-order topological information for the diagnosis of brain diseases. In addition, we observed that quadruplet-level interaction signatures are significantly more important than two-dimensional void descriptors and edge features, with their average attention scores exceeding 0.5 on both datasets.

Next, we also conducted a visualization analysis of the attention scores $\alpha_{1}$, $\alpha_{2}$, $\beta_{1}$, $\beta_{2}$ derived from the signed higher-order feature decoupling mechanism for all patients across each disease type, using ADNI and ABIDE as examples. Figure \ref{fig_4}(a), (b) show that across both datasets, the importance of negatively synergistic quadruplets is generally greater than that of positively synergistic information in most cases, while the importance of positively synergistic voids exceeds that of negatively synergistic information. This new phenomenon may provide a clearer direction for research into the pathological mechanisms of brain diseases.

In order to conduct a more in-depth investigation into the aforementioned phenomena in the attention mechanism, group comparisons were conducted of the signed higher-order topological features, the quadruplet-level interaction signatures and two-dimensional void descriptors using Mann–Whitney U tests \citep{83}. Results were corrected using Benjamini–Hochberg FDR correction \citep{82} for the 3 pairwise group comparisons. The ADNI dataset was used as an example. As demonstrated in Figure \ref{fig_4}(c), which presents a comparison of group differences in the number of signed higher-order topological features, it was observed that the group differences in the quadruplet-level interaction signatures were found to be highly significant. This finding is consistent with the results of our model visualisation. This phenomenon may be attributed to the disruption of HOIs within the brain during the progression of the disease, which in turn exerts an influence on the high-dimensional hole structures present within the brain. Furthermore, as the disease progresses, the number of positive quadruplets exhibits a gradual decrease, while the number of negative quadruplets initially decreases and subsequently increases. This finding suggests that during the transition from CN to MCI, there is a decline in synergistic interactions in the brain, whether positive or negative, as brain communication becomes progressively disrupted. Conversely, as the condition progresses from MCI to AD, brain regions have been observed to exhibit more negatively synergistic interactions, concurrently reducing their excitability to preserve the functionality of diverse neural modules. This finding aligns with the reference stating that excitatory neurons in the brains of Alzheimer's patients suffer severe genomic damage \citep{75}. With regard to the number of voids, although no clear trend was observed in the progression of the disease, it was found that positive voids exhibited more significant inter-group differences compared to negative voids. This finding lends further support to the hypothesis that they exhibit higher levels of attention scores. Consequently, CN appears to utilise positive information to construct the whole-brain topological structure, whereas AD relies more heavily on negative information. These findings may offer a novel perspective on the pathological mechanisms underlying Alzheimer's disease.

\subsubsection{In-depth Analysis of Cluster Mechanism}
To better understand how our model leverages modular-level similarities between ROIs across different brain network patterns, we first provide a visualization of three attention matrices at the first layer of the multi-head self-attention module on the ADNI and ABIDE datasets in Figures \ref{fig_4}(d), (e). The attention scores are averaged across all subjects in the ADNI or ABIDE test set. These figures demonstrate that the learned attention scores reveal varying degrees of modularity in both lower-order and higher-order patterns, highlighting the interpretability of our model. Furthermore, we provide a visualization of three cluster assignment matrices on the ADNI or ABIDE dataset in Figure \ref{fig_4}(d), (e). The cluster assignment matrices are multiplied across layers and averaged across all subjects in the ADNI and ABIDE test set. Each row corresponds to an ROI and each column is a cluster.  We observe that the assignment matrix, which is calculated based on the similarity between nodes and clusters, contains only one high-value entry per row in most cases, despite the fact that we did not impose any regularization constraints on the assignment matrix. This indicates that our model can effectively partition nodes into functional modules. Additionally, there are differences between the assignment matrices of low-order and higher-order patterns, which further demonstrates that there are varying degrees of modularity in both low-order and higher-order patterns. This result aligns with the visualization of three attention matrices, further indicating the interpretability of our model.

\subsubsection{The Important Brain Regions and Interactions of Brain Regions}
In this section, we further investigate whether the learned ROIs attention scores are interpretable and consistent with previous findings. Here, we take the ADNI, PPMI and ABIDE datasets as examples for discussion of three diseases. These three datasets contain larger sample sizes, enabling the reflection of more generalizable phenomena for brain diseases.
First, we analyze the important brain regions and interactions of brain regions associated with brain diseases using HOI-Brain from a more comprehensive perspective rather than a low-level perspective as in previous studies \citep{102,20,22}. For each subject, HOI-Brain generates three attention matrices at the first layer of multi-head self attention module. Higher attention scores indicate greater discriminative power of the regions for AD, PD, and ASD. For each disease, we obtain a comprehensive attention matrix that simultaneously captures both lower-order and higher-order patterns by performing a weighted summation of the three attention matrices for each individual, with the weights derived from the attention scores of each channel. At the regional and regional interaction level, we average the integrated attention matrices across all individuals to construct a matrix representing the importance of interactions between regions. At the regional level, we further compute the row-wise mean of this matrix to obtain the final regional importance scores. More details about the  distributions of attention scores for all ROIs for each dataset
(ADNI, PPMI, and ABIDE) can be found in \ref{The Attention Scores for ROIs}. We visualize the top ten ROIs and interactions of ROIs with the highest attention scores using BrainNet Viewer \citep{56}.

\begin{figure}
\centering
\includegraphics[scale=0.7]{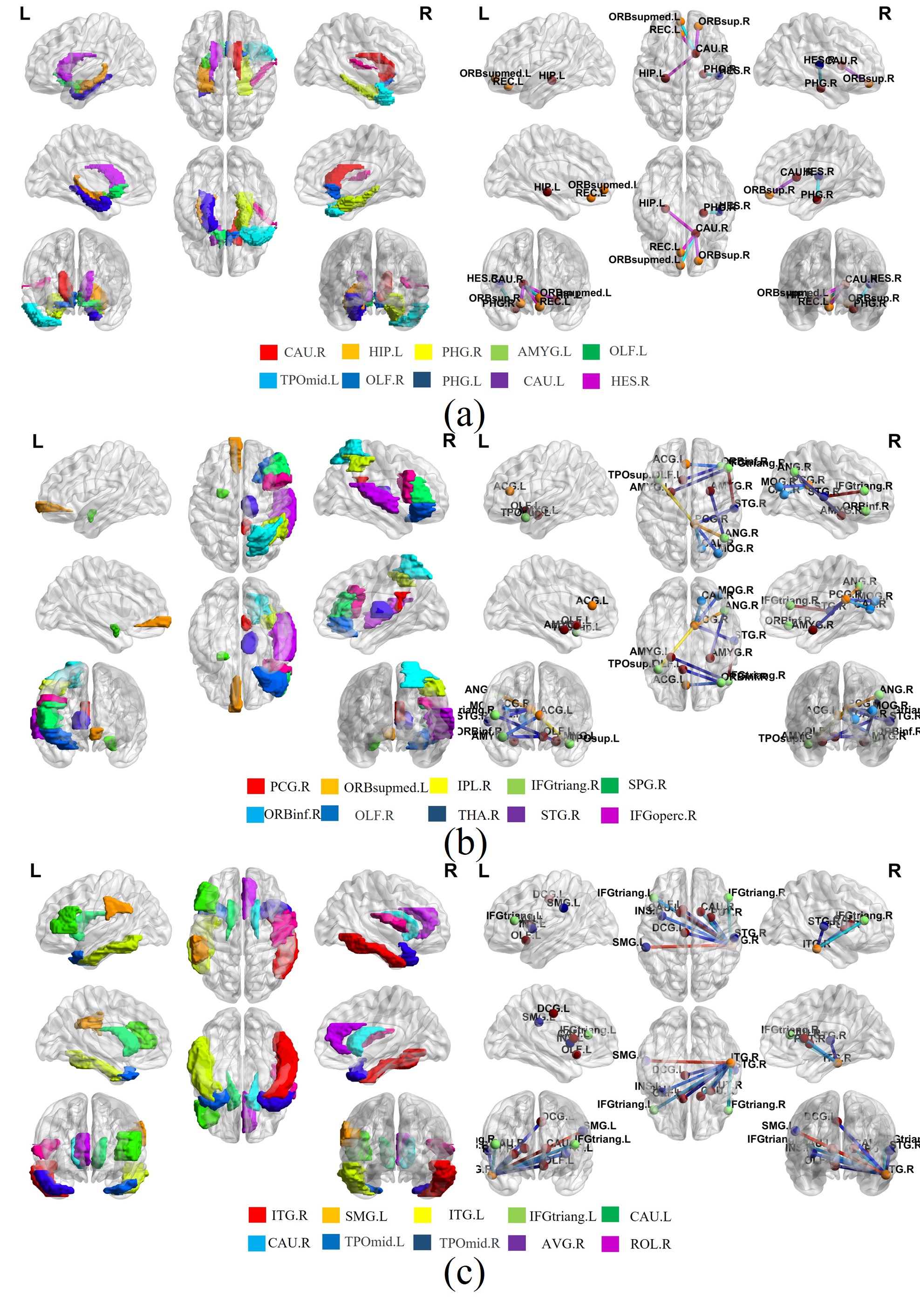}
\caption{(a) shows the visualization of top 10 important brain regions (left) and top 10 important
interactions between brain regions (right) associated with ADNI. The following two lines display the colors of the brain regions and their corresponding names. (b) shows the visualization of top 10 important brain regions (left) and top 10 important interactions of brain regions (right) associated with PPMI. (c) shows the visualization of top 10 important brain regions (left) and top 10 important interactions of brain regions (right) associated with ABIDE.}
\label{fig_5}
\end{figure}

\textbf{Alzheimer’s disease (AD).}
As shown in Figure \ref{fig_5}(a), the top attention ROIs include
CAU.R, HIP.L, PHG.R, AMYG.L, OLF.L, TPOmid.L, OLF.R, PHG.L, CAU.L, and HES.R.
Converging resting-state fMRI evidence indicates that AD/MCI is characterized by disrupted large-scale functional connectivity, prominently involving medial temporal and limbic circuitry (e.g., HIP, PHG and AMYG) and its interactions with other networks \citep{84}.
Consistent with this network perspective, our high-attention interaction motifs concentrate on limbic-striatal coupling patterns, with CAU-related edges repeatedly emerging and in line with prior reports of altered caudate-centered functional connectivity in AD \citep{85}, suggesting that HOI-Brain primarily captures higher-order co-fluctuation signatures at the circuit level rather than isolated regional effects.
We also observe sensory-network nodes (OLF and HES) among the top-ranked ROIs: olfactory dysfunction is well documented along the AD continuum \citep{86}, and altered olfactory-network functional connectivity with medial temporal regions (including the hippocampus) has been reported in AD \citep{87}.
In addition, intrinsic connectivity between auditory and reward systems differs across CN/MCI/AD in ADNI resting-state fMRI \citep{88}, and hearing impairment has been recognized as an important dementia-related factor at the population level \citep{89};
therefore, the prominence of auditory- and olfactory-related ROIs in our attention map is more plausibly interpreted as informative sensory-network coupling patterns that merit targeted validation in independent cohorts.

\textbf{Parkinson’s disease (PD).}
As illustrated in Figure \ref{fig_5}(b), the top attention ROIs include PCG.R, ORBsupmed.L, IPL.R, AMYG.L, IFGtriang.R, SPG.R, ORBinf.R, THA.R, STG.R, and IFGoperc.R.
Converging evidence from resting-state connectomics suggests that PD is characterized by abnormal functional connectivity within the sensorimotor system and the cortico--basal ganglia-thalamo-cortical loop \citep{90,91}.
Accordingly, PCG together with parietal nodes (IPL and SPG) points to disrupted coupling within sensorimotor and fronto-parietal/dorsal-attention networks supporting motor control and action monitoring \citep{90,91}.
Also, orbitofrontal and inferior frontal regions (ORBsupmed, ORBinf, IFGtriang, and IFGoperc) are core components of fronto-striatal control circuits, for which altered resting-state connectivity and network reorganization have been repeatedly reported in PD \citep{93}.
In addition, limbic involvement (AMYG) is consistent with reports of abnormal amygdala-centered functional connectivity in PD (e.g., disrupted cortico-limbic coupling) \citep{94}.
We also observe THA and STG among the high-attention ROIs, which may reflect that thalamo-cortical integration (THA) and temporal/auditory network nodes (STG) can provide discriminative connectivity signals in PD \citep{95}.
Notably, ROI-level importance and interaction-level hubs capture complementary aspects of network organization: the core hubs in the interaction map (e.g., PCG.R, STG.R, IFGtriang.R, and ANG.R) are not necessarily the top-ranked regions by ROI importance, and vice versa. Together, these findings highlight two complementary notions of importance in functional brain networks: global centrality versus local strong coupling.

\textbf{Autism Spectrum Disorder (ASD).}
As illustrated in Figure \ref{fig_5}(c), the top attention ROIs include ITG.R, SMG.L, ITG.L, IFGtriang.L, CAU.L, CAU.R, TPOmid.L, TPOmid.R, ACG.R, and ROL.R.
Converging evidence from resting-state fMRI reviews suggests that ASD is characterized by widespread atypical functional connectivity across intrinsic brain networks rather than isolated regional deficits \citep{96}.
Also, temporal regions along the social-perceptual pathway (ITG and TPOmid) show reproducible connectivity abnormalities in ASD, consistent with altered temporo-occipital and temporo-frontal coupling reported previously \citep{97}.
In addition, abnormal cortico-striatal functional coupling involving the caudate (CAU) has been repeatedly linked to restricted and repetitive behaviors, supporting a striatal-circuit interpretation of our high-attention motifs \citep{98}.
Moreover, reduced anterior cingulate connectivity with the rolandic operculum and temporal regions has been reported in ASD, which aligns with the co-occurrence of ACG and ROL among our highly weighted ROIs and suggests disrupted salience--sensorimotor-temporal integration \citep{99}.
Finally, many of the top interaction motifs center on ITG.R, suggesting that temporal hubs may play an important role in the discriminative connectome patterns captured by HOI-Brain, consistent with prior ASD connectivity findings involving temporal systems \citep{62}.
Altogether, these results indicate that HOI-Brain highlights discriminative network-level coupling patterns spanning temporo-frontal social and language circuits, cortico-striatal loops, and sensorimotor integration systems in ASD.

\subsubsection{The Higher-Order Organizations that Differ between Patients and Healthy Individuals}
\begin{figure}
\centering
\subfigure{
\includegraphics[scale=0.50]{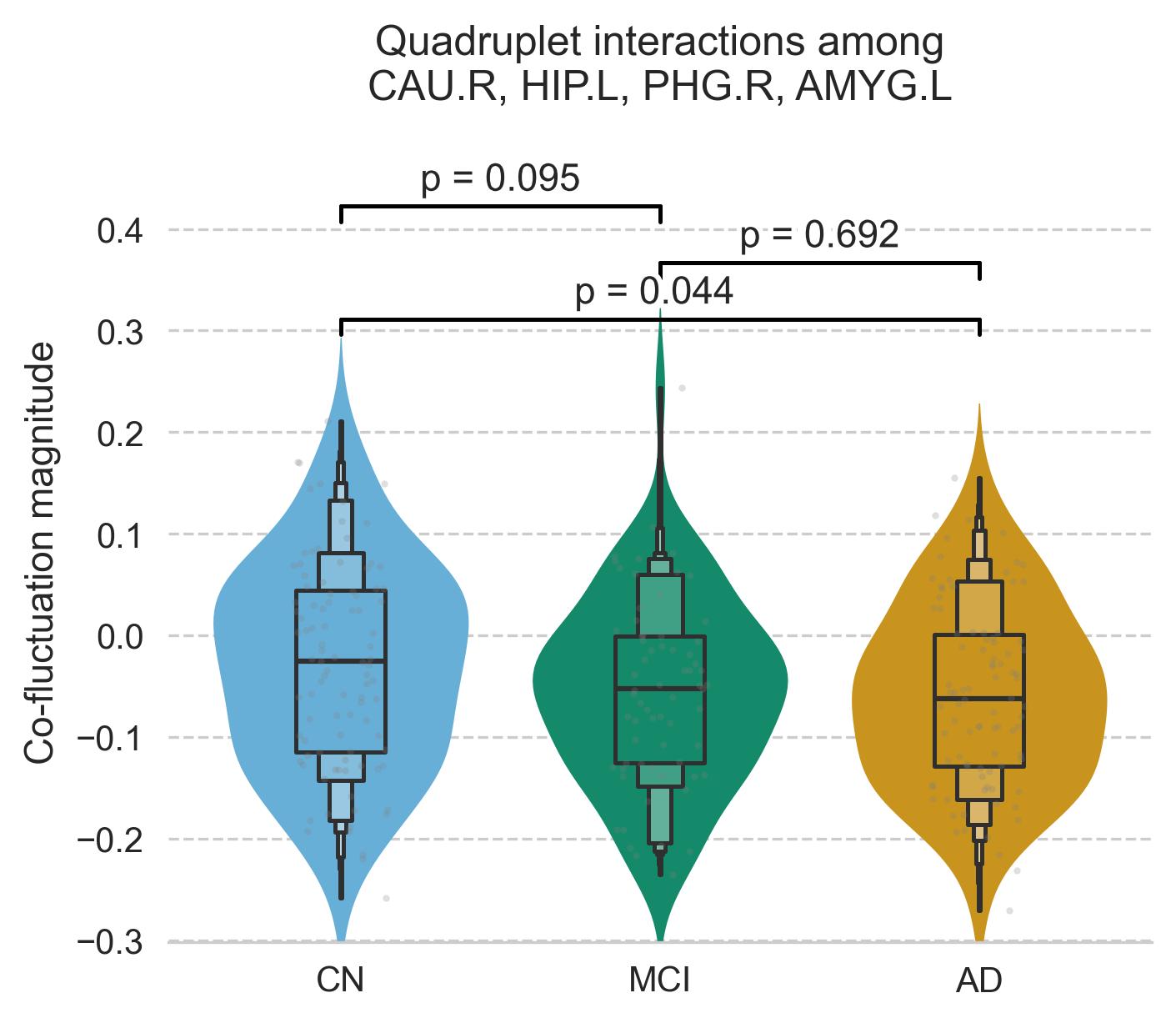}
}
\subfigure{
\includegraphics[scale=0.50]{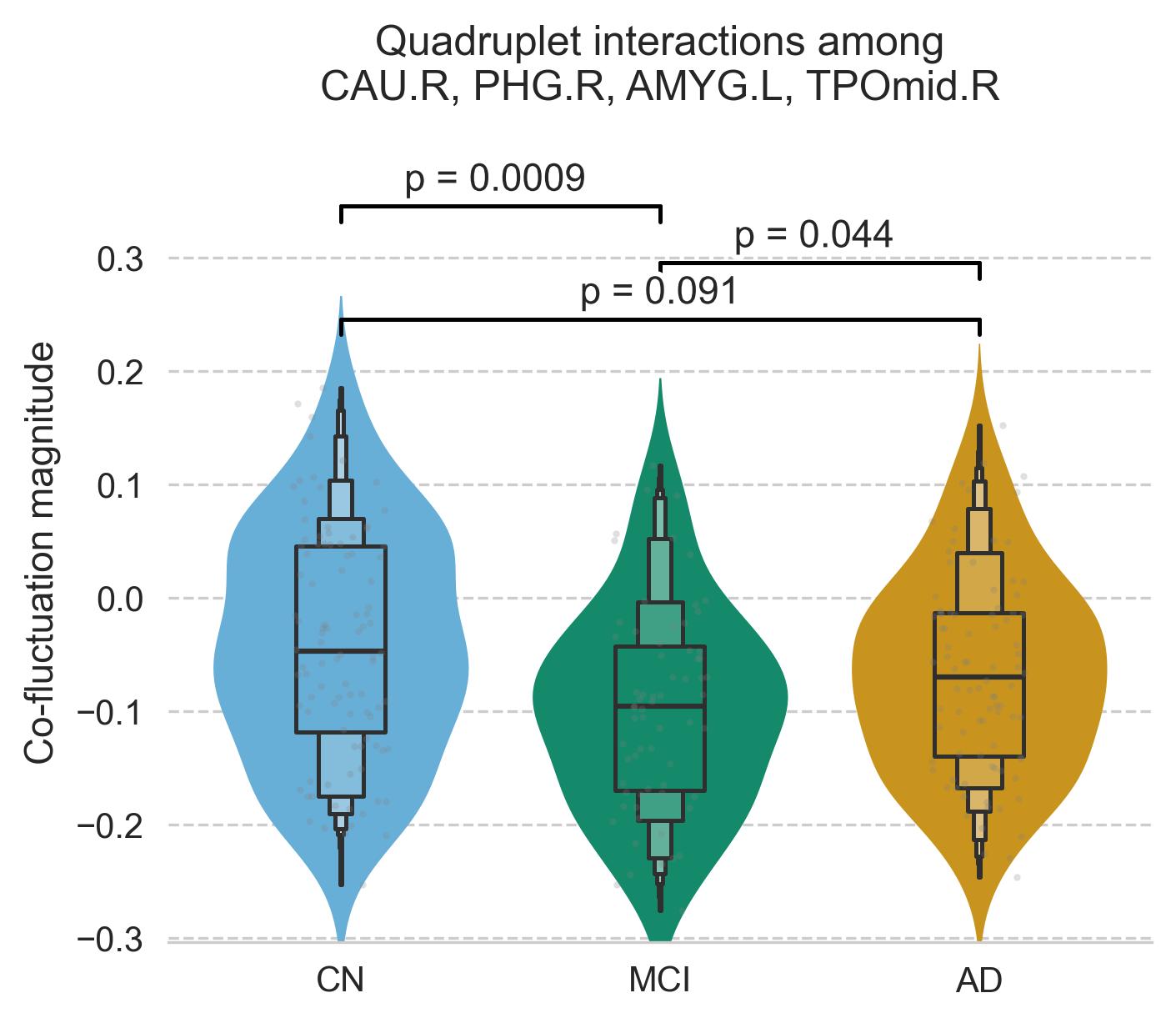}
}
\subfigure{
\includegraphics[scale=0.50]{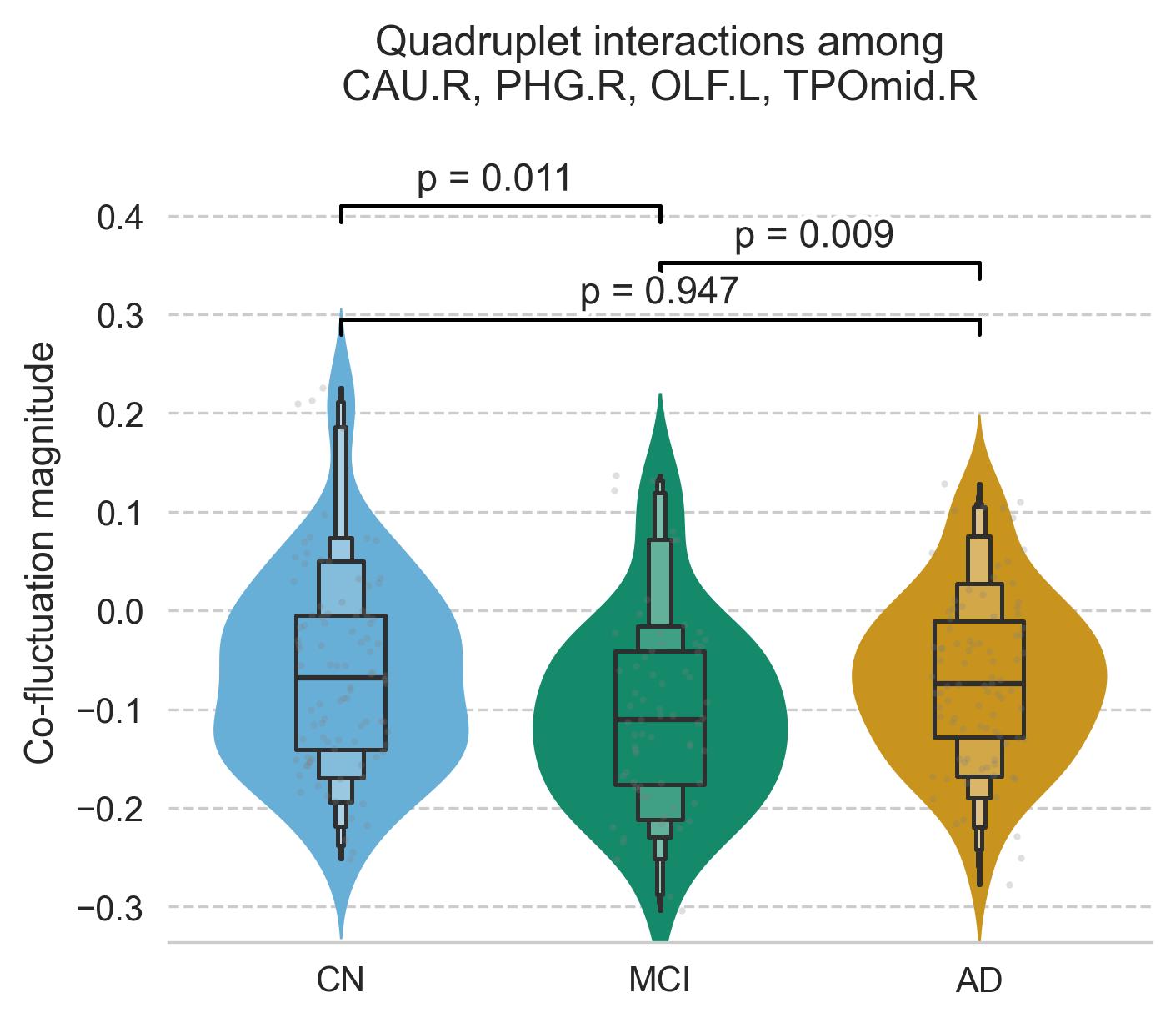}
}
\subfigure{
\includegraphics[scale=0.50]{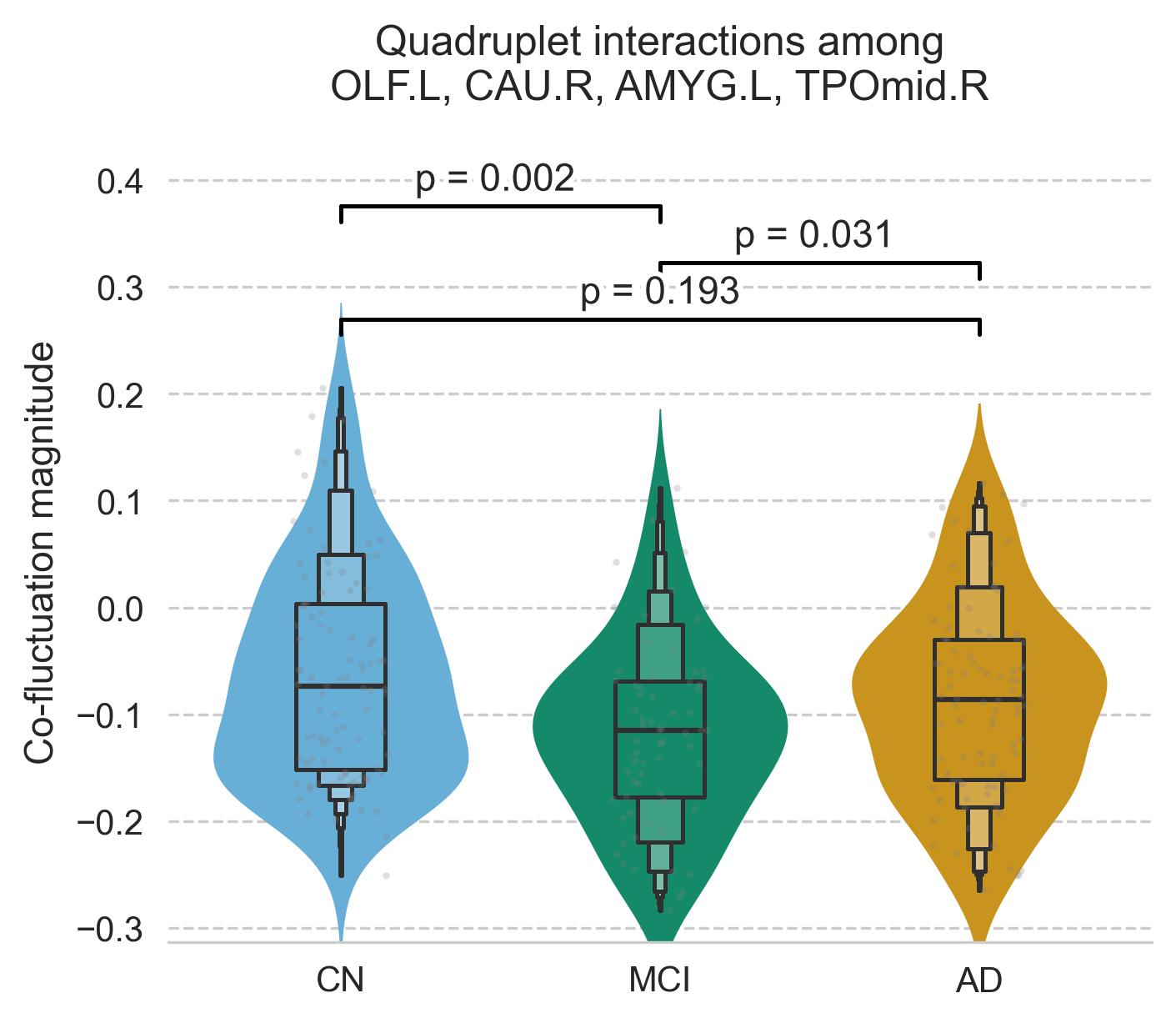}
}
\subfigure{
\includegraphics[scale=0.50]{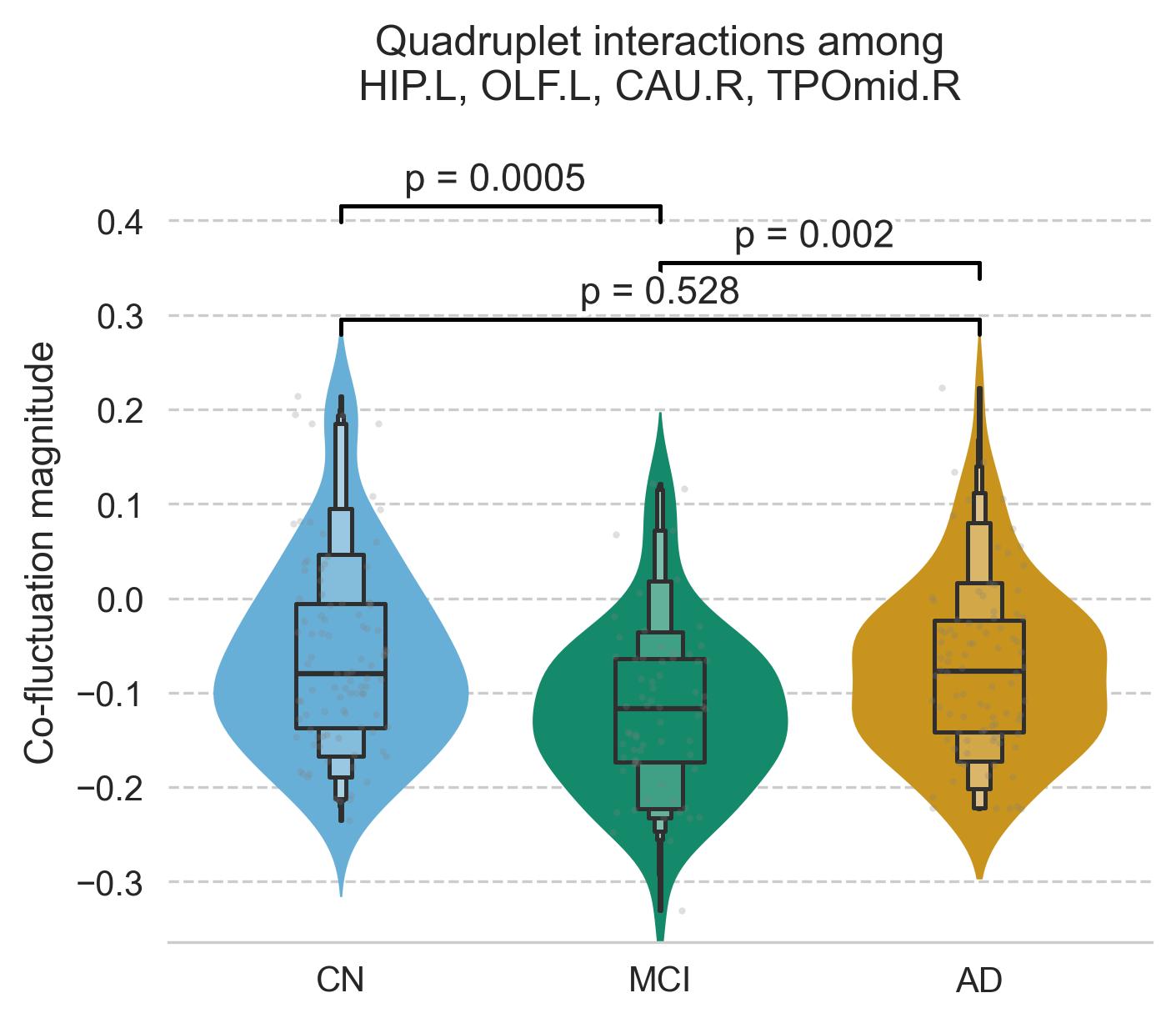}
}
\subfigure{
\includegraphics[scale=0.50]{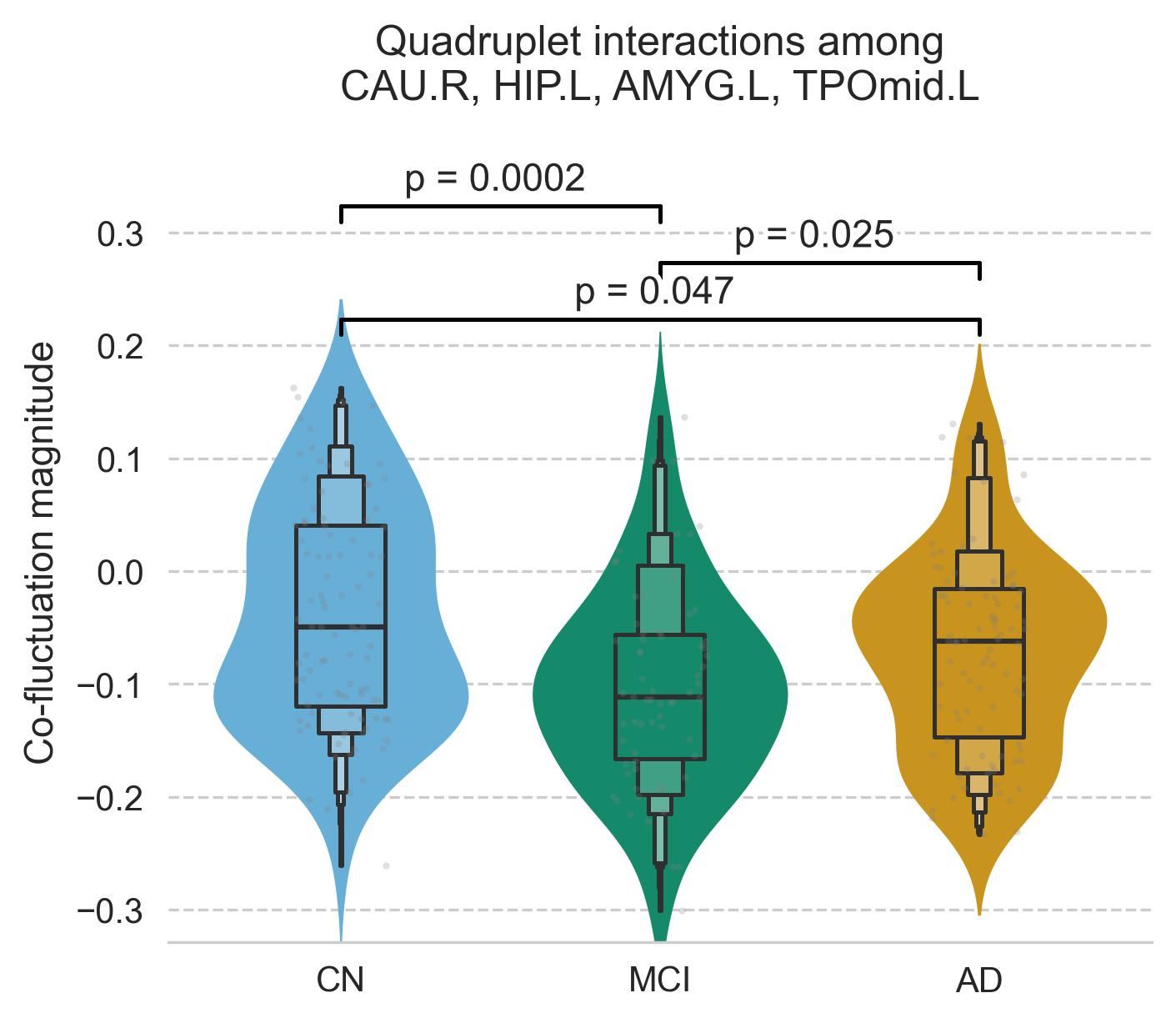}
}
\caption{Group difference comparison of quadruplet interactions among the relevant brain regions (CAU.R, HIP.L, PHG.R, AMYG.L, OLF.L, TPOmid.L) on ADNI. These figures show that the positive HOIs between relevant brain regions gradually weaken and the negative HOIs between these brain regions gradually strengthen during the transition from CN to MCI or AD.}
\label{quadruplet interactions ADNI}
\end{figure}

\begin{figure}
\centering
\subfigure{
\includegraphics[scale=0.50]{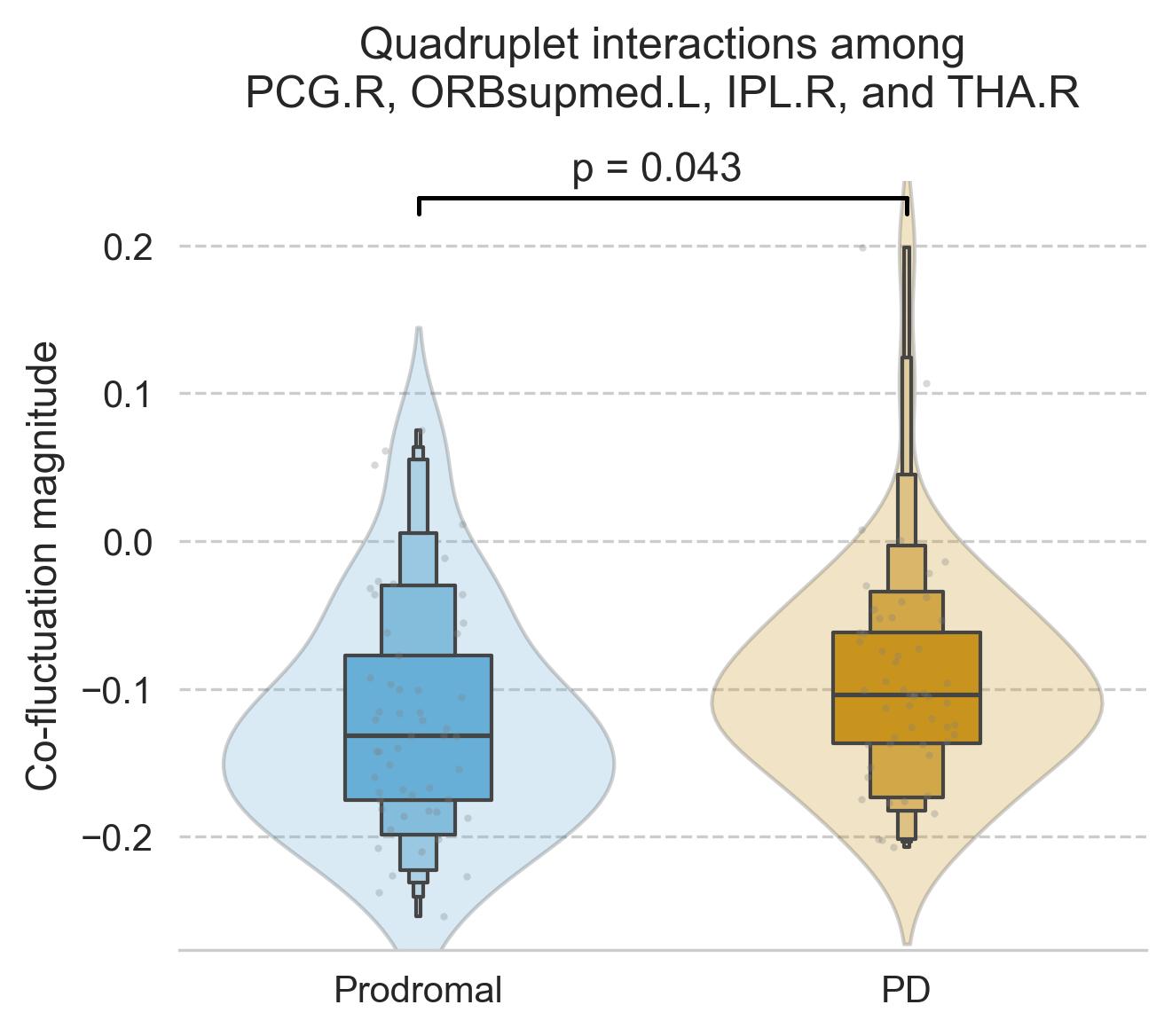}
}
\subfigure{
\includegraphics[scale=0.50]{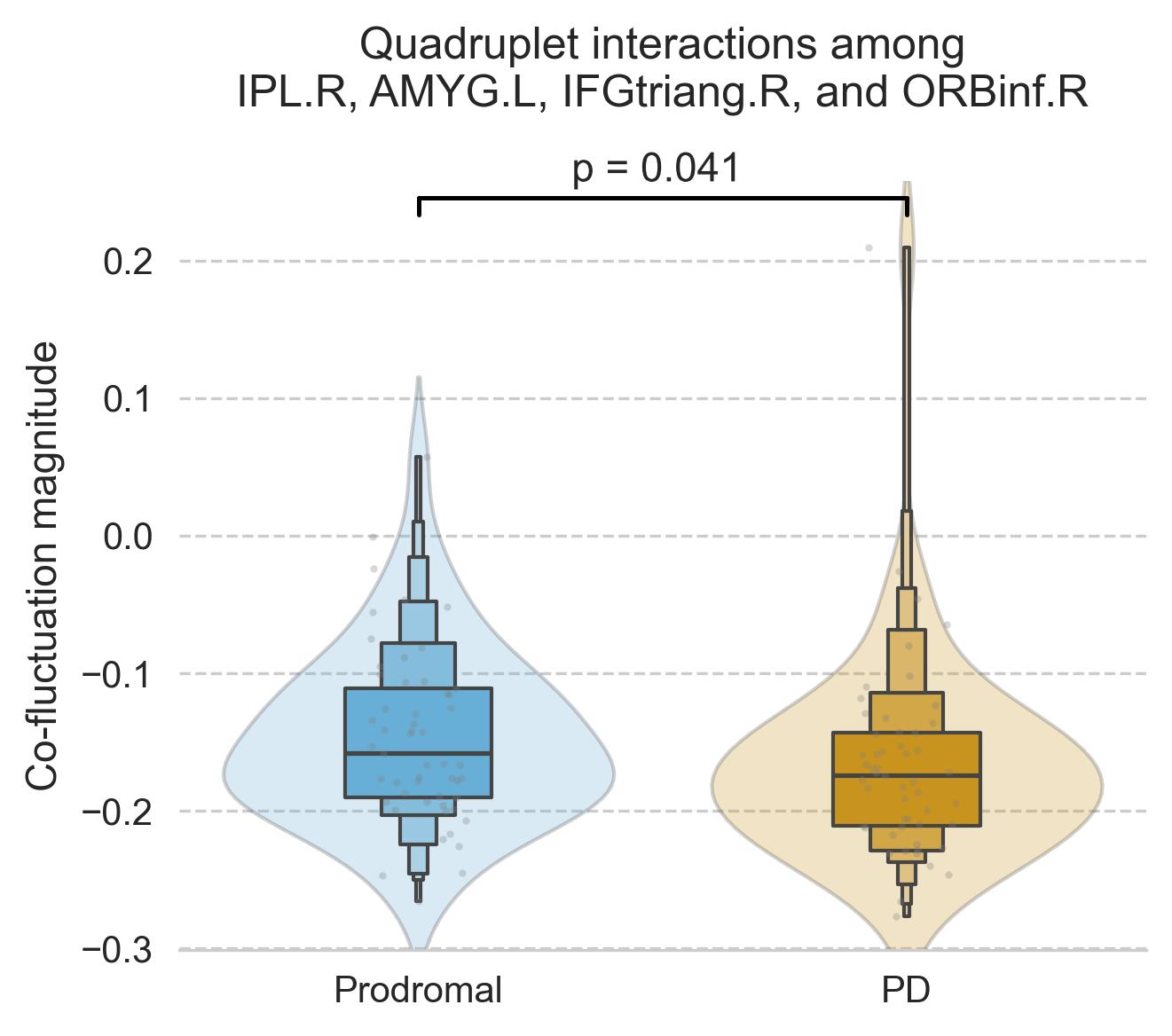}
}
\subfigure{
\includegraphics[scale=0.50]{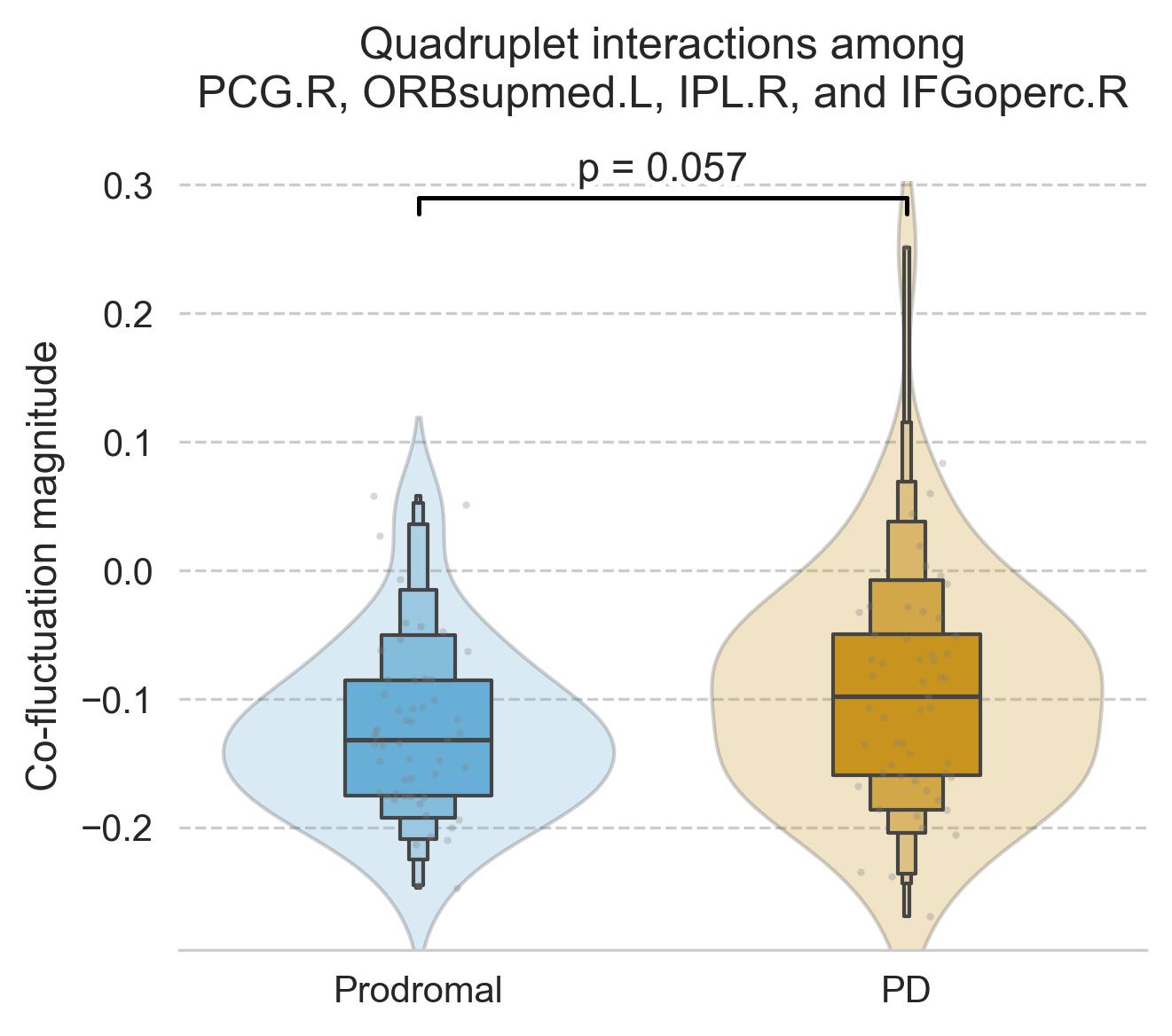}
}
\subfigure{
\includegraphics[scale=0.50]{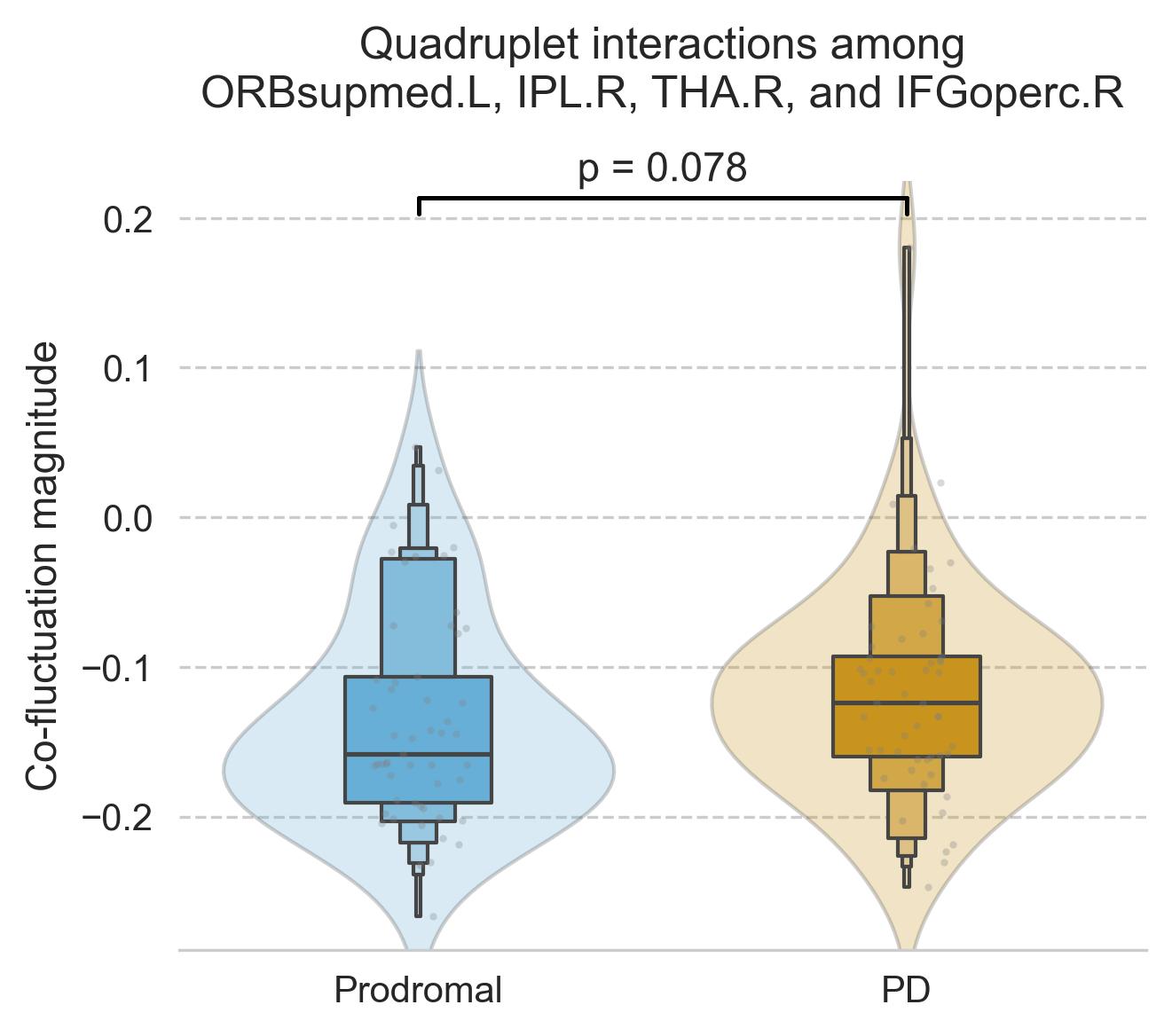}
}
\subfigure{
\includegraphics[scale=0.50]{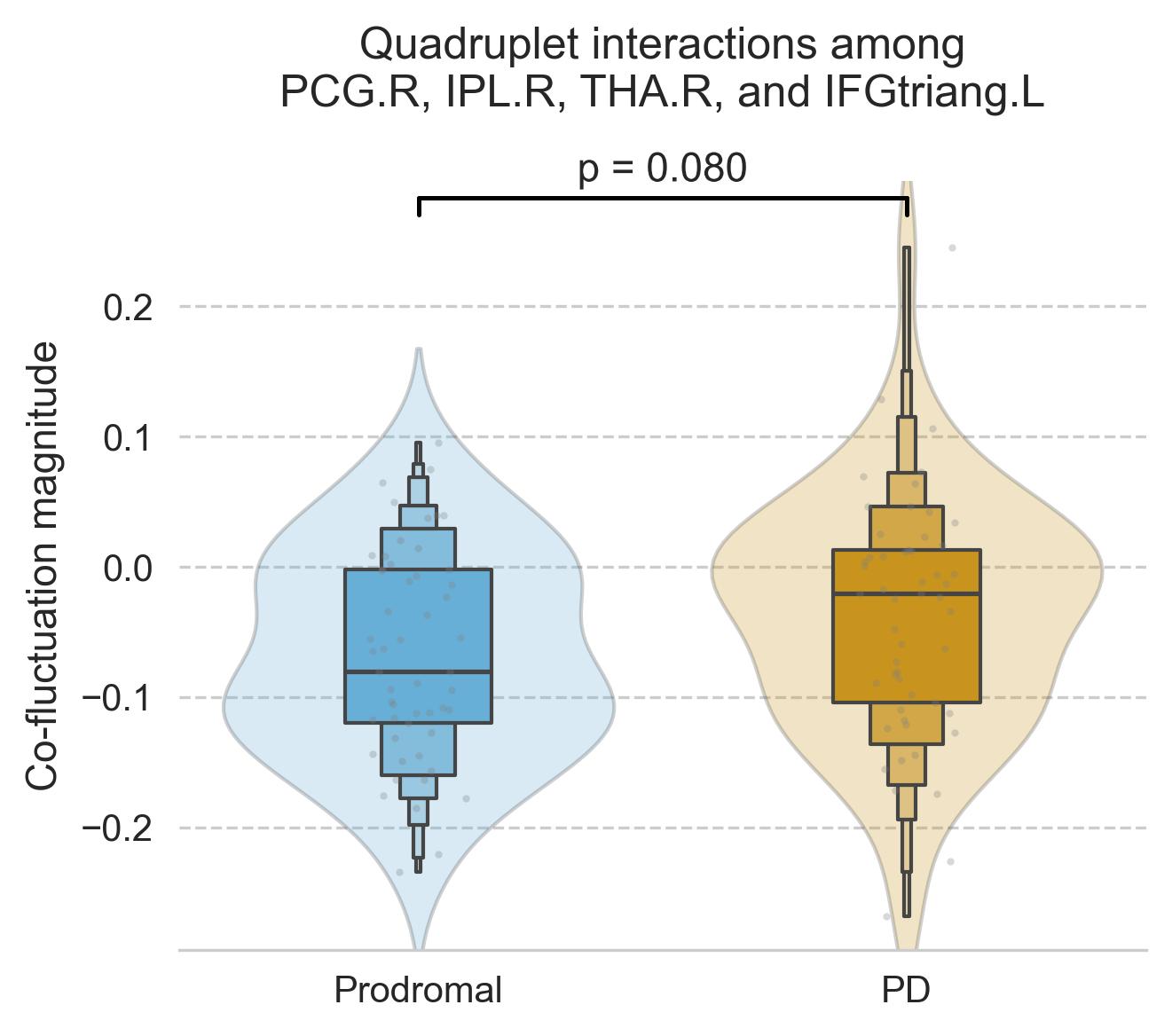}
}
\subfigure{
\includegraphics[scale=0.50]{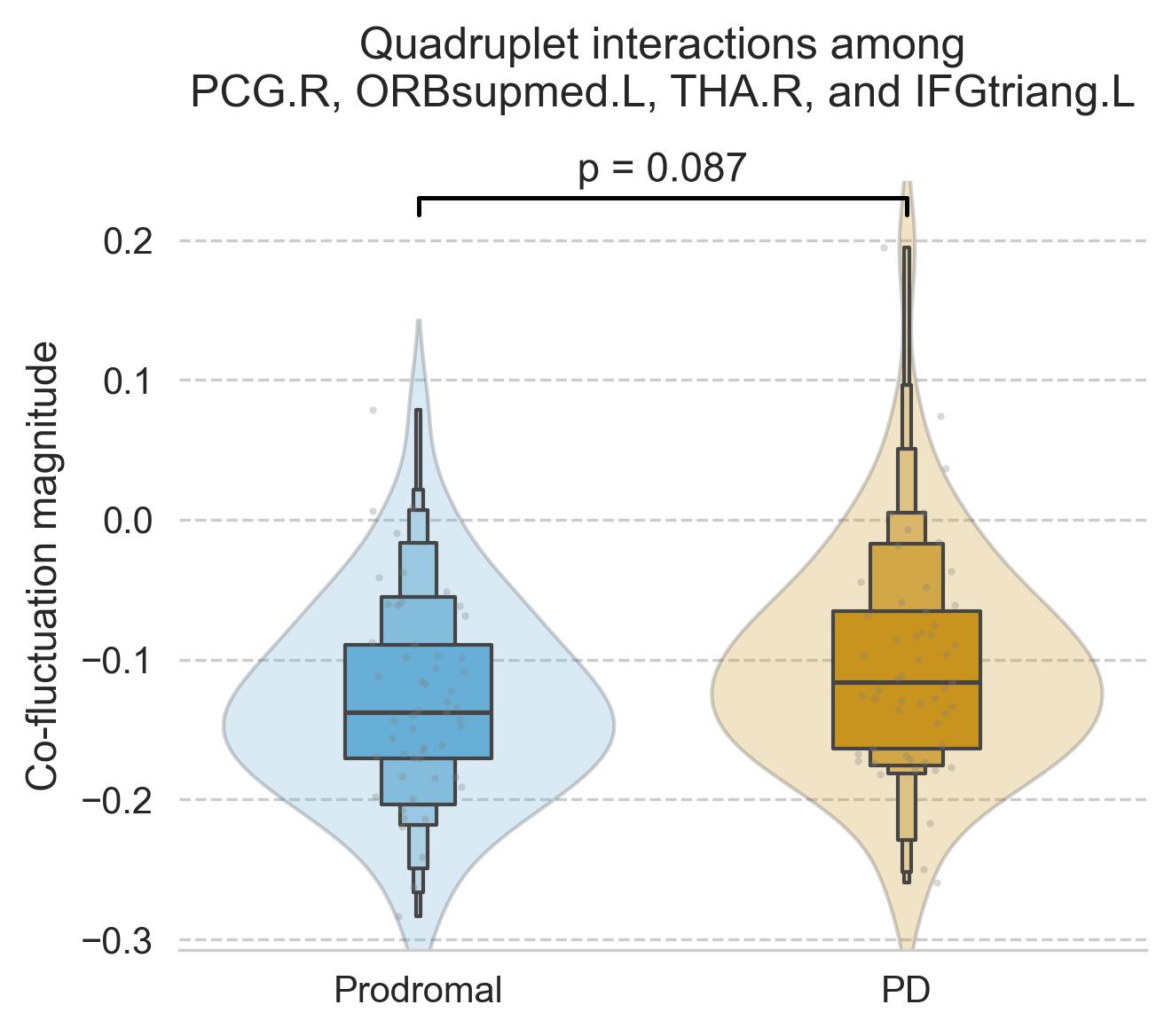}
}
\caption{Group difference comparison of quadruplet interactions among the relevant brain regions (PCG.R, ORBsupmed.L, IPL.R, THA.R, AMYG.L, IFGoperc.R, IFGtriang.R, ORBinf.R) on PPMI. These figures show that the positive HOIs between relevant brain regions gradually strengthen and the negative HOIs between these brain regions gradually weaken during the transition from prodromal to PD.}
\label{quadruplet interactions PPMI}
\end{figure}

\begin{figure}
\centering
\subfigure{
\includegraphics[scale=0.50]{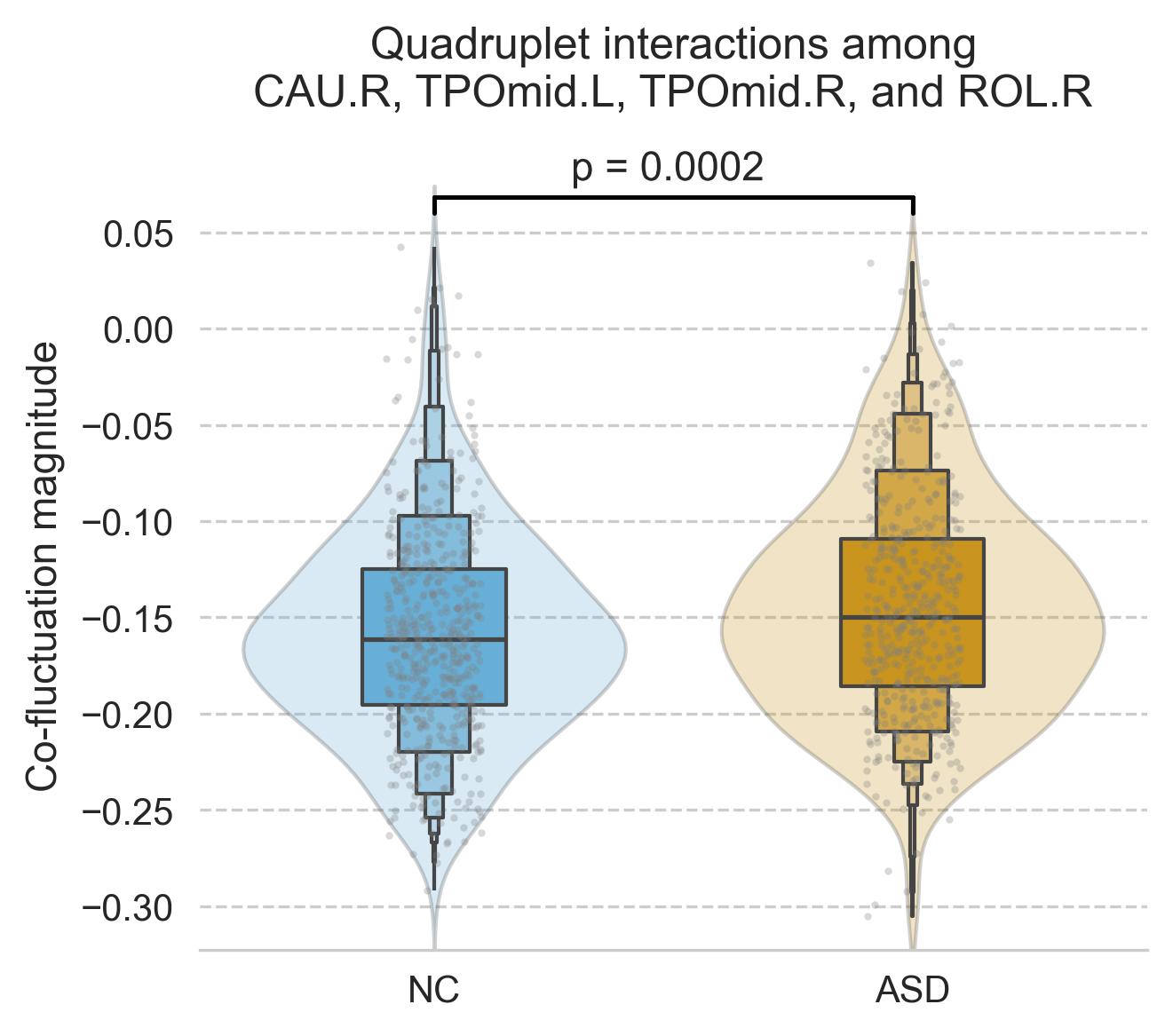}
}
\subfigure{
\includegraphics[scale=0.50]{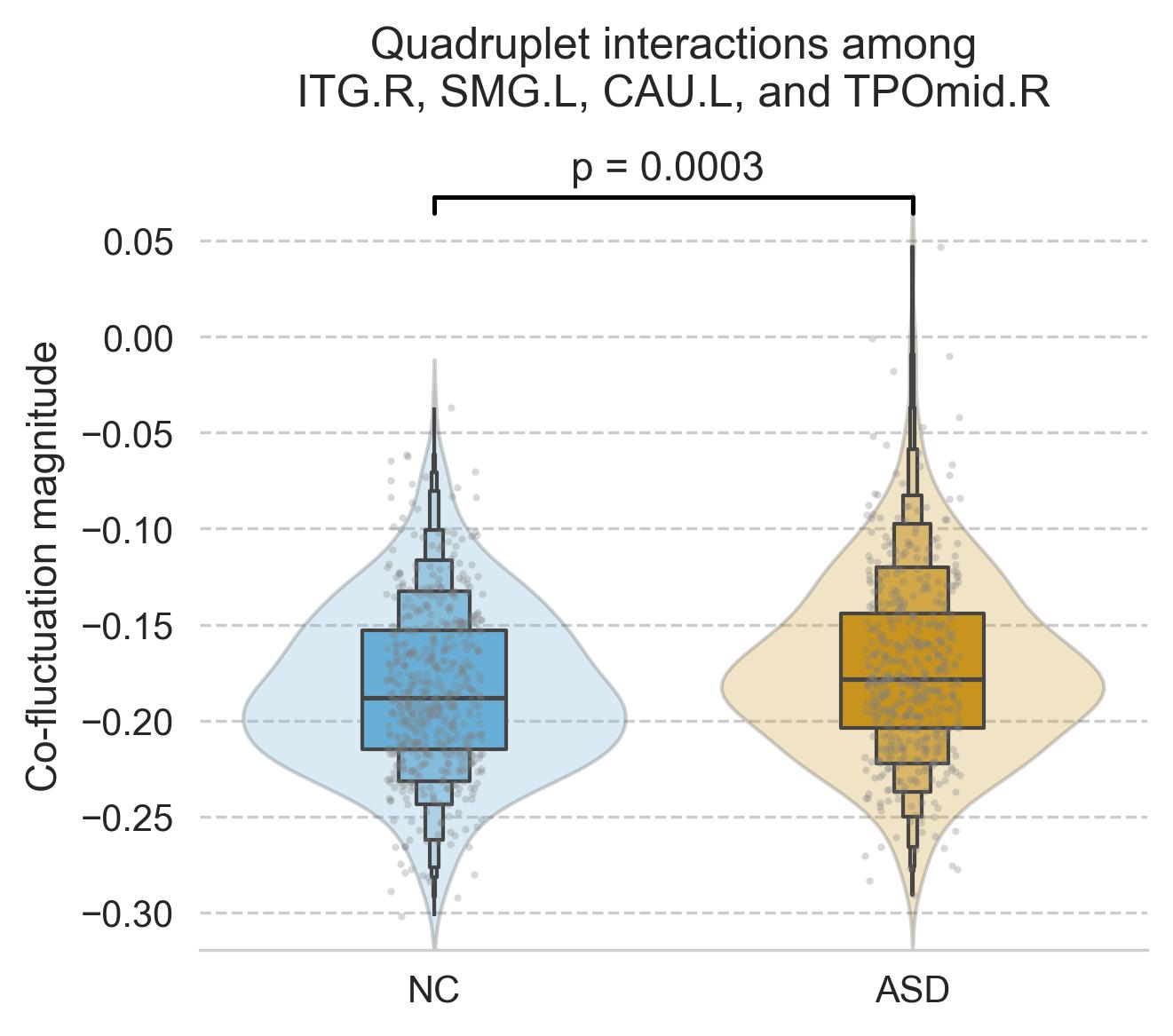}
}
\subfigure{
\includegraphics[scale=0.50]{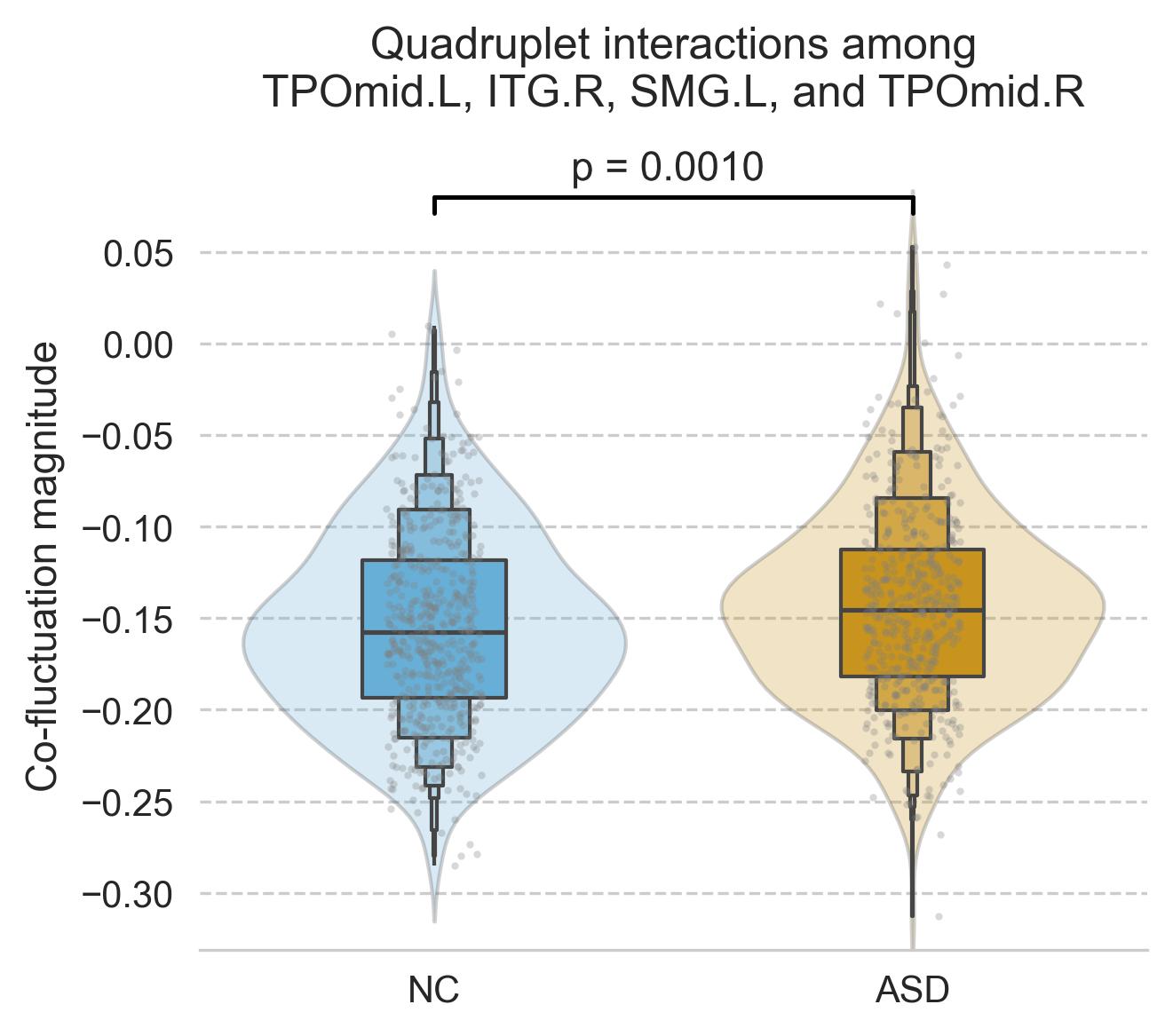}
}
\subfigure{
\includegraphics[scale=0.50]{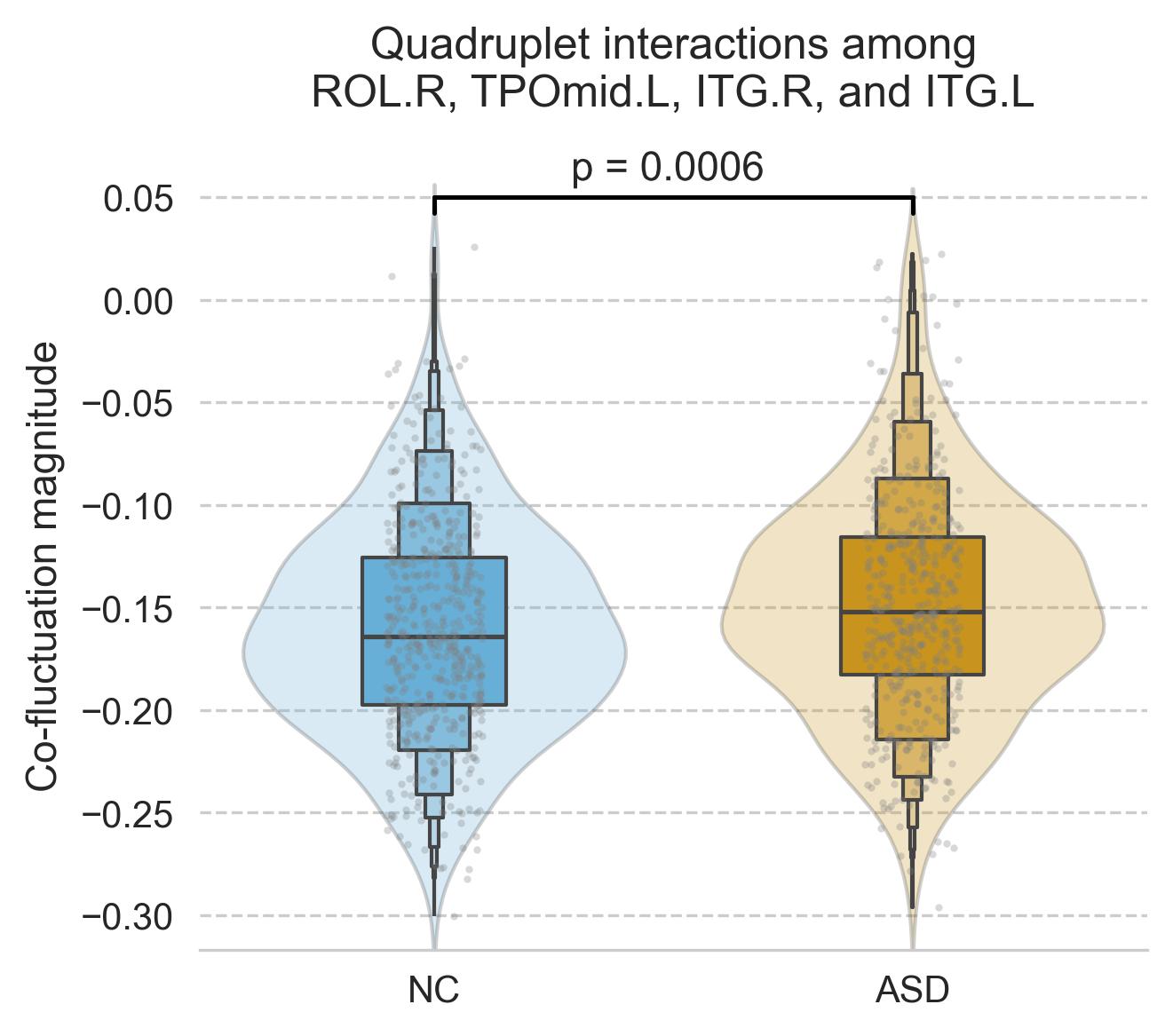}
}
\subfigure{
\includegraphics[scale=0.50]{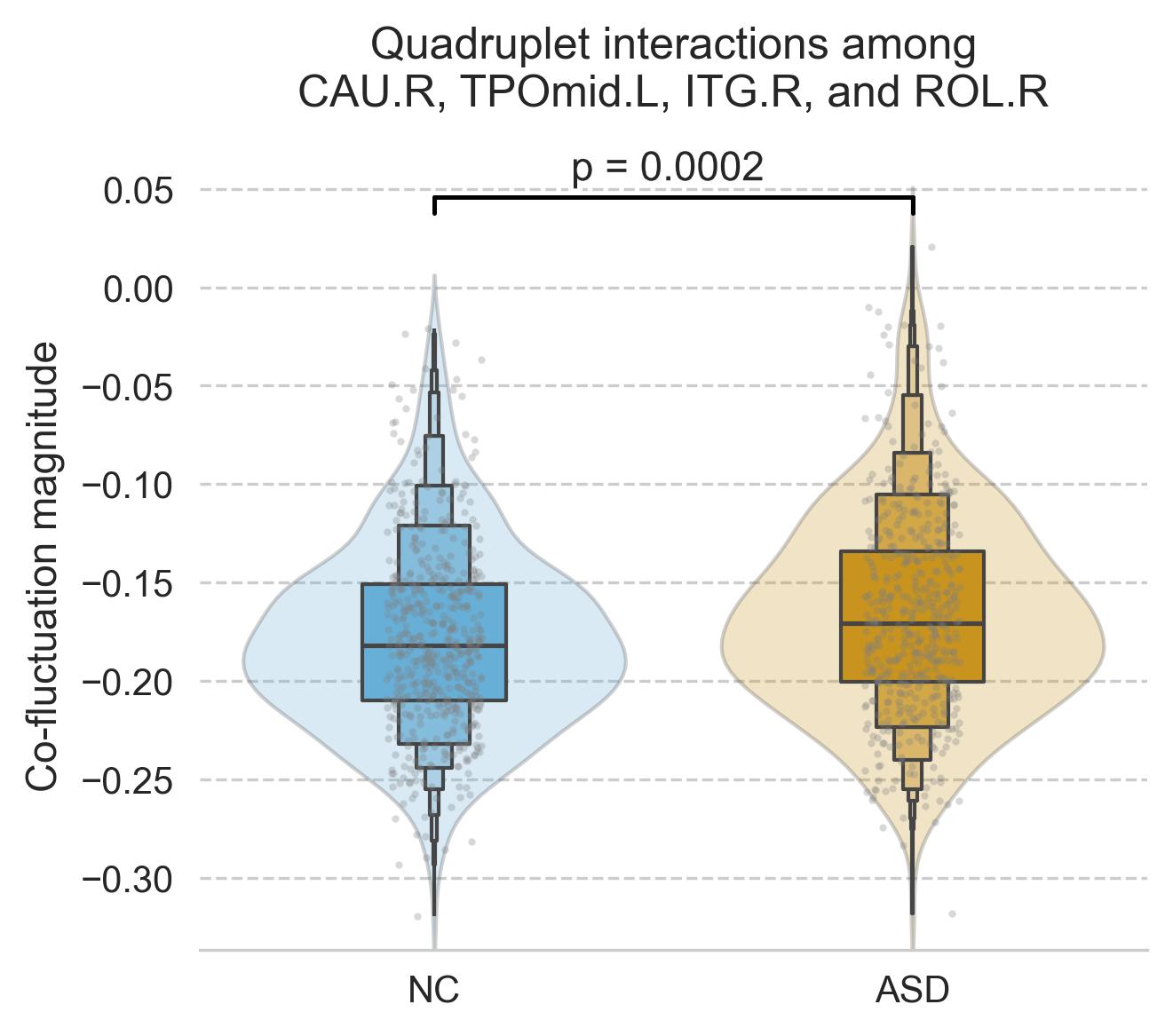}
}
\subfigure{
\includegraphics[scale=0.50]{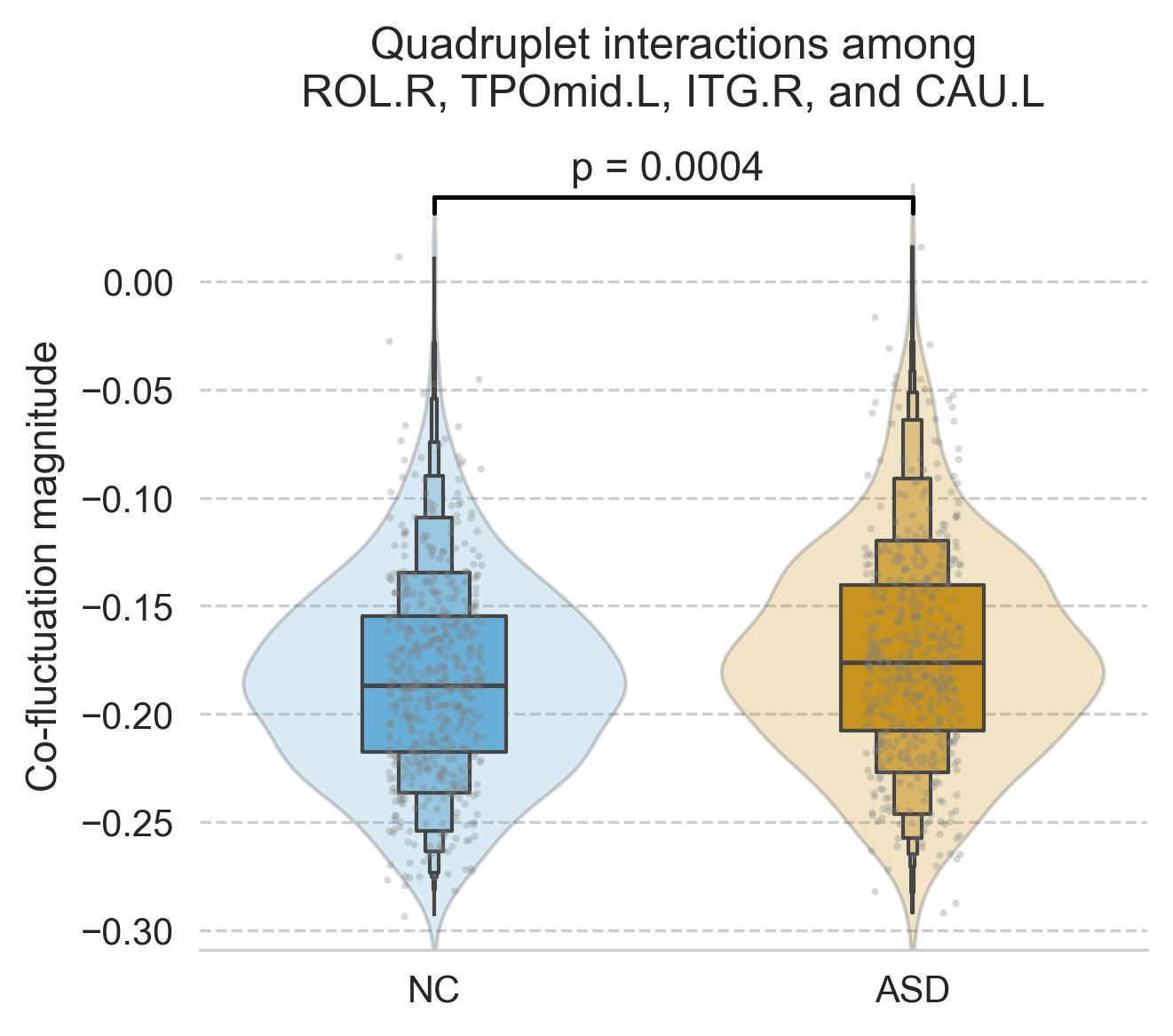}
}
\caption{Group difference comparison of quadruplet interactions among the relevant brain regions (ITG.R, SMG.L, ITG.L, CAU.L, CAU.R, TPOmid.L, TPOmid.R, and ROL.R) on ABIDE. These figures show that the positive HOIs between relevant brain regions gradually strengthen and the negative HOIs between these brain regions gradually weaken during the transition from NC to ASD.}
\label{quadruplet interactions ABIDE}
\end{figure}

Next, we further investigate whether the HOIs among key brain regions identified by HOI-Brain show significant differences during disease progression, which could offer additional therapeutic insights for brain disease research. For ease of calculation, we selected four out of the top ten key brain regions identified for each disease. Using our proposed Multiplication of Temporal Derivatives (MTD)-based co-fluctuation calculation method, we computed the quadruplet interactions among the relevant brain regions and subsequently compared the group differences. Group comparisons were conducted using Mann–Whitney U tests \citep{83}. Results were Benjamini–Hochberg FDR correction \citep{82} for the 3 pairwise group comparisons. For more details regarding the statistical analyses of higher-order organizations (including p-values, statistical measures, effect sizes, etc.), please refer to \ref{The Statistical Analyses of The Higher-Order Organizations}. As shown in Figures \ref{quadruplet interactions ADNI}, \ref{quadruplet interactions PPMI}, \ref{quadruplet interactions ABIDE}, compared to other brain regions, although these brain regions all demonstrate significant importance in the diagnosis of AD or PD—indicating that they may have sustained varying degrees of damage—their HOIs exhibit notable differences across various stages of disease progression. This suggests a dissociation between HOIs in the brain and the functions of the individual brain regions themselves, which is consistent with the literature \citep{43}. In addition, we observed that during the transition from CN to those with MCI or AD, the positive HOIs between relevant brain regions gradually weaken, indicating a progressive impairment of higher-order synergistic functions among brain regions. On the other hand, the negative HOIs between these brain regions gradually strengthen, reflecting an increasing degree of internal functional disruption in the brain. This phenomenon is consistent with findings reported in previous literature \citep{34}. 
However, this phenomenon shows the opposite pattern in PD and ASD. 
For PD, this may be related to the higher prevalence of hallucinations reported in Parkinson’s disease. 
Hallucinations in Parkinson's disease may result from excessive influence of higher-order brain regions on early sensory processing, which is consistent with the enhanced functional integration between sensory and higher-order networks \citep{74}.
In the case of ASD, this phenomenon is also the opposite of what is observed in AD; this fact is  consistent with the observation that certain areas of the brains of children with autism experience excessive connectivity \citep{100}.

\textbf{Alzheimer’s disease (AD).} As shown in Figure \ref{quadruplet interactions ADNI}, in particular, the quadruplet interaction among CAU.R, HIP.L, PHG.R, and AMYG.L, which involves the medial temporal–limbic–striatal circuit, exhibits significant intergroup differences during the progression from CN to MCI, with p-values $< 0.05$. However, these differences become less pronounced during the progression from MCI to AD, with p-values $\ge 0.05$. This suggests that these HOIs may serve as potential biomarkers for the early diagnosis of Alzheimer’s disease. In contrast, the quadruplet interaction among CAU.R, PHG.R, AMYG.L and TPOmid.R; the quadruplet interaction among CAU.R, PHG.R, OLF.L, and TPOmid.R; the quadruplet interaction among HIP.L, OLF.L, CAU.R, and TPOmid.R exhibit significant differences ($p \le 0.05$)  from CN to MCI as well as significant differences from MCI to AD ($p \le 0.05$), while the differences from CN to AD are not significant in these HOIs ($p \ge 0.05$).
This pattern suggests that these higher-order co-fluctuation signatures are particularly sensitive to stage transitions (CN to MCI and MCI to AD), rather than following a strictly monotonic CN to AD separation. Notably, the quadruplet interaction among CAU.R, TPOmid.L, HIP.L, and AMYG.L, which involves the social-reward-emotion-memory integrated circuit, shows significant intergroup differences throughout the entire AD progression, with p-values $< 0.05$. This indicates that this HOI may be persistently disrupted across the entire disease course, playing a critical role in AD pathogenesis.

\textbf{Parkinson's disease (PD).}  As shown in Figure \ref{quadruplet interactions PPMI}, two quadruplet interaction patterns show significant inter-group differences between Prodromal and PD: (i) the interaction among PCG.R, ORBsupmed.L, IPL.R, and THA.R ($p=0.043$), and (ii) the interaction among IPL.R, AMYG.L, IFGtriang.R, and ORBinf.R ($p=0.041$). These HOIs jointly involve a sensorimotor node (PCG), parietal association cortex (IPL), thalamic relay (THA), and frontal/orbitofrontal regions (IFG and ORB), suggesting that higher-order co-fluctuation coupling within cortico-thalamo-frontal and fronto-limbic control circuits is already detectably altered at the prodromal to PD transition.
In addition, four related HOIs show consistent but weaker group-separation trends. Overall, these results suggest a systematic reconfiguration of HOI magnitudes across these circuits during the transition from prodromal to PD.

\textbf{Autism Spectrum Disorder (ASD).} From Figure \ref{quadruplet interactions ABIDE}, six quadruplet interaction patterns show significant inter-group differences between NC and ASD, with all p-values $\le 0.001$.
These HOIs consistently involve temporal regions (ITG and TPOmid) together with parietal (SMG), striatal (CAU), and opercular (ROL) nodes, suggesting that the NC to ASD contrast is characterized by altered higher-order co-fluctuation coupling across temporo-parietal and cortico-striatal/sensorimotor integration circuits.
Across these HOIs, the ASD group shows a systematic shift in co-fluctuation magnitudes relative to NC.

\section{Discussion}
\subsection{The model}
Recent conceptual and technical advances in fMRI data analysis provide an opportunity for the field to not only improve the effectiveness of brain disease diagnosis but also deepen our understanding of brain disease mechanisms \citep{14, 15, 20,22, 25,26,32,33}. However, current methods overlook higher-order patterns with signs, limiting an integrated understanding of brain-wide communication for brain disease diagnosis. 

In this study, we focused on the importance of signed higher-order (group) interactions in fMRI signals, which were inferred using a recently developed topological approach \citep{29}. Our proposed HOI-Brain incorporates: 
(i) a novel higher-order interaction representation strategy via Multiplication of Temporal Derivatives (MTD) to quantify dynamic functional co-fluctuations of group ROIs;
(ii) a new definition of weighted simplicial complex is proposed and a novel higher-order feature extraction method is employed on them focusing on quadruplet-level interaction signatures and two-dimensional void descriptors through Persistent Homology;
(iii) a pioneering effort to distinguish between positively and negatively synergistic higher-order interactions in the brain;
(iv) a novel multi-channel brain network Transformer that integrates lower-order edge features with higher-order topological invariants. This new framework emphasizes the potential of quadruplet or higher-order region interactions representing critical advancements for precision medicine, a deeper understanding of neurological disorders, and broader contributions to neuroimaging research. It demonstrates superior prediction accuracy for classifying Alzheimer's Disease, Parkinson's Disease, and Autism Spectrum Disorder compared to traditional machine learning models, graph neural network (GNN)-based models, Transformer-based models, hypergraph neural network (HGNN)-based models, and the Persistent Homology (PH)-based models. Additionally, ablation study also indicates that the effectiveness of quadruplet interactions compared to triplet interactions, the effectiveness of distinguishing between positively and negatively synergistic HOIs, and the effectiveness of different components in model.

The findings of our investigation revealed intriguing patterns. In the process of designing an orthonormal clustering readout for the identification of clusters of functionally similar nodes through soft clustering with orthonormal constraints, it was found that the assignment of nodes from different modes to distinct functional modules significantly improves classification accuracy. This finding indicates that both low-level and high-level feature patterns, which are frequently situated within different functional modular structures, are of paramount importance. This observation is in alignment with the conclusions reported in previous studies \citep{43}. This novel finding also provides a new opportunity for brain disease diagnostic models, specifically how to effectively identify and integrate low-order and higher-order patterns across different functional modules in the brain.

Furthermore, an investigation was conducted into the most discriminative neural patterns using an attention mechanism. This investigation revealed that quadruplet-level interaction signatures exhibit substantially greater importance than two-dimensional void descriptors and edge features in both the ADNI and ABIDE datasets. It is noteworthy that negative synergistic quadruplets tend to exhibit a higher degree of significance in comparison to their positive synergistic counterparts. Furthermore, positive synergistic voids have been observed to exceed negative synergistic voids in terms of their impact. This novel observation may offer critical insights into the pathological mechanisms underlying brain disorders. In order to provide further validation of these findings, inter-group comparisons were performed of signed higher-order topological signatures in the ADNI dataset. The findings demonstrated a substantial discrepancy in the quantity of quadruplet-level interaction signatures among the groups in comparison to void descriptors. Furthermore, positive voids demonstrated more pronounced inter-group disparities than negative voids. The findings suggest that disease progression may primarily disrupt higher-order neural interactions, which subsequently influence high-dimensional hole structures. An intriguing pattern emerged in the comparative analysis of CN individuals and those with AD. CN individuals appear to rely more on positive information for whole-brain topological organisation, whereas AD patients show a greater dependence on negative information. Furthermore, longitudinal trends indicate that the number of positive quadruplets gradually declines with disease progression, while negative quadruplets follow a biphasic trajectory decreasing initially and then increasing. This suggests that during the transition from CN to MCI, there is a decline in both positive and negative synergistic interactions.  However, as the condition progresses from MCI to AD, there is an increasing adoption of negative synergy by neural networks, alongside a suppression of excitability. This may be a compensatory mechanism employed by the brain to preserve functional integrity across neural modules. This finding aligns with the reference stating that excitatory neurons in the brains of Alzheimer's patients suffer severe genomic damage \citep{75}.

In the context of investigating the significant brain regions and their interactions in the context of brain diseases, the utilisation of HOI-Brain facilitates the identification of these regions from a more comprehensive perspective, a departure from the low-level perspective that has been adopted in previous studies \citep{11, 12}. The findings, which function as potential biomarkers, are consistent with a substantial corpus of extant literature \citep{59,60,61,62}. Furthermore, they furnish novel insights with regard to the diagnosis of brain diseases. The right caudate has been identified as a potential hub node within the brain network. Previous literature has indicated that disruptions in this region are consistent with the progression of Alzheimer's disease \citep{60}. In a similar manner, the ITG.R has been observed to function as an additional pivotal hub within the brain network. It has been demonstrated that this region is selectively disrupted during the progression of ASD \citep{62}.

Furthermore, the higher-order organisational patterns identified by HOI-Brain offer novel insights into the diagnosis of brain diseases. Specifically, the
quadruplet interaction among CAU.R, HIP.L, PHG.R, and AMYG.L, which involves the medial temporal–limbic–striatal circuit, may serve as potential biomarkers for the early diagnosis of AD with p-values $< 0.05$ during the progression from CN to MCI. It is noteworthy that the brain regions identified by HOI-Brain via the analysis of resting-state fMRI data are deemed to be of significant importance in the diagnosis of AD or ASD. This finding indicates that these brain regions may have sustained varying degrees of damage. However, HOIs exhibited by these regions demonstrate notable differences across various stages of disease progression. This finding is consistent with the conclusions of previous studies \citep{43}, suggesting a dissociation between HOIs in the brain and the functions of the individual brain regions themselves. Furthermore, observations were made concerning the transition from healthy individuals to those with AD or ASD. It was noted that the positive HOIs between relevant brain regions gradually weaken, indicating a progressive impairment of higher-order synergistic functions among brain regions. Conversely, the negative HOIs between these brain regions gradually strengthen, reflecting an increasing degree of internal functional disruption in the brain. This phenomenon is consistent with the findings reported in previous literature \citep{34}. However, this phenomenon is exactly the opposite in PD and ASD. For PD, this may be due to the fact that patients suffering from Parkinson's disease, due to their motor impairments, are prone to hallucinations. Hallucinations in Parkinson's disease may be attributable to an excessive influence of higher-order brain regions on early sensory processing. This is consistent with enhanced functional integration between sensory and higher-order networks \citep{74}.  In the case of ASD,
this fact is consistent with the observation that certain areas of the brains of children with autism experience excessive connectivity \citep{100}.

From an application perspective, HOI-Brain provides a unified pipeline that maps resting-state fMRI time series to subject-level predictions while simultaneously returning interpretable higher-order interaction motifs and ROI-importance maps.
This makes the framework potentially useful for (i) assisting early screening or risk stratification (e.g., CN vs MCI in ADNI) using non-invasive rs-fMRI, (ii) providing network-level candidate biomarkers (higher-order co-fluctuation signatures) that can be followed up in independent cohorts or longitudinal settings, and (iii) supporting hypothesis generation about circuit-level dysconnectivity by highlighting disease-relevant HOIs beyond pairwise connectivity. Importantly, we view HOI-Brain as a complementary decision-support and biomarker-discovery tool rather than a stand-alone clinical diagnostic system; rigorous prospective validation and multi-site deployment studies are required before clinical translation.

\subsection{Limitations and Future work}
Although HOI-Brain achieves strong performance and promising interpretability for brain disease diagnosis, several limitations and future directions remain. 
First, for modeling convenience we mainly focus on concordant signed effects. Our exploratory analysis suggests that discordant sign patterns may encode redundant or competing information across multiple regions; future work will develop principled strategies to model both concordant and discordant patterns explicitly. 
Second, we currently obtain subject-level representations by averaging higher-order features across all time points. While this improves robustness to fMRI noise and stabilizes estimation, it may obscure temporally evolving higher-order co-fluctuation patterns. Investigating the dynamics of HOIs (e.g., time-resolved or state-dependent modeling) could therefore provide additional insight into disease progression and the temporal organization of brain network topology. 
Finally, the proposed framework is general and could be extended to other neurophysiological time series (e.g., EEG/MEG), enabling broader evaluation of signed higher-order interactions across modalities.

\section{Conclusions}

In summary, we propose a new framework, HOI-Brain (Higher-Order Interactions in Brain Networks), which can accurately capture signed HOIs in brain networks, extract interpretable signed higher-order topological features, and further exploit the information between lower-order and higher-order features for brain disorder diagnosis. We applied HOI-Brain to datasets for Alzheimer's disease, Parkinson's disease, and Autism Spectrum Disorder. With strong interpretability, HOI-Brain not only performs better on classification than 20 baselines, but also detects salient brain regions associated with
classification and discovers important higher-order organizations. Overall, our
model shows superiority over alternative graph learning and traditional machine learning classification models. 
By analyzing the attention maps of our multi-channel brain Transformer, our study identifies salient ROIs and their key interactions from a whole-brain perspective to distinguish brain disorders from healthy controls. Additionally, we uncover higher-order organizational patterns associated with specific disease progression stages.
Notably, our framework is generalizable to the analysis of other neuroimaging time-series data. Our proposed framework demonstrates the potential of quadruplet or higher-order region interactions—critical advancements for precision medicine, enhanced understanding of neurological disorders, and broader contributions to neuroimaging research.

\section{Declaration of Competing Interest}
The authors declare that they have no known competing financial interests or personal relationships that could have appeared to
influence the work reported in this paper.
\section{Acknowledgements}
This work was supported in part by the National Natural Science Foundation of China No.12471330, the Shandong Provincial Natural Science Foundation No.ZR2025MS71 and the National Natural Science Foundation of China No.12231018.



\newpage
\appendix
\section{Data Preprocessing Using fMRIPrep }
\label{Data preprocessing}
\begin{figure}[ht]
\centering
\includegraphics[scale=0.4]{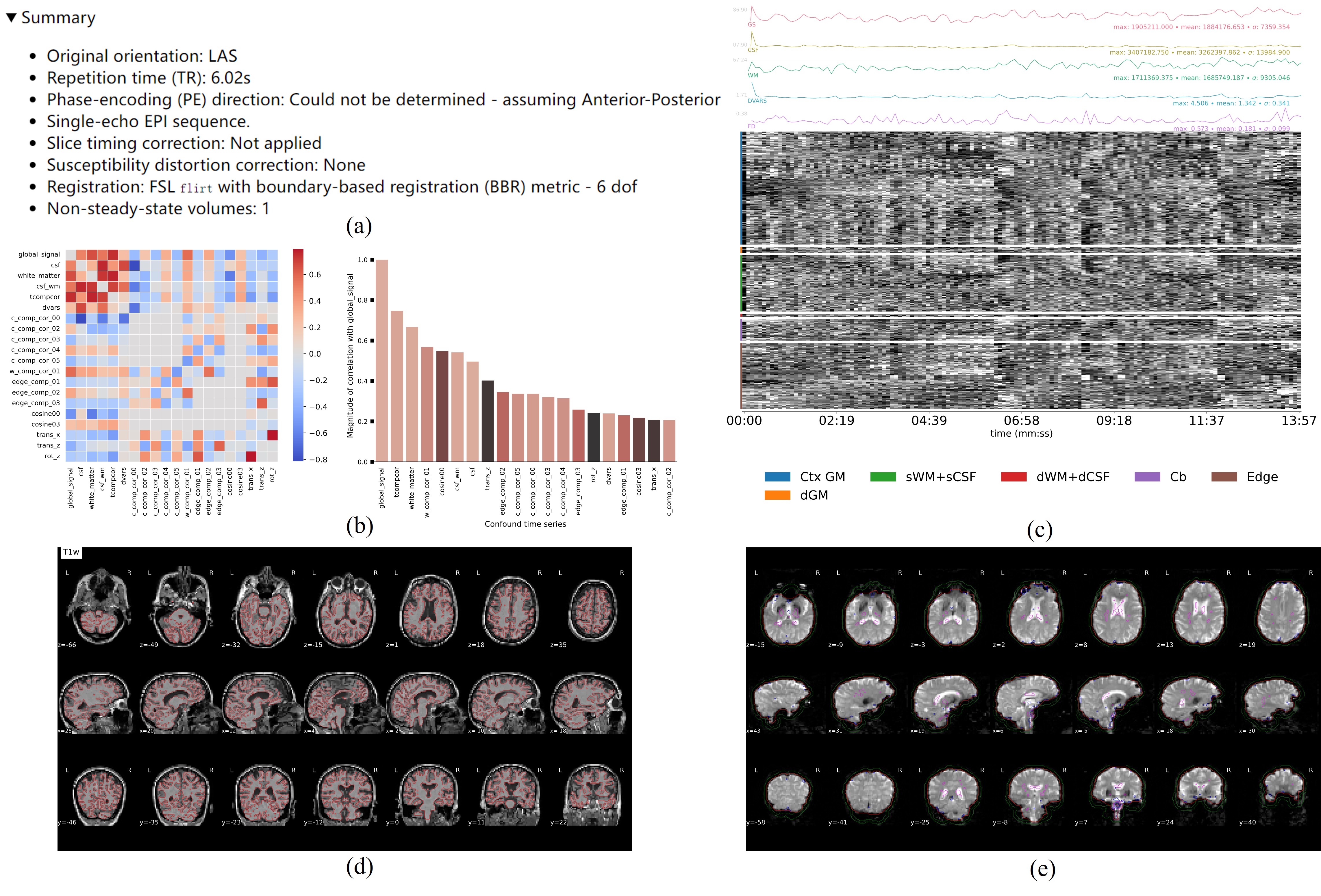}
\caption{The visual reports ease quality control and help a representative subject understand the processing flow. (a) shows the summary of data preprocessed. (b) shows the correlations among nuisance regressors. (c) shows the summary statistics plotted, which may reveal trends or artifacts in the BOLD data. (d) shows the alignment of the BOLD reference image to the anatomical (T1-weighted) image. (e) shows the BOLD-derived brain mask (red contour) and the ROIs used to estimate physiological and motion confounds for nuisance regression.}
\label{fig data}
\end{figure}

Resting-state fMRI data from ADNI were preprocessed using fMRIPrep (v25.2.3), producing outputs in native T1w space and in standard space MNI152NLin2009cAsym, including minimally preprocessed BOLD time series, brain masks, and confound regressors. The command being executed is as follows: 
\textbf{/app/.pixi/envs/fmriprep/bin/fmriprep /data /out participant --participant-label sub-001 --fs-no-reconall --output-spaces MNI152NLin2009cAsym.}
Quality control was performed by visually inspecting the automatically generated fMRIPrep reports for each subject. As shown in Figure \ref{fig data}, we now provide the fMRIPrep outputs for a representative subject, including the automatically generated HTML reports and quality control figures.
For PPMI/ABIDE/TaoWu, we used publicly released fMRIPrep-preprocessed derivatives from the original studies \citep{45}.

\subsection{Full fMRIPrep boilerplate}
The full auto-generated fMRIPrep boilerplate text is provided for transparency and reproducibility.
Results included in this manuscript come from preprocessing performed
using fMRIPrep 25.2.3 (\citep{fmriprep1,fmriprep2};
RRID:SCR\_016216), which is based on \emph{Nipype} 1.10.0
\citep{nipype1,nipype2}.

\textbf{Anatomical data preprocessing}
A total of T1-weighted (T1w) images were found within the input BIDS
dataset. The T1w image was corrected for intensity non-uniformity 
with \texttt{N4BiasFieldCorrection} \citep{n4}, distributed with ANTs
2.6.2 \citep[RRID:SCR\_004757]{ants}, and used as T1w-reference
throughout the workflow. The T1w-reference was then skull-stripped with
a \emph{Nipype} implementation of the \textbf{antsBrainExtraction.sh}
workflow (from ANTs), using OASIS30ANTs as target template. Brain tissue
segmentation of cerebrospinal fluid (CSF), white-matter (WM) and
gray-matter (GM) was performed on the brain-extracted T1w using
\texttt{fast} \citep[FSL (version unknown),
RRID:SCR\_002823,][]{fsl_fast}. Volume-based spatial normalization to
one standard space (MNI152NLin2009cAsym) was performed through nonlinear
registration with \texttt{antsRegistration} (ANTs 2.6.2), using
brain-extracted versions of both T1w reference and the T1w template. The
following template was were selected for spatial normalization and
accessed with \emph{TemplateFlow} \citep[25.0.4,][]{templateflow}:
\emph{ICBM 152 Nonlinear Asymmetrical template version 2009c}
{[}\citet{mni152nlin2009casym}, RRID:SCR\_008796; TemplateFlow ID:
MNI152NLin2009cAsym{]}.

 \textbf{Functional data preprocessing}
For each of the BOLD runs found per subject (across all tasks and
sessions), the following preprocessing was performed. First, a reference
volume was generated, using a custom methodology of \emph{fMRIPrep}, for
use in head motion correction. Head-motion parameters with respect to
the BOLD reference (transformation matrices, and six corresponding
rotation and translation parameters) are estimated before any
spatiotemporal filtering using \texttt{mcflirt} \citep[FSL
,][]{mcflirt}. The BOLD reference was then co-registered to the T1w
reference using \texttt{mri\_coreg} (FreeSurfer) followed by
\texttt{flirt} \citep[FSL ,][]{flirt} with the boundary-based
registration \citep{bbr} cost-function. Co-registration was configured
with six degrees of freedom. Several confounding time-series were
calculated based on the \emph{preprocessed BOLD}: framewise displacement
(FD), DVARS and three region-wise global signals. FD was computed using
two formulations following Power (absolute sum of relative motions,
\citet{power_fd_dvars}) and Jenkinson (relative root mean square
displacement between affines, \citet{mcflirt}). FD and DVARS are
calculated for each functional run, both using their implementations in
\emph{Nipype} \citep[following the definitions by][]{power_fd_dvars}.
The three global signals are extracted within the CSF, the WM, and the
whole-brain masks. Additionally, a set of physiological regressors were
extracted to allow for component-based noise correction
\citep[\emph{CompCor},][]{compcor}. Principal components are estimated
after high-pass filtering the \emph{preprocessed BOLD} time-series
(using a discrete cosine filter with 128s cut-off) for the two
\emph{CompCor} variants: temporal (tCompCor) and anatomical (aCompCor).
tCompCor components are then calculated from the top 2\% variable voxels
within the brain mask. For aCompCor, three probabilistic masks (CSF, WM
and combined CSF+WM) are generated in anatomical space. The
implementation differs from that of Behzadi et al.~in that instead of
eroding the masks by 2 pixels on BOLD space, a mask of pixels that
likely contain a volume fraction of GM is subtracted from the aCompCor
masks. This mask is obtained by thresholding the corresponding partial
volume map at 0.05, and it ensures components are not extracted from
voxels containing a minimal fraction of GM. Finally, these masks are
resampled into BOLD space and binarized by thresholding at 0.99 (as in
the original implementation). Components are also calculated separately
within the WM and CSF masks. For each CompCor decomposition, the
\emph{k} components with the largest singular values are retained, such
that the retained components' time series are sufficient to explain 50
percent of variance across the nuisance mask (CSF, WM, combined, or
temporal). The remaining components are dropped from consideration. The
head-motion estimates calculated in the correction step were also placed
within the corresponding confounds file. The confound time series
derived from head motion estimates and global signals were expanded with
the inclusion of temporal derivatives and quadratic terms for each
\citep{confounds_satterthwaite_2013}. Frames that exceeded a threshold
of 0.5 mm FD or 1.5 standardized DVARS were annotated as motion
outliers. Additional nuisance timeseries are calculated by means of
principal components analysis of the signal found within a thin band
(\emph{crown}) of voxels around the edge of the brain, as proposed by
\citep{patriat_improved_2017}. All resamplings can be performed with
\emph{a single interpolation step} by composing all the pertinent
transformations (i.e.~head-motion transform matrices, susceptibility
distortion correction when available, and co-registrations to anatomical
and output spaces). Gridded (volumetric) resamplings were performed
using \texttt{nitransforms}, configured with cubic B-spline
interpolation.

\section{Model Comparison for ABIDE Using Schaefer-100}
\label{Model Comparison}
Given that different parcellation schemes may lead to varying network representations and downstream results, we further evaluated the robustness of our method to the choice of parcellation. Specifically, we re-ran the experiments on the ABIDE dataset under the Schaefer-100 parcellation setting for all compared models, and include the resulting classification performance. As summarized in Table \ref{tab:performance_comparison on ABIDE_sch100}, our method remains the best-performing (or highly competitive) compared with representative baselines, suggesting good robustness to parcellation choice.

\begin{table}[h]
  \centering
  \caption{Performance comparison with five categories of best baselines on the ABIDE dataset under Schaefer-100 parcellation (\%). }
  \scalebox{0.74}{
  \begin{tabular}{@{}cccccc@{}}
    \toprule
   Type & Method & Accuracy & Precision & Recall & F1-score \\
    \midrule
    \multirow{1}{*}{\parbox{5cm}{\centering  Machine Learning Models}}
   
    & RandomForest  & 61.0$\pm$2.2 & 61.7$\pm$2.3 & 66.4$\pm$3.4 & 64.0$\pm$2.5   \\
    \midrule
    \multirow{1}{*}{\parbox{5cm}{\centering GNN-based models}}
    & BrainGNN & 60.5$\pm$2.4 & 61.2$\pm$3.1 & 63.6$\pm$2.7 & 62.7$\pm$3.5  \\
    \midrule
    \multirow{1}{*}{\parbox{5cm}{\centering Transformer-based models}}
    & BrainnetTransformer & 62.4$\pm$2.8 & 65.6$\pm$3.8 &  58.9$\pm$2.0 &  62.1$\pm$3.3   \\
   \midrule
    \multirow{1}{*}{\parbox{5cm}{\centering HGNN-based models}} 
    & HGAT & 61.0$\pm$2.3 & 61.7$\pm$3.7 & 66.4$\pm$2.7 & 64.0$\pm$3.3   \\
      \midrule      
    \multirow{1}{*}{\parbox{5cm}{\centering PH-based models}} 
     & Brain-HORS & 62.4$\pm$2.7 & 65.6$\pm$2.3& \textbf{68.9$\pm$1.4} & 62.1$\pm$2.5   \\
    \midrule
    \multirow{1}{*}{Our Framework} 
    & HOI-Brain & \textbf{64.9$\pm$1.6} & \textbf{66.7$\pm$2.2} & 65.4$\pm$2.4 & \textbf{66.9$\pm$1.9}  \\
    \bottomrule
  \end{tabular}
  \label{tab:performance_comparison on ABIDE_sch100} 
  }
\end{table}

\section{Persistent Homology Theory}
\label{Persistent Homology}
\subsection{Basic Concepts in Algebraic Topology}
For related basic concepts about Persistent Homology theory, please refer to the following references \citep{78, 77}.

\noindent
\begin{definition}(\textbf{Simplex})
Given affinely independent points $v_0, v_1, \dots, v_k$ in $\mathbb{R}^d$, 
a \textit{k-simplex} $\sigma = [v_0, v_1, \dots, v_k]$ is the convex hull of these points. A \textit{face} of $\sigma$ is any simplex $\tau$ spanned by a nonempty subset of its vertices.
Geometrically, a 0-simplex represents a point, a 1-simplex represents an edge, a 2-simplex represents a triangle (including interior), and a 3-simplex represents a quadruplet (solid). 
\end{definition}

\noindent
\begin{definition}(\textbf{Simplicial Complex})
\textit{An abstract simplicial complex} $\mathcal{K}$ on a finite vertex set $V$ is a collection of some non-empty subsets of $V$ satisfying:
\begin{enumerate}
    \item  For every $v \in V$, the singleton $\{v\}$ belongs to $\mathcal{K}$.
    \item If $\sigma \in \mathcal{K}$ and $\tau \subseteq \sigma$ is a non-empty subset, then $\tau \in \mathcal{K}$.
\end{enumerate}
Each element $\sigma \in \mathcal{K}$ is called a simplex and $\tau$ is called the face of $\sigma$. A simplex with $|\sigma| = k + 1$ vertices is a $k$-simplex, representing a $k$-order interaction. 
\end{definition}

\noindent
\begin{definition}(\textbf{Chain Group})
The $k$-\textit{chain group} $C_k$ is the free abelian group generated by oriented $k$-simplices. 
An element $c \in C_k$ is a formal sum $c = \sum a_i \sigma_i$ with $a_i \in \mathbb{Z}$.

\end{definition}
\noindent
\begin{definition}(\textbf{Boundary Operator}) 
The linear operator $\partial_k: C_k \to C_{k-1}$ acts on a $k$-simplex $\sigma = [v_0, \dots, v_k]$ as:
\[
\partial_k(\sigma) = \sum_{i=0}^{k} (-1)^i [v_0, \dots, \hat{v_i}, \dots, v_k]
\]
where $\hat{v_i}$ denotes vertex removal. The fundamental property is $\partial_k \circ \partial_{k+1} = 0$.
\end{definition}
\noindent
\begin{definition}(\textbf{Homology Group})
The $k$-\textit{homology group} is the quotient:
\[
H_k = \ker \partial_k / \operatorname{im} \partial_{k+1}
\]
Elements of $\ker \partial_k$ are \textit{cycles} (closed structures). 
Elements of $\operatorname{im} \partial_{k+1}$ are \textit{boundaries} (trivially closed structures).
\end{definition}
Homology groups provide a mathematical characterization of topological features in a shape. Formally, for each dimension $k$, the $k$-th Betti number $\beta_k = \text{rank}\,H_k(\mathcal{K})$ enumerates the independent non-boundary $k$-dimensional cycles in the simplicial complex $\mathcal{K}$. These Betti numbers have concrete geometric interpretations: $\beta_0$ counts the connected components; $\beta_1$ counts independent 1-dimensional loops (e.g., holes/tunnels); $\beta_2$ counts enclosed cavities bounded by 2D surfaces.

\subsection{Persistent Homology}
\noindent \textbf{Persistent Homology:} Persistent homology is an important tool in topological shape analysis, which aims at overcoming intrinsic limitations of classical homology by allowing for a multi-scale approach to shape description. In a nutshell, persistent homology describes the changes in homology that occur to an object which evolves with respect to a parameter. 

Given a simplicial complex $\mathcal{K}$, a \textit{filtration} of $\mathcal{K}$ is a finite sequence of subcomplexes $\mathcal{K}^f := \{\mathcal{K}^p \mid 0 \leq p \leq m\}$ of $\mathcal{K}$ such that:
\[
\emptyset = \mathcal{K}^0 \subseteq \mathcal{K}^1 \subseteq \cdots \subseteq \mathcal{K}^m = \mathcal{K}
\]

For indices $p,q \in \{0,\ldots,m\}$ with $p \leq q$, the \textit{$(p,q)$-persistent $k$-homology group} $H_k^{p,q}(\mathcal{K}^f)$ consists of the $k$-cycles included from $C_k(\mathcal{K}^p)$ into $C_k(\mathcal{K}^q)$ modulo boundaries. Formally, it is defined as:
\[
H_k^{p,q}(\mathcal{K}^f) := \operatorname{Im}(i_k^{p,q})
\]
where $i_k^{p,q}: H_k( \mathcal{K}^p) \to H_k(\mathcal{K}^q)$ denotes the linear map induced by the inclusion $\mathcal{K}^p \hookrightarrow \mathcal{K}^q$.

Crucially, persistent homology enriches classical homology by quantifying feature persistence across scales. Whereas classical homology identifies non-boundary cycles at a single scale, persistent homology captures cycles that emerge as non-boundaries at filtration stage $p$ and subsequently become boundaries at stage $q > p$. The interval $(p,q)$, termed the \textit{persistence} of the cycle, provides a quantitative measure of its topological significance within the shape.

\noindent
\begin{definition}(2-dimensional void) A \textit{2-dimensional void} corresponds to a non-boundary 2-cycle in $H_2$. 
Geometrically, it is an enclosed void bounded by a surface of triangles. 
For a simplicial complex $\mathcal{K}$:
\begin{itemize}
    \item \textbf{Birth:} When a set of triangular faces forms a closed surface.
    \item \textbf{Death:} When the void is filled by a chain of 3-simplices (quadruplet).
\end{itemize}
\end{definition}
The persistence $\text{pers}(b,d) = d - b$ quantifies topological significance, where \textit{b} denotes the birth (or appearance) weight of a topological feature (e.g., the weight of last triangular face forming a closed surface), and \textit{d} denotes the death (or disappearance) weight of that feature (e.g., when a quadruplet fills the void).

\section{Ablation Study of Higher-Order Features on PPMI and Taowu Datasets}
\label{app1}

To evaluate the effectiveness of more complex signed quadruplet structures and voids (2D holes) extracted by MTD methods, as shown in Tables \ref{tab:performance_comparison_Taowu} and \ref{tab:performance_comparison_PPMI},  we report a performance comparison with varying feature combinations on the TaoWu and PPMI datasets to have more insight into how HOI-Brain performs on different tasks.
\begin{table}[ht]
  \centering
  \caption{Performance comparison with varying combinations of features on the TaoWu datasets (\%). The best results are marked in bold.}
  \scalebox{0.74}{
  \begin{tabular}{@{}cccccc@{}}
    \toprule
  \multirow{2}{*}{Method} & \multicolumn{4}{c}{Taowu} \\
    \cmidrule(lr){2-5} 
     & Accuracy & Precision & Recall & F1-score &   \\
    \midrule
     edge   & 65.0$\pm$5.0 & \underline{84.8$\pm$18.9} & 55.0$\pm$36.7 & 54.6$\pm$17.9 \\
     edge+violating triangles+1D loops  & 70.0$\pm$16.9 & 63.3$\pm$37.1 & 60.0$\pm$33.9 & 60.2$\pm$32.9\\
    edge+violating triangles+good quadruplets   & 67.5$\pm$16.9 & 57.4$\pm$33.7 & \underline{75.0$\pm$38.7} & 62.8$\pm$32.4 \\
     edge+1D loops+2D voids  & 67.5$\pm$17.0 & 79.8$\pm$17.5 & 60.0$\pm$25.5 & 62.3$\pm$12.8\\
     edge+good quadruplets+2D voids & \textbf{77.5$\pm$9.3} & \textbf{89.3$\pm$13.7} & 70.0$\pm$29.1 & \underline{72.2$\pm$17.8}\\
     edge+signed higher-order features+ extended Pearson & {72.0$\pm$18.6} & {70.0$\pm$27.9} & {65.0$\pm$19.1} & {70.0$\pm$19.2} \\
     edge+signed good quadruplets+signed 2D voids & \underline{77.5$\pm$12.3} & 82.4$\pm$9.3 &  \textbf{77.5$\pm$12.3} & \textbf{75.9$\pm$13.9}  \\
    \bottomrule
  \end{tabular}
  \label{tab:performance_comparison_Taowu} 
   }
\end{table}

\begin{table}[ht]
  \centering
  \caption{Performance comparison with varying combinations of features on the PPMI datasets (\%). The best results are marked in bold.}
  \scalebox{0.74}{
  \begin{tabular}{@{}cccccc@{}}
    \toprule
  \multirow{2}{*}{Method} & \multicolumn{4}{c}{PPMI} \\
    \cmidrule(lr){2-5} 
     & Accuracy & Precision & Recall & F1-score &   \\
    \midrule
     edge   & 61.4$\pm$7.4 & \underline{71.6$\pm$18.7} & 45.1$\pm$10.1 & 53.5$\pm$8.9  \\
     edge+violating triangles+1D loops  & 57.6$\pm$2.4 & 64.0$\pm$9.2 & 43.8$\pm$13.9 & 49.4$\pm$8.1   \\
     edge+violating triangles+good quadruplets  & 60.5$\pm$9.2 & 66.5$\pm$19.7 & 56.4$\pm$12.0 & 58.7$\pm$8.3  \\
     edge+1D loops+2D voids  & 63.3$\pm$4.2 & 69.8$\pm$11.2 & 56.9$\pm$18.4 & 59.6$\pm$6.1   \\
     edge+good quadruplets+2D voids & \underline{65.2$\pm$4.3} & \textbf{73.1$\pm$12.2} & \underline{58.7$\pm$11.1} & \underline{63.7$\pm$4.5}  \\
      edge+signed higher-order features+ extended Pearson & {60.2$\pm$3.6} & {60.0$\pm$8.5} & {56.6$\pm$9.1} & {62.5$\pm$9.2} \\
     edge+signed good quadruplets+signed 2D voids  & \textbf{66.1$\pm$4.0} & 69.0$\pm$4.0 & \textbf{66.3$\pm$4.2} & \textbf{64.7$\pm$4.7}  \\
    \bottomrule
  \end{tabular}
  \label{tab:performance_comparison_PPMI} 
   }
\end{table}

\section{Results of Hyperparameter Analysis on Other Metrics}
\begin{figure}[ht]
\centering

\subfigure{
\includegraphics[scale=0.6]{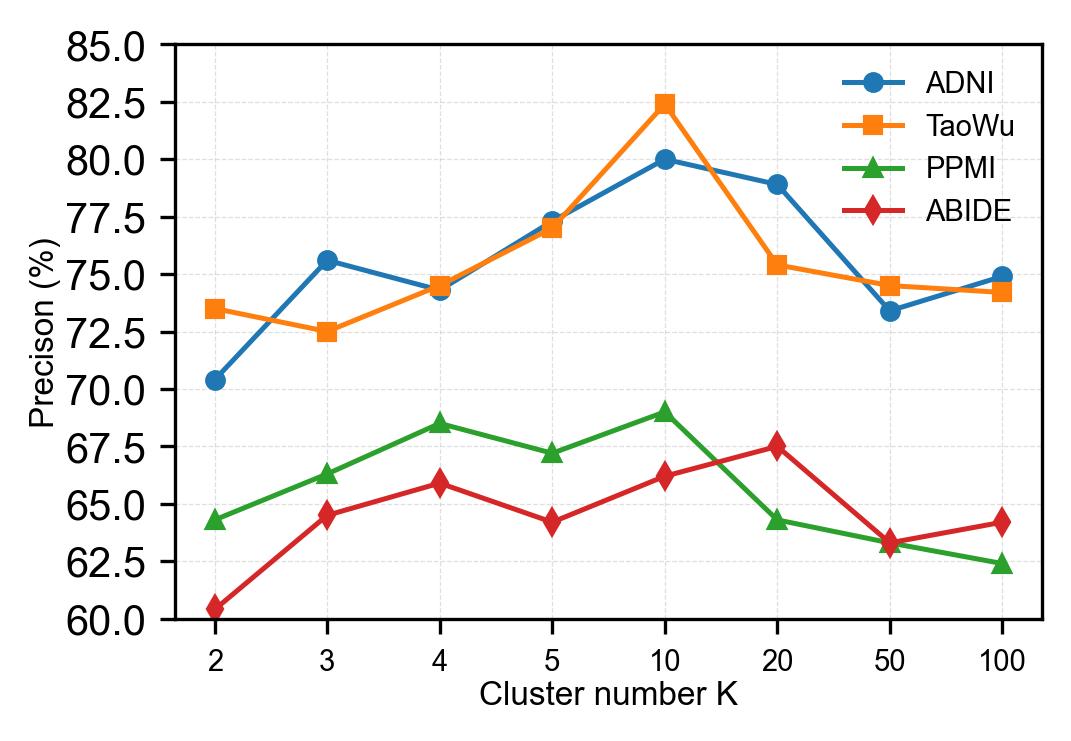}
}
\subfigure{
\includegraphics[scale=0.6]{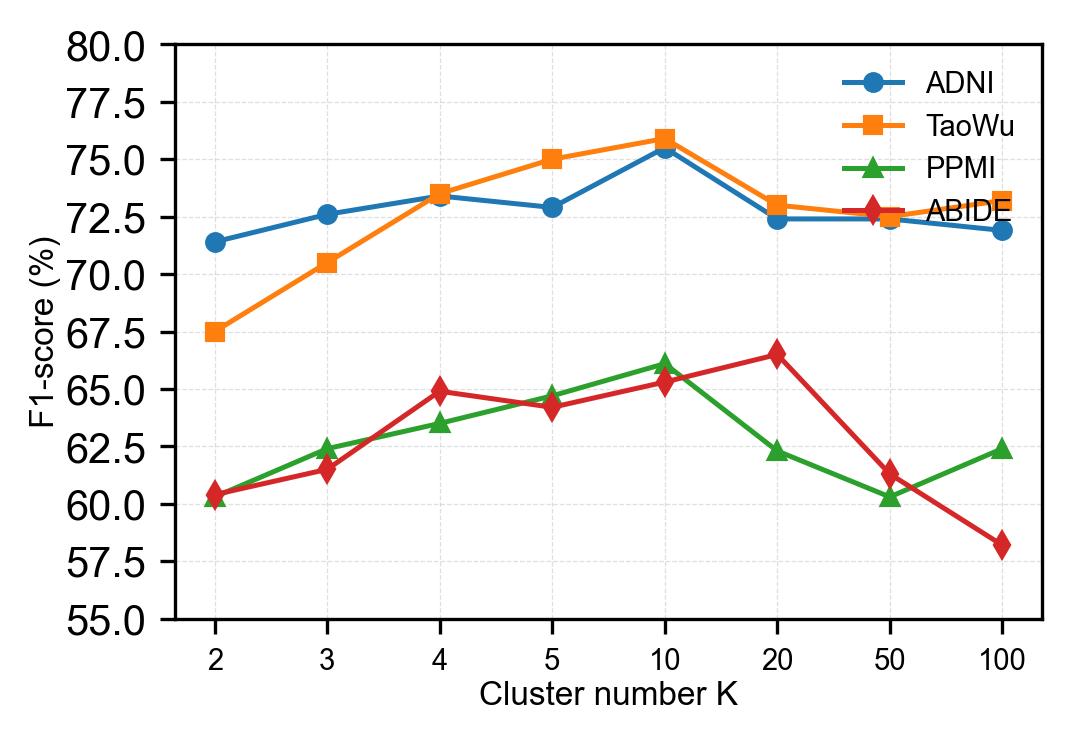}
}
\caption{Influence of the key hyper-parameter, the number of clusters, for model performance on other two metrics.}
\label{fig:enter-label1}
\end{figure}

In the hyperparameter analysis section, we only reported the influence of the key hyperparameter, the number of clusters, based on the accuracy metric. To better evaluate the impact of this hyperparameter, we additionally present the results of hyperparameter analysis in Figure \ref{fig:enter-label1} for the other two metrics: precision, and F1 score.

\section{The Attention Scores for ROIs}
\label{The Attention Scores for ROIs}
To provide a comprehensive view of the attention distribution across brain regions, we report the attention scores for all ROIs for each dataset (ADNI, PPMI, and ABIDE) in the following tables. In the main text, we present only the top-10 ROIs with the highest attention scores for brevity; here, the complete ranked lists are included to help readers better understand the overall score distribution across the whole brain. As shown in Tables  \ref{F_11}, \ref{F_12}, \ref{F_13}, each table is sorted in descending order and includes the ROI rank, node index, anatomical label, and the corresponding (normalized) attention score.
\begin{table*}

\centering
\scriptsize
\setlength{\tabcolsep}{3.2pt}
\renewcommand{\arraystretch}{1.08}
\caption{Attention scores across all ROIs on ADNI (sorted in descending order). The table is sorted in descending order and includes the ROI rank, node index, anatomical label, and the corresponding (normalized) attention score.}
\label{tab:adni_attention_all_rois}
\begin{tabular}{r r l S[table-format=1.4]
                r r l S[table-format=1.4]
                r r l S[table-format=1.4]}
\toprule
\multicolumn{4}{c}{Column 1} & \multicolumn{4}{c}{Column 2} & \multicolumn{4}{c}{Column 3} \\
\cmidrule(lr){1-4}\cmidrule(lr){5-8}\cmidrule(lr){9-12}
Rank & Node & ROI & {Score} &
Rank & Node & ROI & {Score} &
Rank & Node & ROI & {Score} \\
\midrule
1  & 72 & \texttt{CAU.R} & 0.0153 & 31 & 18 & \texttt{ROL.R} & 0.0114 & 61 & 20 & \texttt{SMA.R} & 0.0104 \\
2  & 37 & \texttt{HIP.L} & 0.0147 & 32 & 31 & \texttt{ACG.L} & 0.0113 & 62 & 49 & \texttt{SOG.L} & 0.0104 \\
3  & 40 & \texttt{PHG.R} & 0.0137 & 33 & 13 & \texttt{IFGtriang.L} & 0.0113 & 63 & 68 & \texttt{PCUN.R} & 0.0103 \\
4  & 41 & \texttt{AMYG.L} & 0.0136 & 34 & 11 & \texttt{IFGoperc.L} & 0.0112 & 64 & 03 & \texttt{SFGdor.L} & 0.0103 \\
5  & 21 & \texttt{OLF.L} & 0.0135 & 35 & 47 & \texttt{LING.L} & 0.0111 & 65 & 55 & \texttt{FFG.L} & 0.0103 \\
6  & 88 & \texttt{TPOmid.R} & 0.0133 & 36 & 45 & \texttt{CUN.L} & 0.0111 & 66 & 65 & \texttt{ANG.L} & 0.0102 \\
7  & 22 & \texttt{OLF.R} & 0.0132 & 37 & 60 & \texttt{SPG.R} & 0.0110 & 67 & 73 & \texttt{PUT.L} & 0.0102 \\
8  & 39 & \texttt{PHG.L} & 0.0132 & 38 & 36 & \texttt{PCG.R} & 0.0110 & 68 & 14 & \texttt{IFGtriang.R} & 0.0102 \\
9  & 71 & \texttt{CAU.L} & 0.0132 & 39 & 62 & \texttt{IPL.R} & 0.0110 & 69 & 59 & \texttt{SPG.L} & 0.0102 \\
10 & 80 & \texttt{HES.R} & 0.0131 & 40 & 77 & \texttt{THA.L} & 0.0109 & 70 & 85 & \texttt{MTG.L} & 0.0102 \\
11 & 06 & \texttt{ORBsup.R} & 0.0130 & 41 & 12 & \texttt{IFGoperc.R} & 0.0109 & 71 & 86 & \texttt{MTG.R} & 0.0101 \\
12 & 35 & \texttt{PCG.L} & 0.0129 & 42 & 24 & \texttt{SFGmed.R} & 0.0108 & 72 & 82 & \texttt{STG.R} & 0.0101 \\
13 & 27 & \texttt{REC.L} & 0.0128 & 43 & 75 & \texttt{PAL.L} & 0.0108 & 73 & 54 & \texttt{IOG.R} & 0.0101 \\
14 & 09 & \texttt{ORBmid.L} & 0.0126 & 44 & 83 & \texttt{TPOsup.L} & 0.0108 & 74 & 52 & \texttt{MOG.R} & 0.0101 \\
15 & 87 & \texttt{TPOmid.L} & 0.0125 & 45 & 74 & \texttt{PUT.R} & 0.0108 & 75 & 17 & \texttt{ROL.L} & 0.0100 \\
16 & 05 & \texttt{ORBsup.L} & 0.0123 & 46 & 43 & \texttt{CAL.L} & 0.0107 & 76 & 64 & \texttt{SMG.R} & 0.0099 \\
17 & 29 & \texttt{INS.L} & 0.0123 & 47 & 53 & \texttt{IOG.L} & 0.0107 & 77 & 46 & \texttt{CUN.R} & 0.0098 \\
18 & 79 & \texttt{HES.L} & 0.0122 & 48 & 23 & \texttt{SFGmed.L} & 0.0106 & 78 & 32 & \texttt{ACG.R} & 0.0098 \\
19 & 30 & \texttt{INS.R} & 0.0122 & 49 & 70 & \texttt{PCL.R} & 0.0106 & 79 & 48 & \texttt{LING.R} & 0.0097 \\
20 & 19 & \texttt{SMA.L} & 0.0122 & 50 & 69 & \texttt{PCL.L} & 0.0106 & 80 & 50 & \texttt{SOG.R} & 0.0096 \\
21 & 89 & \texttt{ITG.L} & 0.0122 & 51 & 51 & \texttt{MOG.L} & 0.0106 & 81 & 90 & \texttt{ITG.R} & 0.0096 \\
22 & 67 & \texttt{PCUN.L} & 0.0122 & 52 & 07 & \texttt{MFG.L} & 0.0106 & 82 & 33 & \texttt{DCG.L} & 0.0096 \\
23 & 28 & \texttt{REC.R} & 0.0122 & 53 & 78 & \texttt{THA.R} & 0.0105 & 83 & 44 & \texttt{CAL.R} & 0.0096 \\
24 & 42 & \texttt{AMYG.R} & 0.0121 & 54 & 26 & \texttt{ORBsupmed.R} & 0.0105 & 84 & 57 & \texttt{PoCG.L} & 0.0096 \\
25 & 10 & \texttt{ORBmid.R} & 0.0117 & 55 & 58 & \texttt{PoCG.R} & 0.0105 & 85 & 81 & \texttt{STG.L} & 0.0096 \\
26 & 84 & \texttt{TPOsup.R} & 0.0116 & 56 & 66 & \texttt{ANG.R} & 0.0105 & 86 & 04 & \texttt{SFGdor.R} & 0.0095 \\
27 & 63 & \texttt{SMG.L} & 0.0116 & 57 & 15 & \texttt{ORBinf.L} & 0.0105 & 87 & 34 & \texttt{DCG.R} & 0.0094 \\
28 & 38 & \texttt{HIP.R} & 0.0115 & 58 & 01 & \texttt{PreCG.L} & 0.0104 & 88 & 02 & \texttt{PreCG.R} & 0.0093 \\
29 & 76 & \texttt{PAL.R} & 0.0115 & 59 & 16 & \texttt{ORBinf.R} & 0.0104 & 89 & 56 & \texttt{FFG.R} & 0.0093 \\
30 & 25 & \texttt{ORBsupmed.L} & 0.0115 & 60 & 61 & \texttt{IPL.L} & 0.0104 & 90 & 08 & \texttt{MFG.R} & 0.0090 \\
\bottomrule
\end{tabular}
\label{F_11}
\end{table*}

\begin{table*}

\centering
\scriptsize
\setlength{\tabcolsep}{3.2pt}
\renewcommand{\arraystretch}{1.08}
\caption{Attention scores across all ROIs on PPMI (sorted in descending order). The table is sorted in descending order and includes the ROI rank, node index, anatomical label, and the corresponding (normalized) attention score.}
\label{tab:ppmi_attention_all_rois}
\begin{tabular}{r r l S[table-format=1.4]
                r r l S[table-format=1.4]
                r r l S[table-format=1.4]}
\toprule
\multicolumn{4}{c}{Column 1} & \multicolumn{4}{c}{Column 2} & \multicolumn{4}{c}{Column 3} \\
\cmidrule(lr){1-4}\cmidrule(lr){5-8}\cmidrule(lr){9-12}
Rank & Node & ROI & {Score} &
Rank & Node & ROI & {Score} &
Rank & Node & ROI & {Score} \\
\midrule
1  & 36 & \texttt{PCG.R} & 0.0130 & 31 & 65 & \texttt{ANG.L} & 0.0113 & 61 & 22 & \texttt{OLF.R} & 0.0108 \\
2  & 25 & \texttt{ORBsupmed.L} & 0.0123 & 32 & 21 & \texttt{OLF.L} & 0.0113 & 62 & 52 & \texttt{MOG.R} & 0.0108 \\
3  & 62 & \texttt{IPL.R} & 0.0122 & 33 & 76 & \texttt{PAL.R} & 0.0113 & 63 & 51 & \texttt{MOG.L} & 0.0108 \\
4  & 41 & \texttt{AMYG.L} & 0.0122 & 34 & 33 & \texttt{DCG.L} & 0.0113 & 64 & 73 & \texttt{PUT.L} & 0.0108 \\
5  & 14 & \texttt{IFGtriang.R} & 0.0121 & 35 & 43 & \texttt{CAL.L} & 0.0113 & 65 & 69 & \texttt{PCL.L} & 0.0108 \\
6  & 60 & \texttt{SPG.R} & 0.0120 & 36 & 28 & \texttt{REC.R} & 0.0113 & 66 & 64 & \texttt{SMG.R} & 0.0107 \\
7  & 16 & \texttt{ORBinf.R} & 0.0120 & 37 & 24 & \texttt{SFGmed.R} & 0.0112 & 67 & 86 & \texttt{MTG.R} & 0.0107 \\
8  & 78 & \texttt{THA.R} & 0.0119 & 38 & 79 & \texttt{HES.L} & 0.0112 & 68 & 85 & \texttt{MTG.L} & 0.0107 \\
9  & 82 & \texttt{STG.R} & 0.0119 & 39 & 09 & \texttt{ORBmid.L} & 0.0112 & 69 & 04 & \texttt{SFGdor.R} & 0.0106 \\
10 & 12 & \texttt{IFGoperc.R} & 0.0119 & 40 & 59 & \texttt{SPG.L} & 0.0112 & 70 & 54 & \texttt{IOG.R} & 0.0106 \\
11 & 83 & \texttt{TPOsup.L} & 0.0118 & 41 & 55 & \texttt{FFG.L} & 0.0112 & 71 & 49 & \texttt{SOG.L} & 0.0106 \\
12 & 17 & \texttt{ROL.L} & 0.0118 & 42 & 87 & \texttt{TPOmid.L} & 0.0111 & 72 & 77 & \texttt{THA.L} & 0.0106 \\
13 & 32 & \texttt{ACG.R} & 0.0117 & 43 & 47 & \texttt{LING.L} & 0.0111 & 73 & 02 & \texttt{PreCG.R} & 0.0106 \\
14 & 34 & \texttt{DCG.R} & 0.0117 & 44 & 03 & \texttt{SFGdor.L} & 0.0111 & 74 & 10 & \texttt{ORBmid.R} & 0.0105 \\
15 & 81 & \texttt{STG.L} & 0.0117 & 45 & 15 & \texttt{ORBinf.L} & 0.0111 & 75 & 72 & \texttt{CAU.R} & 0.0105 \\
16 & 45 & \texttt{CUN.L} & 0.0117 & 46 & 11 & \texttt{IFGoperc.L} & 0.0111 & 76 & 89 & \texttt{ITG.L} & 0.0105 \\
17 & 26 & \texttt{ORBsupmed.R} & 0.0117 & 47 & 27 & \texttt{REC.L} & 0.0110 & 77 & 20 & \texttt{SMA.R} & 0.0105 \\
18 & 18 & \texttt{ROL.R} & 0.0117 & 48 & 06 & \texttt{ORBsup.R} & 0.0110 & 78 & 84 & \texttt{TPOsup.R} & 0.0105 \\
19 & 31 & \texttt{ACG.L} & 0.0116 & 49 & 19 & \texttt{SMA.L} & 0.0110 & 79 & 53 & \texttt{IOG.L} & 0.0105 \\
20 & 67 & \texttt{PCUN.L} & 0.0116 & 50 & 68 & \texttt{PCUN.R} & 0.0110 & 80 & 46 & \texttt{CUN.R} & 0.0105 \\
21 & 88 & \texttt{TPOmid.R} & 0.0116 & 51 & 70 & \texttt{PCL.R} & 0.0110 & 81 & 63 & \texttt{SMG.L} & 0.0104 \\
22 & 80 & \texttt{HES.R} & 0.0115 & 52 & 61 & \texttt{IPL.L} & 0.0110 & 82 & 75 & \texttt{PAL.L} & 0.0104 \\
23 & 66 & \texttt{ANG.R} & 0.0115 & 53 & 13 & \texttt{IFGtriang.L} & 0.0109 & 83 & 58 & \texttt{PoCG.R} & 0.0103 \\
24 & 30 & \texttt{INS.R} & 0.0115 & 54 & 71 & \texttt{CAU.L} & 0.0109 & 84 & 37 & \texttt{HIP.L} & 0.0103 \\
25 & 90 & \texttt{ITG.R} & 0.0115 & 55 & 08 & \texttt{MFG.R} & 0.0109 & 85 & 01 & \texttt{PreCG.L} & 0.0102 \\
26 & 38 & \texttt{HIP.R} & 0.0114 & 56 & 74 & \texttt{PUT.R} & 0.0109 & 86 & 05 & \texttt{ORBsup.L} & 0.0102 \\
27 & 42 & \texttt{AMYG.R} & 0.0114 & 57 & 35 & \texttt{PCG.L} & 0.0109 & 87 & 56 & \texttt{FFG.R} & 0.0102 \\
28 & 29 & \texttt{INS.L} & 0.0114 & 58 & 23 & \texttt{SFGmed.L} & 0.0109 & 88 & 40 & \texttt{PHG.R} & 0.0101 \\
29 & 57 & \texttt{PoCG.L} & 0.0114 & 59 & 44 & \texttt{CAL.R} & 0.0109 & 89 & 07 & \texttt{MFG.L} & 0.0101 \\
30 & 39 & \texttt{PHG.L} & 0.0113 & 60 & 48 & \texttt{LING.R} & 0.0108 & 90 & 50 & \texttt{SOG.R} & 0.0100 \\
\bottomrule
\end{tabular}
\label{F_12}
\end{table*}

\begin{table*}

\centering
\scriptsize
\setlength{\tabcolsep}{3.2pt}
\renewcommand{\arraystretch}{1.08}
\caption{Attention scores across all ROIs on ABIDE (sorted in descending order). The table is sorted in descending order and includes the ROI rank, node index, anatomical label, and the corresponding (normalized) attention score.}
\label{tab:abide_attention_all_rois}
\begin{tabular}{r r l S[table-format=1.4]
                r r l S[table-format=1.4]
                r r l S[table-format=1.4]}
\toprule
\multicolumn{4}{c}{Column 1} & \multicolumn{4}{c}{Column 2} & \multicolumn{4}{c}{Column 3} \\
\cmidrule(lr){1-4}\cmidrule(lr){5-8}\cmidrule(lr){9-12}
Rank & Node & ROI & {Score} &
Rank & Node & ROI & {Score} &
Rank & Node & ROI & {Score} \\
\midrule
 1 & 90 & \texttt{ITG.R} & 0.0133 & 31 & 39 & \texttt{PHG.L} & 0.0112 & 61 & 61 & \texttt{IPL.L} & 0.0109 \\
 2 & 63 & \texttt{SMG.L} & 0.0123 & 32 & 53 & \texttt{IOG.L} & 0.0112 & 62 & 86 & \texttt{MTG.R} & 0.0109 \\
 3 & 89 & \texttt{ITG.L} & 0.0122 & 33 & 67 & \texttt{PCUN.L} & 0.0112 & 63 & 73 & \texttt{PUT.L} & 0.0108 \\
 4 & 13 & \texttt{IFGtriang.L} & 0.0119 & 34 & 59 & \texttt{SPG.L} & 0.0112 & 64 & 49 & \texttt{SOG.L} & 0.0108 \\
 5 & 71 & \texttt{CAU.L} & 0.0119 & 35 & 37 & \texttt{HIP.L} & 0.0111 & 65 & 51 & \texttt{MOG.L} & 0.0108 \\
 6 & 72 & \texttt{CAU.R} & 0.0118 & 36 & 17 & \texttt{ROL.L} & 0.0111 & 66 & 45 & \texttt{CUN.L} & 0.0108 \\
 7 & 87 & \texttt{TPOmid.L} & 0.0118 & 37 & 44 & \texttt{CAL.R} & 0.0111 & 67 & 36 & \texttt{PCG.R} & 0.0108 \\
 8 & 88 & \texttt{TPOmid.R} & 0.0118 & 38 & 01 & \texttt{PreCG.L} & 0.0111 & 68 & 28 & \texttt{REC.R} & 0.0108 \\
 9 & 32 & \texttt{ACG.R} & 0.0117 & 39 & 05 & \texttt{ORBsup.L} & 0.0111 & 69 & 47 & \texttt{LING.L} & 0.0107 \\
10 & 18 & \texttt{ROL.R} & 0.0117 & 40 & 03 & \texttt{SFGdor.L} & 0.0111 & 70 & 78 & \texttt{THA.R} & 0.0107 \\
11 & 09 & \texttt{ORBmid.L} & 0.0117 & 41 & 79 & \texttt{HES.L} & 0.0111 & 71 & 35 & \texttt{PCG.L} & 0.0107 \\
12 & 56 & \texttt{FFG.R} & 0.0117 & 42 & 76 & \texttt{PAL.R} & 0.0111 & 72 & 64 & \texttt{SMG.R} & 0.0107 \\
13 & 74 & \texttt{PUT.R} & 0.0117 & 43 & 84 & \texttt{TPOsup.R} & 0.0111 & 73 & 20 & \texttt{SMA.R} & 0.0107 \\
14 & 33 & \texttt{DCG.L} & 0.0116 & 44 & 55 & \texttt{FFG.L} & 0.0110 & 74 & 31 & \texttt{ACG.L} & 0.0107 \\
15 & 21 & \texttt{OLF.L} & 0.0116 & 45 & 22 & \texttt{OLF.R} & 0.0110 & 75 & 16 & \texttt{ORBinf.R} & 0.0107 \\
16 & 10 & \texttt{ORBmid.R} & 0.0116 & 46 & 60 & \texttt{SPG.R} & 0.0110 & 76 & 25 & \texttt{ORBsupmed.L} & 0.0107 \\
17 & 83 & \texttt{TPOsup.L} & 0.0115 & 47 & 41 & \texttt{AMYG.L} & 0.0110 & 77 & 11 & \texttt{IFGoperc.L} & 0.0107 \\
18 & 24 & \texttt{SFGmed.R} & 0.0115 & 48 & 02 & \texttt{PreCG.R} & 0.0110 & 78 & 15 & \texttt{ORBinf.L} & 0.0106 \\
19 & 54 & \texttt{IOG.R} & 0.0115 & 49 & 04 & \texttt{SFGdor.R} & 0.0110 & 79 & 23 & \texttt{SFGmed.L} & 0.0105 \\
20 & 14 & \texttt{IFGtriang.R} & 0.0115 & 50 & 65 & \texttt{ANG.L} & 0.0110 & 80 & 68 & \texttt{PCUN.R} & 0.0105 \\
21 & 69 & \texttt{PCL.L} & 0.0115 & 51 & 48 & \texttt{LING.R} & 0.0110 & 81 & 26 & \texttt{ORBsupmed.R} & 0.0105 \\
22 & 19 & \texttt{SMA.L} & 0.0115 & 52 & 43 & \texttt{CAL.L} & 0.0110 & 82 & 12 & \texttt{IFGoperc.R} & 0.0105 \\
23 & 42 & \texttt{AMYG.R} & 0.0115 & 53 & 34 & \texttt{DCG.R} & 0.0110 & 83 & 07 & \texttt{MFG.L} & 0.0104 \\
24 & 27 & \texttt{REC.L} & 0.0114 & 54 & 52 & \texttt{MOG.R} & 0.0110 & 84 & 81 & \texttt{STG.L} & 0.0104 \\
25 & 70 & \texttt{PCL.R} & 0.0114 & 55 & 77 & \texttt{THA.L} & 0.0109 & 85 & 50 & \texttt{SOG.R} & 0.0104 \\
26 & 82 & \texttt{STG.R} & 0.0114 & 56 & 85 & \texttt{MTG.L} & 0.0109 & 86 & 30 & \texttt{INS.R} & 0.0103 \\
27 & 06 & \texttt{ORBsup.R} & 0.0114 & 57 & 66 & \texttt{ANG.R} & 0.0109 & 87 & 75 & \texttt{PAL.L} & 0.0103 \\
28 & 29 & \texttt{INS.L} & 0.0114 & 58 & 62 & \texttt{IPL.R} & 0.0109 & 88 & 46 & \texttt{CUN.R} & 0.0102 \\
29 & 80 & \texttt{HES.R} & 0.0113 & 59 & 08 & \texttt{MFG.R} & 0.0109 & 89 & 57 & \texttt{PoCG.L} & 0.0102 \\
30 & 40 & \texttt{PHG.R} & 0.0113 & 60 & 38 & \texttt{HIP.R} & 0.0109 & 90 & 58 & \texttt{PoCG.R} & 0.0102 \\
\bottomrule
\end{tabular}
\label{F_13}
\end{table*}
\newpage
\section{The Statistical Analyses of The Higher-Order Organizations}
\label{The Statistical Analyses of The Higher-Order Organizations}
To facilitate a comprehensive interpretation of the model’s statistical interpretability analysis, this appendix reports the complete non-parametric test results for the identified higher-order organizations across datasets. Specifically, we provide (i) the three-group comparison results on ADNI using the Kruskal–Wallis test (with effect size $\varepsilon^2$) and post-hoc Mann–Whitney tests, and (ii) the two-group comparison results on PPMI and ABIDE using two-sided Mann–Whitney $U$ tests. For all pairwise comparisons, we additionally report effect sizes quantified by the rank-biserial correlation ($r_{rb}$). When multiple post-hoc tests are performed, the corresponding $p$-values are adjusted using the Benjamini–Hochberg FDR correction. The results mentioned above are shown in the Tables  \ref{F_14}, \ref{F_15}, \ref{F_16}.
\begin{table*}[t]
\centering
\scriptsize
\setlength{\tabcolsep}{4.0pt}
\renewcommand{\arraystretch}{1.12}
\caption{Non-parametric group comparisons for top quadruplet interaction patterns on ADNI (CN vs MCI vs AD).
Overall differences are assessed by the Kruskal--Wallis test (statistic $H$) with effect size $\varepsilon^2$.
Post-hoc comparisons use two-sided Mann--Whitney $U$ tests with effect size given by rank-biserial correlation ($r_{rb}$).
Post-hoc $p$-values are reported after Benjamini--Hochberg FDR correction. Bold indicates $p<0.05$.}
\label{tab:adni_nonparam_quadruplets_updown}

\begin{tabular}{
>{\ttfamily\raggedright\arraybackslash}p{5.2cm}
S[table-format=1.5]
S[table-format=2.4]
S[table-format=1.4]
S[table-format=1.5]
l
}
\toprule
& \multicolumn{3}{c}{Overall (CN vs MCI vs AD)} &
\multicolumn{2}{c}{AD vs CN} \\
\cmidrule(lr){2-4}\cmidrule(lr){5-6}
ROIs &
{$p$} & {$H$} & {$\varepsilon^2$} &
{$p_{\mathrm{FDR}}$} & {$U\,(r_{rb})$} \\
\midrule
CAU.R, HIP.L, PHG.R, AMYG.L &
0.0875 & 4.8721 & 0.0115 &
{\bfseries 0.0444} & \num{3.58e3} (\num{0.1708}) \\
CAU.R, PHG.R, AMYG.L, TPOmid.R &
{\bfseries 0.00277} & 11.7766 & 0.0391 &
0.0914 & \num{3.70e3} (\num{0.1435}) \\
CAU.R, PHG.R, OLF.L, TPOmid.R &
{\bfseries 0.0153} & 8.3607 & 0.0254 &
0.9470 & \num{4.30e3} (\num{0.0058}) \\
CAU.R, AMYG.L, OLF.L, TPOmid.R &
{\bfseries 0.00561} & 10.3670 & 0.0335 &
0.1930 & \num{3.84e3} (\num{0.1106}) \\
CAU.R, HIP.L, OLF.L, TPOmid.R &
{\bfseries 0.000777} & 14.3196 & 0.0493 &
0.5280 & \num{4.09e3} (\num{0.0537}) \\
CAU.R, HIP.L, AMYG.L, TPOmid.R &
{\bfseries 0.000497} & 15.2153 & 0.0529 &
{\bfseries 0.0471} & \num{3.59e3} (\num{0.1687}) \\
\bottomrule
\end{tabular}
\end{table*}
\begin{table*}[t]
\centering
\ContinuedFloat
\scriptsize
\setlength{\tabcolsep}{4.0pt}
\renewcommand{\arraystretch}{1.12}

\captionsetup{list=no}
\caption[]{\textbf{Table~\ref{tab:adni_nonparam_quadruplets_updown} (continued).}}

\begin{tabular}{
>{\ttfamily\raggedright\arraybackslash}p{5.2cm}
S[table-format=1.5]
l
S[table-format=1.5]
l
}
\toprule
& \multicolumn{2}{c}{AD vs MCI} &
\multicolumn{2}{c}{MCI vs CN} \\
\cmidrule(lr){2-3}\cmidrule(lr){4-5}
ROIs &
{$p_{\mathrm{FDR}}$} & {$U\,(r_{rb})$} &
{$p_{\mathrm{FDR}}$} & {$U\,(r_{rb})$} \\
\midrule
CAU.R, HIP.L, PHG.R, AMYG.L &
0.6920 & \num{2.90e3} (\num{0.0371}) &
0.0947 & \num{3.71e3} (\num{-0.1542}) \\
CAU.R, PHG.R, AMYG.L, TPOmid.R &
{\bfseries 0.0436} & \num{3.58e3} (\num{-0.1887}) &
{\bfseries 0.000877} & \num{4.20e3} (\num{-0.3069}) \\
CAU.R, PHG.R, OLF.L, TPOmid.R &
{\bfseries 0.00943} & \num{3.75e3} (\num{-0.2428}) &
{\bfseries 0.0111} & \num{3.97e3} (\num{-0.2341}) \\
CAU.R, AMYG.L, OLF.L, TPOmid.R &
{\bfseries 0.0308} & \num{3.62e3} (\num{-0.2020}) &
{\bfseries 0.00190} & \num{4.14e3} (\num{-0.2864}) \\
CAU.R, HIP.L, OLF.L, TPOmid.R &
{\bfseries 0.00176} & \num{3.90e3} (\num{0.0537}) &
{\bfseries 0.000455} & \num{4.26e3} (\num{-0.3234}) \\
CAU.R, HIP.L, AMYG.L, TPOmid.R &
{\bfseries 0.0252} & \num{3.65e3} (\num{-0.2093}) &
{\bfseries 0.000478} & \num{4.34e3} (\num{-0.3483}) \\
\bottomrule
\end{tabular}
\label{F_14}
\end{table*}

\begin{table*}[t]
\centering
\scriptsize
\setlength{\tabcolsep}{4.0pt}
\renewcommand{\arraystretch}{1.12}
\caption{Non-parametric group comparisons for top quadruplet interaction patterns on PPMI (two-group setting).
Two-sided Mann--Whitney $U$ tests are reported together with effect sizes measured by rank-biserial correlation ($r_{rb}$).
Bold indicates $p<0.05$.}
\label{tab:ppmi_nonparam_quadruplets}
\begin{tabular}{
l
>{\ttfamily\raggedright\arraybackslash}p{6.2cm}
S[table-format=1.4]
S[table-format=1.2e+3]
S[table-format=1.4]
}
\toprule
Nodes & ROIs & {$p$} & {$U$} & {$r_{rb}$} \\
\midrule
36, 25, 62, 78 &
PCG.R, ORBsupmed.L, IPL.R, THA.R &
{\bfseries 0.0432} & \num{1.72e3} & 0.2282 \\
62, 41, 14, 16 &
IPL.R, AMYG.L, IFGtriang.R, ORBinf.R &
{\bfseries 0.0413} & \num{1.08e3} & -0.2303 \\
36, 25, 62, 12 &
PCG.R, ORBsupmed.L, IPL.R, IFGoperc.R &
0.0572 & \num{1.71e3} & 0.2147 \\
25, 62, 78, 12 &
ORBsupmed.L, IPL.R, THA.R, IFGoperc.R &
0.0779 & \num{1.68e3} & 0.1990 \\
36, 62, 78, 12 &
PCG.R, IPL.R, THA.R, IFGtriang.L &
0.0801 & \num{1.68e3} & 0.1976 \\
36, 25, 78, 12 &
PCG.R, ORBsupmed.L, THA.R, IFGtriang.L &
0.0868 & \num{1.68e3} & 0.1933 \\
\bottomrule
\end{tabular}
\label{F_15}
\end{table*}

\begin{table*}[t]
\centering
\scriptsize
\setlength{\tabcolsep}{4.0pt}
\renewcommand{\arraystretch}{1.12}
\caption{Non-parametric group comparisons for top quadruplet interaction patterns on ABIDE (two-group setting).
Two-sided Mann--Whitney $U$ tests are reported together with effect sizes measured by rank-biserial correlation ($r_{rb}$).
Bold indicates $p<0.05$.}
\label{tab:abide_nonparam_quadruplets}
\begin{tabular}{
l
>{\ttfamily\raggedright\arraybackslash}p{6.4cm}
S[table-format=1.6]
S[table-format=1.2e+5]
S[table-format=1.4]
}
\toprule
Nodes & ROIs & {$p$} & {$U$} & {$r_{rb}$} \\
\midrule
72, 87, 88, 18 &
CAU.R, TPOmid.L, TPOmid.R, ROL.R &
{\bfseries 0.000209} & \num{1.49e5} & 0.1339 \\
90, 63, 71, 88 &
ITG.R, SMG.L, CAU.L, TPOmid.R &
{\bfseries 0.000334} & \num{1.48e5} & 0.1296 \\
90, 63, 87, 88 &
ITG.R, SMG.L, TPOmid.L, TPOmid.R &
{\bfseries 0.000993} & \num{1.47e5} & 0.1189 \\
90, 89, 87, 18 &
ITG.R, ITG.L, TPOmid.L, ROL.R &
{\bfseries 0.000557} & \num{1.47e5} & 0.1247 \\
90, 72, 88, 18 &
ITG.R, CAU.R, TPOmid.R, ROL.R &
{\bfseries 0.000230} & \num{1.48e5} & 0.1331 \\
90, 71, 87, 18 &
ITG.R, CAU.L, TPOmid.L, ROL.R &
{\bfseries 0.000435} & \num{1.48e5} & 0.1271 \\
\bottomrule
\end{tabular}
\label{F_16}
\end{table*}




\bibliographystyle{elsarticle-harv.bst} 
\bibliography{Reference.bib}

\end{document}